# A Generic Workforce Scheduling and Routing Problem with the Vehicle Sharing and Drop-off and Pick-up Policies

Ömer Öztürkoğlu[1], Gökberk Özsakallı[2]

[1,2]Department of Business Administration, Yasar University, Bornova, Izmir, 35100, Turkey

**Abstract**

This paper introduces a new generic problem to the literature of Workforce Scheduling and Routing Problem. In this problem, multiple workers are assigned to a shared vehicle based on their qualifications and customer demands, and then the route is formed so that a traveler may be dropped off and picked up later to minimize total flow time. We introduced a mixed-integer linear programming model for the problem. To solve the problem, an Adaptive Large Neighborhood Search (ALNS) algorithm was developed with problem-specific heuristics and a decomposition-based constructive upper bound algorithm (UBA). To analyze the impact of newly introduced policies, service area, difficulty of service, distribution of care, and number of demand nodes type instance characteristics are considered. The empirical analyses showed that the ALNS algorithm presents solutions with up to 35% less total flow time than the UBA. The implementation of the proposed drop-off and pick-up (DP) and vehicle sharing policies present up to 24% decrease in total flow time or provide savings on the total cost of service especially when the demand nodes are located in small areas like in urban areas.

*Keywords*: routing; workforce scheduling; vehicle sharing; drop-off and pick-up.

## 1. Introduction

This study introduces a new problem to the literature of Workforce Scheduling and Routing Problem (WSRP) with Vehicle Sharing (VS). WSRP refers to the scenarios in which the employees of the industry have to travel across different places through diverse modes of transportation to perform the assigned field operations. The term vehicle and path sharing refer to a way of transportation in which individual travelers who have a rather similar route and schedules that share a vehicle and a cost for their travel.

Some real-life scenarios of WSRP include the visits of caregivers to patients' homes for providing their required treatment (Cheng and Rich, 1998; Eveborn et al., 2006; Liu et al., 2013), companies with technical services that perform repair tasks at various customers with technicians (Chen et al., 2016; Cordeau et al., 2010), energy distribution companies perform maintenance operations at different sites (Çakırgil et al., 2020), night security patrols at various premises within the area (Alfares and Alzahrani, 2020; Misir et al., 2011), etc. All of these studies involve the scheduling and routing of the workers to ensure that they reach the places on time where the tasks assigned to them need to be performed.

[1] omer.ozturkoglu@yasar.edu.tr (Ö. Öztürkoğlu)
[2] *Corresponding author*: gokberk.ozsakalli@gmail.com (G. Özsakallı)

During the COVID-19 pandemic, we observed that truck drivers delivering parcels in Turkey shared their vehicles with a courier in order to complete the deliveries on time due to the increasing number of e-commerce. After the driver dropped off the courier in a location with a high density of customers, such as high-rise offices or residential buildings, or in pedestrian areas, he/she went to another location for delivery. The driver then returned and picked up the courier to travel to other customer locations or back to the depot. We made similar observations in home healthcare services during the pandemic period. Due to the limited number of vehicles equipped with special healthcare equipment and the protection of employees, we observed that two healthcare teams, each consisting of two employees, shared a single van to visit people with confirmed COVID-19 for delivering their medication or people with suspected COVID-19 for getting a nasal swab COVID-19 test as part of the contact tracing program. The van driver had to commute between patients to drop off and pick up crews while visiting patient locations. These observations therefore led us to consider the potentials of vehicle sharing in the workforce scheduling and routing problem. Although the generic WSRP we introduced in this study can be applied in any field, we have defined it specifically in the context of the home healthcare system due to the complexity and growing popularity of home healthcare. Hence, the problem is called Home Healthcare Scheduling and Routing Problem with Vehicle Sharing (HHSRP-VS), hereafter.

Nowadays, home healthcare (HHC) services have been becoming popular services in the world due to the growing aging population, increasing congestion, and medical costs in hospitals. Even in the current pandemic (COVID-19), HHC workers are indispensable workers at the front line in many countries. In addition to some routine and private health services required by patients who either cannot go or do not prefer to go to hospitals due to the COVID-19 global pandemic, the HHC workers are providing various healthcare services to COVID-19 patients who were either treated at home or discharged from hospitals and need healthcare at their homes within isolation. All HHC workers are responsible for performing multiple tasks in a planning period. The numbers of tasks might sometimes be higher than the available workers. Therefore, scheduling a workforce with the availability of transportation on proper routes is essential in many cases to ensure that all of the required tasks are completed within the stipulated time. The mismanagement of scheduling and routing HHC workers or the lack of resources such as appropriate vehicles with proper equipment may cause unserved, delayed, and unsatisfied patients, high working times of caregivers, and high travel and service costs. For instance, according to Weerdt and Baratta (2015), HHC workers spent 43% of their working time in their vehicles for travel in the U.S. and they are at high risk for motor vehicle-related injuries or losing their productivity because of driving their vehicles. In this study, the implementation of the proposed policies, efficient solutions to the scheduling and routing of HHC workers and allowing them to share the same vehicle with a driver increase patients' satisfaction, HHC workers' safety and satisfaction, better service and cost savings to institutions. Hence, this research aims to convey the following main contributions to literature and practice.

● To the best of our knowledge, none of the existing studies in the literature has studied the Workforce Scheduling and Routing Problem with Vehicle Sharing. Thus, a new generic problem has been introduced to the literature.

● In the context of home healthcare, a vehicle sharing policy has been introduced that allows multiple independent caregivers to be assigned a shared vehicle to visit patients.

● A new policy called "drop-off and pick-up (DP)" has been proposed to reduce patient and caregiver waiting times and increase vehicle utilization, thus reducing total caregivers' flow time.

● The Mixed Integer Linear Programming (MILP) model of the new HHSRP-VS has been introduced.

● A constructive matheuristic upper-bound algorithm and an Adaptive Large Neighborhood Search (ALNS) optimization algorithm with problem-specific local search heuristics have been proposed to develop efficient solutions.

● In-depth insights on the implementation of vehicle sharing and DP policies in the HHSRP have been presented after empirical analyses. To gain insights, we also developed two reduced models of the HHSRP called HHSRP-M and HHSRP-STD. The HHSRP-M consists only of vehicle sharing policy, not DP policy. The HHSRP-STD is a typical HHSRP where every caregiver travels in their own vehicle.

The structure of this paper is designed in a way that we begin by providing a review of the existing literature in Section 2. Section 3 introduces the HHSRP-VS and its formulation as a MILP model. The details of the proposed algorithms for solving the HHSRP-VS are given in Section 4, whereas Section 5 focuses on the computational experiments, findings, and insights. The paper is ended by some concluding remarks in Section 6, where we also comment on the possible future direction of the study.

## 2. Literature Review

This section aims to provide the most relevant studies in WSRP and VS literature within two subsections. The presented discussions aim to elucidate the limitations of the existing WSRP and the scope of the introduced problem.

### 2.1 Workforce scheduling and routing problem (WSRP)

WSRP refers to a class of problems in which employees must perform a set of tasks at different locations using various transportation modes. WSRP has a broad range of application areas such as home healthcare, production and/or maintenance, forestry and, telecommunications. In the literature, the problems that are studied in WSRP are classified as home healthcare scheduling and routing problem (HHSRP), technician and task scheduling problem (TTSP), and manpower allocation problem (MAP) (Castillo-Salazar et al., 2016). HHSRP involves the scheduling and routing of caregivers (nurses, general physicians, therapists, etc.) to perform healthcare related tasks at patients (Eveborn et al., 2006). TTSP and MAP are the scheduling of technicians or servicemen to execute a set of installation and maintenance services (Cordeau et al., 2010; Lim et al., 2004).

According to Castillo-Salazar et al. (2016), HHSRP, TTSP, and MAP are considered as WSRP because they contain a combination of scheduling of employees and vehicle routing problem. In addition, the authors also state that the pick-up and delivery problem (PDP) cannot be considered as a WSRP because in terms of time no significant "work" is done within the premises of the customer in the PDP. In PDP, vehicles must transport loads directly from one location to another (Savelsbergh and Sol, 1995). HHSRP, TTSP, and MAP share a similar set of constraints and objective functions. For a more detailed discussion on WSRP, we refer readers to the survey published by Castillo-Salazar et al. (2016). In addition, we refer readers to the recent literature reviews of Cissé et al. (2017) and Fikar and Hirsch (2017) for further reading on HHSRP. In the following, the main characteristics of the constraints are discussed.

Home healthcare (HHC) service providers offer a wide range of services to a person in need. For this purpose, staff (caregivers) with different qualifications like general physicians, therapists, nurses, social workers, dietitians, psychologists, etc. are employed by the providers. To travel between different patients' locations and HHC centers, these caregivers either have their own vehicles or use a vehicle provided by the HHC service providers. The patients require certain types of services, which must be performed by suitably qualified staff members. One of the main problems facing the management of the HHC service provider is the daily scheduling and routing task (Borchani et al., 2019). This scheduling and routing determine which visit will be performed, by which caregivers, and by which vehicles (if personal vehicle is not the case).

Time windows of customers are one of the most common constraints in the WSRP literature. The time windows can be imposed as either soft time windows or hard time windows. In the hard time windows, arriving at the customers after the latest possible arrival time is not allowed (Cheng and Rich, 1998; Bredström and Rönnqvist, 2008; Zamorano and Stolletz, 2017; Mathlouthi et al., 2021; etc.). On the other hand, soft time windows allow late arrivals with a penalty cost (Trautsamwieser and Hirsch, 2011; Mankowska et al., 2014).

Maximum working time regulation for employees defines the maximum amount of time a worker is allowed to work in a shift, which is usually implemented by setting a time window. Maximum working time constraint can be a hard constraint (Rasmussen et al.,2012; Trautsamwieser and Hirsch, 2014; Xie et al., 2017; Frifita et al., 2017; Pereira et al., 2020; Guastaroba et al., 2021) with a penalty cost for unvisited or missed patients, or a soft constraint which allows overtime with an additional cost in the objective function (Cheng and Rich, 1998; Trautsamwieser and Hirsch, 2011; Rest and Hirsch, 2016). In this study, a hard maximum working time constraint is considered.

The planning horizon determines the period in which scheduling and routing plan is made. According to Cisse et al. (2017), the length of the plan depends on the availability of demand information. In the literature, mostly a single-day planning horizon is considered due to the quality of the information (Eveborn et al., 2006; Redjem and Marcon, 2016; Rest and Hirsch, 2016; Pinheiro et al., 2016). On the other hand, studies that consider multi-period planning horizon generally consider one week planning horizon (Begur et al., 1997;

Trautsamwieser and Hirsch, 2014; Qin et al., 2015; Wirnitzer et al., 2016; Chen et al., 2017; Pereira et al., 2020). In addition, continuity of care is an important service quality indicator for HHSRP environment. Continuity of care constraint is often considered in multi-period HHSRP in which patients receive service by the same caregiver and should be visited during the same time (Wirnitzer et al., 2016). In HHSRP context, this constraint builds a relationship of confidence between the patient and caregiver (Cisse et al., 2017).

Qualification or skills of employees is one of the most used characteristics of WSRP. This is because the service providers must match the customers’ varying requests with the employees’ expertise. In the literature qualification of workers is considered in two ways. In the first one, more than one qualification can be assigned to a worker (Eveborn et al. 2006; Rasmussen et al. 2012; Pillac et al., 2013; Bard et al. 2014; Mankowska et al. 2014; Liu et al. 2017; Mathlouthi et al., 2021). Whereas in the second approach, it can be determined based on the hierarchical level of qualification (Cordeau et al., 2010; Nickel et al. 2012; Rest and Hirsch 2016; Trautsamwieser and Hirsch 2011) in which each customer's demand has a minimum required level of qualification and each worker is associated with some qualification level. In our study, we considered the first type of qualification approach in which the demand and the skill should be matched.

Temporal dependency constraints define relations between different tasks to be performed at customers. In general, two types of temporal dependency constraints are considered: synchronization and precedence. Synchronized/shared services require visits of different customers at the same time (Eveborn et al., 2006; Issabakhsh et al., 2018; Frifita et al., 2017). The precedence constraint prioritizes multiple services Cordeau et al., 2010; Liu et al., 2013; Bard et al., 2014), which are very necessary in the case when one of the two services of a customer should be performed before the other. Several studies considered both type of temporal dependency constraints (Bredström and Rönnqvist, 2008; Rasmussen et al., 2012; Mankowska et al., 2014; Pereira et al., 2020).

Finally, teaming is an important feature for TTSP and MAP because of the nature of the services to be performed. In TTSP and MAP, tasks require a different set of skills that can be performed by different workers, in general. Thus, assigning different workers to a team is often applicable in these problems (Li et al., 2005; Cordeau et al., 2010; Kovacs et al., 2012; Pereiraa et al., 2020). Although these studies require multiple workers, they travel as a single entity to perform the same task. Thus, according to the best of our knowledge, it can be said that no study considers routing of multiple independent workers in a shared vehicle. We see this feature in the field of vehicle and ride-sharing literature.

## 2.2 Vehicle and path sharing problem

Different terms have been used for vehicle and path sharing in the literature such as, dial-a-ride, ride-matching, ride-sharing, taxi-sharing, carpooling, etc. All of these problems are the special cases of pickup and delivery problems (PDP) (Agatz et al., 2012) that are the special case of VRP, therefore have NP-Hard complexity. Recker (1995) proposed an interesting extension of pickup and delivery for household activity pattern problems which involves ride-sharing along with vehicle-switching options. The dial-a-ride problem (DARP), which was firstly proposed by Cordeau and Laporte (2003), focuses on planning the routes of vehicles and their schedules for the transportation of multiple passengers who request to travel from a specific place to some destination. Baldacci et al. (2004) interpret the carpooling problem (CPP) based on DARP. Lin et al. (2012) formulated a taxi ride-sharing system based on DARP for picking up and dropping the customers off at different locations. Moreover, dynamic ride-sharing problems intend to bring together travelers with similar itineraries and time schedules on short notice (Agatz et al., 2012). For a comprehensive review of the literature on DARP and ride-sharing, we refer to Molenbruch et al. (2017) and Mourad et al. (2019), respectively.

In all of the abovementioned studies, mainly commuters or their vehicles are routed. Unlike in WSRP or HHSRP, there is no such constraint or requirement as a service time of a job, workers' qualifications, worker-to-task (caregiver-to-patient) assignment, etc. However, in this study, more than one passenger is assigned to a shared vehicle according to demand and their characteristics, and then a route is formed such that a passenger can be dropped off and picked up later to minimize total route length. Thus, the problem introduced in this study combines the characteristics of WSRP and VS.

## 3. Home Healthcare Scheduling and Routing Problem with Vehicle Sharing (HHSRP-VS)

Two distinct features of the HHSRP-VS are (1) multiple independent caregivers, who can provide independent services to different patients, traveling in the same vehicle and (2) a drop-off and pick-up (DP) policy implemented on a trip. Hence, the main objective of this research is to answer the following research questions.

i. How effective are variations of the proposed caregiver swap heuristic used in the proposed ALNS algorithm? (Section 5.3)

ii. How effective and efficient are the proposed upper bound and ALNS algorithms compared to each other and to CPLEX solutions? (Section 5.4)

iii. How effective is the DP policy in HHSRP-VS? Under which circumstances does DP policy provide savings on total flow time? (Section 5.5)

iv. How effective is vehicle sharing policy in HHSRP-VS? Under which circumstances does vehicle sharing with DP policy provide savings on total flow time and total service cost? (Section 5.6)

In this section, we first provide a formal description of the problem and then the mixed-integer linear programming (MILP) formulation. The HHSRP-VS is defined as the complete directed graph $G = (V, A)$, where $V = \{0,1, \dots, n, n+1, \dots, 2n, 2n+1\}$ is the set of all nodes in the graph and $A = \{(i,j): i, j \in V, i \neq j\}$ is the set of arcs between every pair of nodes excluding arcs between the same nodes. $n$ is the number of patients, and nodes 0 and $2n+1$ indicate the same beginning and ending HHC. The set of caregivers and illnesses (types of cares) are denoted by $L = \{1,2,3,\dots,l\}$ and $S = \{1,2,3,\dots,s\}$, where $l$ and $s$ are the numbers of available caregivers and illnesses, respectively. Last, $K = \{1,2,3,\dots,k\}$ indicates the set of $k$ vehicles.

The sub-tour elimination constraint, which is one of the typical constraints in VRP, cannot be enforced due to the implementation of the DP policy that leaves a caregiver to a node and then picks up from the same node. Therefore, we proposed a two-layer modeling approach to easily adapt DP policy and avoid sub-tour elimination. In this approach, $V_1 = \{1,2,3,\dots,n\}$ is defined as the set of original patient nodes, and $V_2 = \{n+1, n+2, \dots, 2n\}$ is the set of their dummy nodes. For clarification, the two-layer approach is demonstrated in Figure 1.

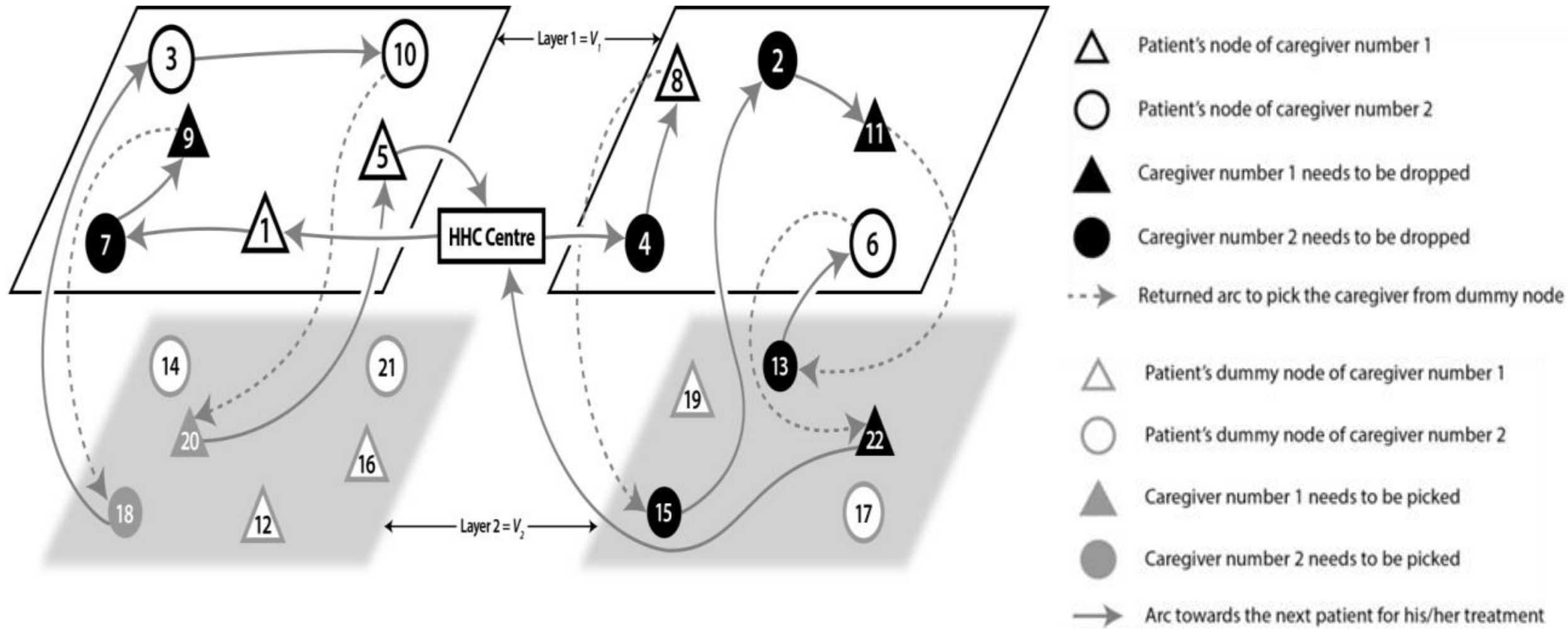

Figure 1. The two-layer representation model of the HHSRP-VS problem. The left and right routes describe the vehicles 1 and 2's routes, respectively.

As seen in Figure 1, the original patient nodes are placed in the first layer, and their projections are in the second layer. A vehicle can visit a dummy node in the second layer for picking up a caregiver if and only if its original patient node was visited before and the requested caregiver was dropped off at that node. Hence, this approach could also be considered as one of the methodological contributions of this study. For example, suppose that there are 11 patients, 4 distinct caregivers, and 2 identical vehicles. Each vehicle carries two distinct caregivers. Suppose that the assigned caregiver_1 in vehicle_1 will treat patients (1, 5, 9) while caregiver_2 in the same vehicle is assigned to patients (3, 7, 10). Similarly, suppose that caregivers_1 and _2 in vehicle_2 are assigned to treat patients (8, 11) and (2, 4, 6), respectively. Suppose that the optimal routes of vehicle_1 and vehicle_2 are computed as {0, 1, 7, 9, 18, 3, 10, 20, 5, 23} and {0, 4, 8, 15, 2, 11, 13, 6, 22, 23}, respectively, in which {0} and {23} indicate the start and end nodes of the single HHC. Hence, vehicle_1 starts its travel with two caregivers and visits directly to patient 1 where only caregiver_1 provides care. The vehicle and caregiver_2 wait for caregiver_1 to finish his/her service. Next, they travel to patient 7 where caregiver_2 is being dropped off to serve. The vehicle goes to patient 9 only with caregiver_1. After the vehicle drops off

caregiver_1 at patient 9, it goes back to patient 7 (dummy node 18) to pick up caregiver_2 empty. After caregiver_2 serves patients 3 and 10 respectively, the vehicle with caregiver_2 goes back to patient 9 (dummy node 20) to pick up caregiver_1. Last, before the vehicle goes back to the HHC with both caregivers, it visits patient 5 who requested caregiver_2. A similar route could also be seen for vehicle 2 on the right diagram in Figure 1.

Table 1. Model parameters and decision variables

| **Parameters** | **Definition** |
|---|---|
| $t_{ij}$ | Nonnegative and deterministic travel time between nodes $i$ and $j$, $(i,j) \in A$ |
| $d_{is}$ | 1, if patient $i \in V_1$ needs to be treated for illness $s \in S$; 0, otherwise (patients' demands) |
| $q_{ls}$ | 1, if caregiver $l \in L$ is qualified to treat illness $s \in S$; 0, otherwise (caregivers' qualifications) |
| $p_{is}$ | Deterministic service time for treating illness $s \in S$ of patient $i \in V_1$ |
| $c$ | Maximum number of workers allowed to be transferred by a vehicle in addition to the dedicated driver to the vehicle |
| $wTime$ | Maximum daily working time (hour) of caregivers |
| $unv$ | Penalty cost incurred if a patient is not visited |
| **Variables** | **Definition** |
| $x_{i,j,k}$ | 1, if vehicle $k \in K$ travels through node $i \in V$ to node $j \in V$; 0, otherwise. |
| $z_{i,j,k,l}$ | 1, if caregiver $l \in L$ travels through node $i \in V$ to node $j \in V$ with vehicle $k \in K$; 0, otherwise. |
| $y_{i,k,l}$ | 1, if vehicle $k \in K$ drops caregiver $l \in L$ off at node $i \in V_1$ such that the caregiver should be picked up at node $i + n$ ; 0, otherwise |
| $\alpha_{i,k,l,s}$ | 1, if caregiver $l \in L$ visits patient $i \in V_1$ with vehicle $k \in K$ to treat illness $s \in S$; 0, otherwise |
| $u_i$ | 1, if patient node $i \in V_1$ is not visited; 0, otherwise. |
| $hw_{i,l}$ | Waiting time of caregiver $l \in L$ at node $i \in V$ |
| $w_{i,k}$ | Waiting time of vehicle $k \in K$ in node $i \in V$ |
| $av_{i,k}$ | Arrival time of vehicle $k \in K$ to node $i \in V$ |
| $ah_{i,l}$ | Arrival time of caregiver $l \in L$ to node $i \in V$ |
| $dv_{i,k}$ | Departure time of vehicle $k \in K$ from node $i \in V$ |
| $dh_{i,l}$ | Departure time of caregiver $l \in L$ from node $i \in V$ |
| **Auxiliary Variables** | **Definition** |
| $\psi_{i,k,l}$ | 1, if vehicle $k \in K$ visits patient $i \in V_1$ with caregiver $l \in L$ and the caregiver $l$ is not dropped off at the patient (either serves the patient or waits for the assigned caregiver in the vehicle);<br>0, if either vehicle $k$ visits patient $i$ but does not wait for the service by caregiver $l$ (dropped off) or it never visits $i$. |
| $\gamma_{i,k,l}$ | 1, if caregiver $l \in L$ is assigned to vehicle $k \in K$ and patient $i \in V_1$ for serving the patients' illness $s \in S$ and the caregiver $l$ is not dropped off at the patient (the vehicle waits for the service completion);<br>0, if either caregiver $l$ is assigned but dropped off by the vehicle $k$ or caregiver $l$ is not assigned to patient $i$. |

The HHSRP-VS consists of determining a set of $k$ routes of the minimal total flow time of the caregivers to serve the patients under several constraints and modeling assumptions. Each vehicle consists of a fixed number of caregivers. There is a single HHC where vehicles and caregivers start and end their travel. The skills of the available caregivers are eligible to meet patients' requirements. Each patient requires only one type of service (treatment). Hence a patient is forced to be visited by a single caregiver and a vehicle for the treatment. Every available vehicle and caregiver are required to be utilized. If any of the caregivers need to be dropped at any of his/her assigned patient's home for the treatment, he/she must be picked up from the same patient's home by the same vehicle before either returning to the depot or visiting the next patient who requested the same caregiver. Caregivers are not allowed to work overtime. Table 1 lists the parameters and decision variables that we define to formulate the mixed-integer linear programming model of the HHSRP-VS given below.

**HHSRP-VS MILP Model:**

$$\min \sum_{l\in L} ah_{(2n+1),l} + \sum_{i\in V_1} u_i * unv \qquad (1)$$

$$\sum_{i\in V} \sum_{k\in K} x_{i,j,k} + u_j = 1, \qquad j \in V_1 \qquad (2)$$

$$\sum_{j\in V_1} x_{0,j,k} = 1, \qquad k \in K \qquad (3)$$

$$\sum_{j\in V_1} \sum_{k\in K} z_{0,j,k,l} = 1, \qquad l \in L \qquad (4)$$

$$\sum_{j\in V} x_{i,j,k} - \sum_{j\in V} x_{j,i,k} = 0, \qquad i \in V_1 \cup V_2, k \in K \qquad (5)$$

$$z_{i,j,k,l} \le x_{i,j,k}, \qquad i,j \in V, k \in K, l \in L \qquad (6)$$

$$\sum_{l\in L} \sum_{k\in K} \alpha_{i,k,l,s} * q_{l,s} + u_i = d_{i,s}, \qquad i \in V_1, s \in S \qquad (7)$$

$$\sum_{i\in V} z_{i,j,k,l} \ge \sum_{s\in S} \alpha_{i,k,l,s}, \qquad j \in V_1, k \in K, l \in L \qquad (8)$$

$$y_{j,k,l} \le \sum_{i\in V} x_{i,j,k}, \qquad j \in V_1, k \in K, l \in L \qquad (9)$$

$$\sum_{i\in V} x_{i,(j+n),k} \ge y_{j,k,l}, \qquad j \in V_1, k \in K, l \in L \qquad (10)$$

$$\sum_{i\in V} x_{i,(j+n),k} \le \sum_{l\in L} y_{j,k,l}, \qquad j \in V_1, k \in K \qquad (11)$$

$$\sum_{j\in V} \sum_{k\in K} z_{i,j,k,l} \le 1 - \sum_{k\in K} y_{i,k,l}, \qquad i \in V_1, l \in L \qquad (12)$$

$$\sum_{j\in V} z_{j,i,k,l} = \sum_{j\in V} z_{i,j,k,l} + y_{i,k,l}, \qquad i \in V_1, k \in K, l \in L \qquad (13)$$

$$\sum_{j\in V} z_{j,i,k,l} + y_{(i-n),k,l} = \sum_{j\in V} z_{i,j,k,l}, \qquad i \in V_2, k \in K, l \in L \qquad (14)$$

$$av_{j,k} \ge dv_{i,k} + t_{i,j} - \left(1 - x_{i,j,k}\right) * M_1, \qquad i,j \in V, k \in K \qquad (15)$$

$$ah_{j,l} \ge dh_{i,l} + t_{i,j} - \left(1 - z_{i,j,k,l}\right) * M_1, \qquad i,j \in V, k \in K, l \in L \qquad (16)$$

$$w_{i,k} \ge \sum_{s\in S} \sum_{l\in L} \gamma_{i,k,l} * p_{i,s} - \sum_{j\in V} x_{j,(i+n),k} * M_2, \qquad i \in V_1, k \in K \qquad (17)$$

$$w_{i,k} \ge av_{(i-n),k} + \sum_{s\in S} \sum_{l\in L} \alpha_{(i-n),k,l,s} * p_{(i-n),s} - av_{i,k} - \left(1 - \sum_{j\in V} x_{j,i,k}\right) * M_1, \qquad i \in V_2, k \in K \qquad (18)$$

$$hw_{i,l} \ge \sum_{s\in S} \sum_{l'\in L\setminus\{l\}} \gamma_{i,k,l'} * p_{i,s} - \left(1 - \sum_{k\in K} \psi_{i,k,l}\right) * M_2, \qquad i \in V_1, l \in L \qquad (19)$$

$$hw_{i,l} \ge av_{i,k} - \sum_{s\in S} \sum_{l'\in L} \alpha_{(i-n),k,l',s} * p_{(i-n),s} - av_{(i-n),k} - \left(1 - y_{(i-n),k,l}\right) * M_1, \qquad i \in V_2, k \in K, l \in L \qquad (20)$$

$$dv_{i,k} \ge av_{i,k} + w_{i,k}, \qquad i \in V, k \in K \qquad (21)$$

$$dh_{i,l} \ge ah_{i,l} + hw_{i,l}, \qquad i \in V, l \in L \qquad (22)$$

$$av_{i,k} + +\left(1 - \sum_{i\in V} z_{j,i,k,l}\right) * M_1 \ge ah_{i,l}, \qquad i \in V, k \in K, l \in L \qquad (23)$$

$$av_{i,k} \le ah_{i,l} + \left(1 - \sum_{i\in V} z_{j,i,k,l}\right) * M_1, \qquad i \in V, k \in K, l \in L \qquad (24)$$

$$dv_{i,k} + \left(1 - \sum_{j\in V} z_{i,j,k,l}\right) * M_1 \ge dh_{i,l}, \qquad i \in V, k \in K, l \in L \qquad (25)$$

$$dv_{i,k} \le dh_{i,l} + \left(1 - \sum_{j\in V} z_{i,j,k,l}\right) * M_1, \qquad i \in V, k \in K, l \in L \qquad (26)$$

$$\psi_{i,k,l} = \sum_{j\in V} z_{j,i,k,l} - y_{i,k,l}, \qquad i \in V_1, k \in K, l \in L \qquad (27)$$

$$\gamma_{i,k,l} = \sum_{s\in S} \alpha_{i,k,l,s} - y_{i,k,l}, \qquad i \in V_1, k \in K, l \in L \qquad (28)$$

$$\sum_{j\in V_1}\sum_{l\in L} z_{0,j,k,l} = c, \qquad k \in K \qquad (29)$$

$$ah_{(2n+1),l} \leq wTime, \qquad l \in L \qquad (30)$$

$$av_{(2n+1),k} \leq wTime, \qquad k \in K \qquad (31)$$

$$x_{i,j,k}; z_{i,j,k,l}; y_{i,k,l}; \alpha_{i,k,l,s}; u_i; \psi_{i,k,l}; \gamma_{i,k,l} \in \{0,1\}, hw_{i,l};\ w_{i,k};\ av_{i,k};\ ah_{i,l};\ dv_{i,k};\ dh_{i,l} \geq 0 \qquad (32)$$

The model aims to minimize total flow times of the caregivers until returning to the HHC, which includes their service, travel and waiting times, and the total penalty cost of unvisited patients, if exist. Constraint set (2) guarantees that each patient node is visited exactly once or unvisited. Constraint sets (3) and (4) ensure that every available vehicle and caregiver must depart from the HHC. Moreover, a caregiver must leave with a single vehicle. Constraint set (5) maintains flow conversation in the network. Constraint set (6) aims to relate the travel of vehicles with caregivers. Hence, a caregiver can travel from nodes $i$ to $j$ if his/her assigned vehicle goes that route. Constraint (7) assures that only a single and qualified caregiver is assigned to treat the illness of a patient if being served. Constraint set (8) ensures that the vehicle must visit a patient if the assigned caregiver to the patient is also assigned to that vehicle. Constraint set (9) maintains that a caregiver could be dropped off at the patient node if the assigned vehicle visits that node. Next, constraint set (10) ensures that the vehicle must visit the patient's dummy node if a caregiver was dropped off at the patient node. Although the terms $y_{j,k,l}$ in (9) and (10) could be replaced by $\sum_{l\in L} y_{j,k,l}$, they might be preferred due to computational sakes (tighter constraints). Constraint set (11) guarantees that the dummy node cannot be visited if none of the caregivers were dropped at the patient. Constraint set (12) ensures that when a caregiver is dropped off at a patient node, that caregiver is not allowed to leave the same patient node, instead, the caregiver must leave from its dummy node due to the two-layer approach. For the sake of the caregivers' flow conservation in the network, constraint sets (13) and (14) guarantee that if a caregiver goes to a patient node, that caregiver must depart from either the same patient node or its dummy node only with the initially assigned vehicle.

Constraint sets (15) through (28) are required to track the arrival, departure, and waiting times of both caregivers and vehicles. Because both caregivers and vehicles can take different actions throughout the route, their synchronization should be maintained for the accuracy of the flow. Therefore, a vehicle or a caregiver may have to wait for the other for the continuity of the travel. These waiting times could either appear at the first (original patient node) or the second layer (dummy node). These could be briefly explained as in the following.

- A vehicle could wait at the patient node $i \in V_1$ (first layer) if and only if the vehicle decides to wait for the caregiver until the completion of the service at the patient. The duration of the waiting time is the amount of service time for the patients' requirements (constraint (17)).
- A vehicle could wait at the dummy node $i \in V_2$ (second layer) when the vehicle returns to the patient node to pick up the dropped-off caregiver and the caregiver has not completed the service yet. The duration of the waiting time is the difference between the completion time of the caregivers' service and the arrival time of the vehicle to the dummy node (constraint (18)).

- A caregiver in a vehicle, if there is, could wait at the patient node $i \in V_1$ while the assigned caregiver serves the patient, and the vehicle waits for the completion of the service. The duration of the waiting time is equal to the amount of service time at the patient (constraint (19)).
- The assigned caregiver could wait at the dummy node $i \in V_2$ if the vehicle returns later than the caregivers' service completion. The waiting time is the difference between the arrival time of the vehicle to the patient and the completion time of the caregivers' service (constraint (20)).
- A caregiver could also wait at the dummy node $i \in V_2$, if he/she returns to the patient with the vehicle to pick up the dropped-off caregiver earlier than the assigned caregivers' service completion. This waiting time was not explicitly computed because it is handled by both constraint (20) and the synchronization constraints (23)-(26).

The synchronization constraints (23) through (26) aim to synchronize the arrival and departures of a vehicle and the caregivers within it throughout the nodes. Moreover, constraints (15) and (16) computes the arrival time of vehicles and caregivers to the nodes, respectively. Constraints (21) and (22) determine the departure times of vehicles and caregivers from the nodes, respectively. Constraints (27) and (28) are used to indicate whether a vehicle takes a caregiver to a patient and waits for the service and whether the assigned caregiver to a patient is not dropped off by the vehicle, respectively (see Table 1 for the description of the respective auxiliary variables). Constraint (29) ensures the capacity of vehicles in terms of the number of caregivers. Constraints (30) and (31) specifies the maximum working time of caregivers and vehicles. Even though one of the constraints (30) or (31) is enough, we embedded both to tighten the model. For the same concern, $M_1$ and $M_2$ could be replaced with tighter $wTime$ and $\sum_{i \in V_1} \sum_{s \in S} p_{i,s}$ values, respectively.

On an individual basis, the complexity of WSRP (Algethami et al., 2019) and VPSP (Bei and Zhang, 2018) are both NP-Hard. As seen in its mathematical model, HHSRP-VS can be considered as difficult as these problems, as it can be treated as a combination of WSRP and VPSP in the context of HHSRP. Thus, the following sections elucidate our attempts to tighten the model by an upper bound and obtain close-optimal solutions using a metaheuristic algorithm.

### 3.1 The proposed upper-bound algorithm (UBA): A clustering-based matheuristic approach

In literature, various HHSRP problems were solved by decomposition-based algorithms in two stages in which the patients are either clustered or partitioned based on caregivers' skills, geographical proximity, or some other characteristics at the first stage. Next, the reduced problem is solved as a variant of TSP or VRP using MILP or heuristics (Rasmussen et al., 2012, Hiermann et al., 2015, and Erdem and Bulkan, 2017). Similar to these studies, we presented a multi-stage decomposition-based matheuristic algorithm to develop feasible solutions without a DP policy to HHSRP-VS.

In the first stage, the caregiver clusters are formed based on the geographical closeness of the patients similar to the K-means clustering algorithm. For the problem under consideration, the clusters have been formed based on the caregivers' skills and qualifications and the patients' demands and their locations. Hence, the caregiver clusters are created in such a way that the demand of every patient should be matched to the caregiver's skill(s). This solution is feasible for HHSRP-VS for the following reasons. (a) All caregivers have been utilized. (b) Qualification constraint is satisfied. Furthermore, to deal with working time constraints, the algorithm aims to evenly distribute the total service workload of each caregiver. The detailed pseudocode of the first stage of the proposed algorithm can be seen in Algorithm A.1 in Appendix A.

In the second stage, caregiver clusters are assigned to vehicles according to the capacity of vehicles. The idea behind this caregivers-vehicle assignment is that the caregivers can visit their patients through the same vehicle who are living close to each other. The caregiver clusters which are closer to each other are assigned to the same vehicle. As a result of the second stage, we determine which caregivers are assigned to which vehicle and which patients are going to be treated by which caregiver.

In the third stage, the problem is turned into a multiple TSP where the optimal route of each vehicle is computed sequentially using the IBM ILOG CPLEX 12.6 solver without considering the maximum working time constraints. After the optimal route of the first vehicle is obtained, if the solution exceeds the working time limit, the costliest patients on the route are removed until the working time constraint is maintained. The removed patients of the vehicle are added to the patient list of the next qualifying vehicle. After solving the last vehicle, if there are still removed patients, they are considered as the unvisited patients. The detailed pseudocode of the second and third stages of the proposed algorithm can be seen in Algorithm A.2 in Appendix A.

In the final stage, we applied an inter-route relocate operator to look for better solutions and a repair function to reduce the unvisited number of patients at the end. Since the optimal route of each vehicle is obtained in the previous stage, changing a patient's position on the same route does not improve the solution. Thus, the inter-route relocate operator removes a patient from its vehicle and inserts it in another qualifying vehicle. The feasibility of the solution is conserved at each iteration by satisfying qualification and maximum working time constraints. Finally, a repair function with a greedy heuristic is applied to assign the unvisited patients to vehicles whose total working time is less than the maximum working time. The pseudocode of the inter-route relocate operator and repair function can be seen in Algorithm A.3 in Appendix A.

In order to narrow the solution space and obtain feasible integer solutions in a short time, we used the solution ($\mu$) obtained by the proposed mathematical algorithm as the upper bound for the original mathematical model of HHSRP-VS. This solution can be used as an upper bound, because it does not include the drop-off and pick-up policy but satisfies all other constraints. For this purpose, equations (33) and (34) can be added as valid upper bound inequalities to the HHSRP-VS MILP model. Moreover, we also used solutions provided by the upper-bound algorithm to analyze the effectiveness of the ALNS-VS algorithm developed in the following sections.

$$\sum_{k\in K} av_{(2n+1),k} + \sum_{i\in V_1} u_i * unv \leq \mu, \qquad (33)$$

$$\sum_{l\in L} ah_{(2n+1),l} + \sum_{i\in V_1} u_i * unv \leq \mu, \qquad (34)$$

## 4. The Proposed ALNS-VS Algorithm

This section presents the Adaptive Large Neighborhood Search (ALNS) heuristic algorithm developed to solve HHSRP-VS. The ALNS algorithm was first proposed by Ropke and Pisinger (2006a) by extending the Large Neighborhood Search (LNS) algorithm proposed by Shaw (1997). Unlike the LNS, the ALNS heuristic involves a variety of removal and insertion heuristics which help in obtaining a good quality solution. As far as the other heuristics are concerned, ALNS is relatively fast and has been successfully implemented in different variants of VRP. Therefore, we preferred to adapt ALNS to our problem. To address the DP policy of the problem under consideration, two local search heuristics have been introduced within the proposed ALNS-VS algorithm, the details of which are discussed below.

The algorithm in our study starts with finding an initial solution, and then, at each iteration, chooses a random removal heuristic to deconstruct the existing solution to some degree, and an insertion heuristic to repair it differently. Through these destroy and repair operations, a new neighborhood solution is obtained at the end of each iteration and is adopted as the current solution for the next iteration. These processes continue until the stopping criteria are met. The pseudocode of the ALNS-VS algorithm proposed in this study is given in Algorithm 1. The details of the algorithm with the parameter definitions are explained in the following subsections.

**Algorithm 1.** Pseudocode of the proposed ALNS-VS algorithm.

**input:** Set of removal heuristics $\Psi$, set of insertion heuristics $Z$, initial temperature $T$, cooling rate $c$, solution update iteration $\omega$, caregiver swap iteration $\varphi$, the iteration of the last best-found solution $t_{best}$
**output:** A feasible solution $x_{best}$

**Generate** *an initial solution* $x_{init}$ *using the Regret-3 with noise insertion heuristic*
**Set** *iteration counter* $t$ *with an initial value of* $t \leftarrow 1$ *and* $t_{best} \leftarrow 1$
**Set** *the initial values,* $x_{curr} \leftarrow x_{best} \leftarrow x_{init}$
**repeat**
    **if** $(t - t_{best} \,\%\, \omega = 0)$ **then**
        $x_{curr} \leftarrow x_{best}$
        $\Psi^* \leftarrow Random.Removal$
  **else** **Select** *a removal heuristic,* $\Psi^* \in \Psi$
    **Let** $x_{new}$ *be a partial solution after applying* $\Psi^*$ *to* $x_{curr}$
    **if** $(t \,\%\, \varphi = 0)$ **then**
        **Apply** *the caregiver swap heuristic algorithm to* $x_{new}$
    **Select** *an insertion heuristic* $Z^* \in Z$ *to* $x_{new}$ *to generate* $x_{new}$
    **Let** $x_{new}$ *be a new solution after applying* $Z^*$ *to* $x_{new}$
    **Apply** *the drop-off and pick-up local search heuristic (Algorithm B.1 in Appendix B) to improve* $x_{new}$
    **Apply** *the repair function (Algorithm B.2 in Appendix B) to generate a new feasible solution* $x_{new}$ *and determine the unvisited patients*
    **if** $f(x_{new}) < f(x_{curr})$ **then**
        $x_{curr} \leftarrow x_{new}$

**Algorithm 1.** Pseudocode of the proposed ALNS-VS algorithm.

$f(x_{curr}) \leftarrow f(x_{new})$
**else**
Let $v \leftarrow e^{-\frac{f(x_{new})-f(x_{curr})}{T}}$
**Generate** *a random number* $\epsilon \in [0,1]$
**if** $\epsilon < v$ **then**
$x_{curr} \leftarrow x_{new}$
$f(x_{curr}) \leftarrow f(x_{new})$
**if** $f(x_{new}) < f(x_{best})$ **then**
$f(x_{best}) \leftarrow f(x_{new})$
**Update** *the temperature,* $T \leftarrow c * T$
**Update** *the iteration counter,* $t \leftarrow t + 1$
**until** *the predetermined number of iterations reached and the predetermined number of iterations without any further improvement found in* $x_{best}$

A vehicle route is represented by a list that includes the depot (HHC) and patients' nodes. For example, let $\pi_k$ represents the route of vehicle $k$, $\pi_k = \{v_0, v_1, \dots, v_i, \dots, v_{2n}, v_{2n+1}\}$, where $v_0$ and $v_{2n+1}$ represent the depot nodes, and the rest represents patient and dummy nodes. Throughout the algorithm, these depot nodes remain fixed at their positions and all the other nodes can only be placed between them. If a dummy of any patient's node is present at a route of a vehicle, the patient's node is placed in the same route before its dummy node. In addition to that, since there is more than one caregiver in a vehicle, a caregiver list is also created for each vehicle: $\pi_k^l$ represents the caregivers that are traveling with vehicle $k$. To keep track of the visits through the nodes due to the DP policy, $v_i^l$ is used to record the list of caregivers who left patient node $i$ with the vehicle.

To generate an initial solution at the beginning of the ALNS-VS algorithm, all of the patient nodes are placed in the request bank $R$ and all of the dummy nodes are placed in the dummy request bank $\underline{R}$. Caregivers are assigned to vehicles at random until the capacity of each vehicle is filled. At every successive step, the *Regret-3 heuristic with noise* algorithm (see Section 4.1) is applied to all the vehicles in parallel by assigning each patient $i \in V_1$ from $R$ to one of the existing fleet vehicles. This process is repeated until all patients are assigned to one of the available vehicles $k \in K$ or the remaining patients cannot be assigned to any vehicle due to the maximum working time of caregivers. Once a feasible solution is found, it is set to the current solution and the best solution.

### 4.1 The set of removal and insertion heuristics

At each iteration, a removal heuristic algorithm, sometimes called a destruction heuristic, is applied to remove a predetermined number of patients $q$ from the current solution $x_{curr}$. These removed patients are placed to the request bank $R$. In general, $q$ is set to an integer number at the beginning of the algorithm in the literature (Ropke and Pisinger, 2006a, 2006b; Pisinger and Ropke, 2007). However, varying $q$ may be preferred due to the exploration and exploitation capabilities of the heuristics. For this, linearly decreasing function of $q$ is used in our algorithm to explore the solution space more at the beginning of the iterations than the latter (Öztürkoğlu et al., 2014; Öztürkoğlu and Hoser, 2019). At each iteration, $q$ is computed by $q = \xi * n - n * (\xi - \upsilon)\frac{t}{\theta}$, where $n$ is the total number of patients, $\xi$ and $\upsilon$ are the parameters that control the maximum and minimum number of removed patients, $t$ is the current iteration and $\theta$ is the maximum number of iterations. As Pisinger and Ropke (2007) suggested that the minimum number of removed elements from a solution should be 10% ($\upsilon = 0.1$) of the total number of elements. We adopted five removal heuristics for the ALNS-VS algorithm. The random, worst, route and Shaw removal heuristics algorithms were adapted from Ropke and Pisinger (2006a, 2006b) and Pisinger and Ropke (2007). The last heuristic called "dummy node removal" was developed specifically for our problem.

**Random Removal:** This heuristic algorithm randomly removes $q$ patients from the current solution $x_{curr}$ and adding them to the request bank R.

**Worst Removal:** This heuristic algorithm selects $q$ costliest patients in terms of distance from the current solution. The heuristic removes the selected patient $i \in x_{curr}$ from the current solution $x_{curr}$ and adds them to $R$. After removing patient $i$, the cost of the $x_{curr}$ is calculated as $f_{-i}$, whereas the cost of $i$ can be calculated as $\Delta f_i = f(x_{curr}) - f_{-i}$.

**Shaw Removal:** The main objective of this heuristic algorithm is to remove the most similar patients in terms of their locations and service times. The heuristic starts with selecting a random patient $i \in x_{curr}$ and adding it to the request bank $R$. The similarity measures ($d_{ij}$) between the selected patient $i$ and the rest of the patients $j \in \frac{x_{curr}}{\{i\}}$ in the solution $x_{curr}$ are calculated by $d_{ij} = \alpha * t_{i,j} + \beta * (|p_i - p_j|)$. In our problem, the lower the $d_{ij}$ is the higher the similarity. The most similar patient $j^*$ is selected and added to $R$ such that $j^* = argmin_{j \in x_{curr}} d_{ij}$, where $\alpha$ and $\beta$ are the shaw parameters, $p_i$ and $p_j$ are the service times of patients $i$ and $j$, and $t_{i,j}$ is the travel time between patient nodes $i$ and $j$. This heuristic algorithm is iteratively applied $q$ times to determine the removed patients such that the patient has the maximum similarity measure with the last removed patient.

**Route Removal:** This heuristic algorithm randomly selects a route of a vehicle $\underline{\upsilon}$ from $\upsilon$ (a set of routes of vehicles in $x_{curr}$), removes all the patients from it, and adds them to the $R$. The idea of route removal is to redesign the route to minimize the travel time by diversifying the search.

**Dummy Node Removal:** Within the scope of the drop-off and pick-up policy, patients' dummy nodes are also included in the $x_{curr}$. This heuristic algorithm removed $q$ dummy nodes, where $q$ is a random integer number between $\sigma * d$ and $\phi * d$. $d$ is the total number of dummy nodes in the current solution. $\sigma$ and $\phi$ are the minimum and maximum ratios of the dummy removal constant, respectively. Since the drop-off and pick-up local search algorithm is applied at each iteration, removing a large number of dummy nodes from the solution helps to explore different solutions. Therefore, $\sigma$ and $\phi$ are set to 0.5 and 0.8, respectively. Finally, the removed dummy nodes are added to dummy request bank $\underline{R}$.

In our algorithm we applied the most commonly used insertion heuristics in literature: Greedy and Regret-k, specifically Regret-2 and Regret-3, and their noise versions (Ropke and Pisinger, 2006a, 2006b). The selected insert heuristic assigns patients in R to vehicles to improve the objective function value. In our implementation, all routes of the vehicles were evaluated simultaneously, not sequentially, in order to increase the solution quality.

**Greedy Insertion:** All of the patients from $R$ are assigned to all possible positions of the routes $v$ of caregivers and an insertion cost is calculated for each position through $\Delta_{i,k,j}^{l} = t_{i,k} + t_{k,j} - t_{i,j}$ for $i, j = 1, \dots, n$ and $i \neq j$. In this process, only feasible assignments are considered. After insertion cost is calculated for all patients, the patient with the least insertion cost is assigned to determine the position of the route of the vehicle. This process continues until all patients are assigned to a route or no more insertion is possible. Since at each iteration only one route of a vehicle is changed, the insertion cost for the other routes does not need to be recalculated. This idea improves the computation time for all of the insertion heuristics.

**Greedy Insertion with Noise:** The idea of adding noise to the insertion cost is to provide randomization to the search process. This is done by considering the degree of freedom in determining the best location for a node. The steps of greedy insertion heuristic remain the same while the new insertion cost is calculated by $\Delta_{i,k,j}^{l} = t_{i,k} + t_{k,j} - t_{i,j} + t_{max} * \mu * \varepsilon$, where $t_{max}$ is the maximum time between patients, $\mu$ is the noise parameter which is used for the diversification and set to 0.1, and $\varepsilon$ is a random number between [-1,1].

**Regret-k Insertion**: Regret-k heuristics are proposed by Potvin and Rousseau (1993). Contrary to the greedy insertion, this heuristic considers the $k$ best positions (depending on choice) instead of the best one. Patients are assigned to positions to maximize the regret cost ($cost_i^k$) which is computed as the difference between $k$ best position costs $\Delta_{i,m,j}^{l}$ i.e., change in objective value by inserting patient $m$ between patients $i$ and $j$ in route $v$. In this respect, the greedy heuristic can be seen as a regret-1 heuristic. The proposed algorithm considers regret-2 and regret-3 insertions.

**Regret-k Insertion with Noise:** The steps of this insertion heuristic are similar to the regret-k insertion heuristics but use the same cost function as discussed in the greedy insertion with noise.

### 4.2 Drop-off and Pick-up (DP) local search heuristic algorithm

In addition to the removal and insertion heuristics, we developed a special local search heuristic algorithm to determine whether a caregiver should be dropped off or waited by the vehicle at a patient node during his/her service. This local search is applied to the solution obtained after the removal and insertion heuristics are completed. Because of the complexity of the drop-off decision and its effect on the whole tour, we developed a smart approach for deciding drop-off and pick-up. Hence, this approach consists of the following features. The pseudocode of the DP local search heuristic is also given in Algorithm 2.

- The effect of DP on the route length is computed for decision-making.
- The position where the patient is being picked up is determined.

● When more than one caregiver is eligible to treat a patient, the approach also decides the best caregiver who is being dropped off at the patient node (if applied).

● The feasibility of the solution is maintained when DP is decided to be applied. For example, if a caregiver $l$ is decided to be dropped off at patient node $i$ and to be picked up before visiting patient $j$, then the patients between $i + 1$ and $j$ i.e., $[i + 1, j]$ in the existing route are guaranteed to be treated by the other caregivers in the vehicle.

**Algorithm 2.** The framework of the drop-off and pick-up local search heuristic algorithm.

---

**input:** Route of vehicle $\pi_k, k \in K$ in the $x_{curr}$, and the saving of dropping the caregiver $l$ off at the patient $i$ and picking up after visiting the node $j$ by vehicle $k$ i.e., $dp^l_{i,j,k}$
**output:** A new feasible solution $x_{new}$

---

**forall** route of vehicle in $\pi_k, k \in K$
  **do**
    **forall** *caregivers* $l \in \pi^l_k$ *in vehicle* $k$
      **forall** *patients* $i \in \pi_k$
        **forall** patients $j \in \pi_k$ that are being visited after patient $i$
          *drop caregiver* $l$ *off at patient* $i$*, then add patient* $i$*'s dummy node after patient* $j$*, and calculate* $dp^l_{i,j,\pi_k}$ *using equation (35) in section 4.2.*
        **end for**
      **end for**
    **end for**
    **Update** $\pi_k$ *with the drop-off and picking-up decision where the maximum positive* $dp^l_{i,j,\pi_k}$ *occurs if it exists. Then, update the current solution.*
  **while** $dp^l_{i,j,\pi_k} > 0$
**end for**
**return** A new improved feasible solution $x_{new} \leftarrow x_{curr}$

---

The amount of savings on one caregivers' flow time in a vehicle $dp^l_{i,j,\pi_k}$ is calculated using equation (35). This saving, if exist, is induced by dropping caregiver $l$ off at a patient $i$ and picking up after node $j$ in route $\pi_k$ of vehicle $k$. Note that the notations were previously defined in Table 1.

$$dp^l_{i,j,\pi_k} = \left(t_{j,(j+1)} + p_{is}\right) - \left(t_{j,(i+n)} + \left(0, av_{ik} + p_{is} - \left(dv_{jk} + t_{j,i+n}\right)\right) + t_{i+n,(j+1)}\right), \tag{35}$$

The first term indicates the maximum amount of savings induced by the elimination of waiting for caregiver $l$ at patient $i$ with a duration of $p_{is}$ and the removal of travel from nodes $j$ to $j + 1$ in the existing route because the dummy node $i + n$ must be visited after node $j$. The second term specifies the amount of increase in flow time due to drop-off. Hence, the first and the last terms indicate additional travels from nodes $j$ to $i + n$ and $i + n$ to $j + 1$. The second term includes the waiting time of caregiver $l$, who was dropped off at patient $i$, if the vehicle arrives at the dummy node later than the service completion time of the caregiver If the saving is greater than zero, then the drop-off and pick-up decision is made.

### 4.3 Caregiver swap heuristic algorithm

After the caregivers were randomly assigned to the vehicles in the initial solution, any of the applied insertion or removal heuristics do not change their assignments. To search for the whole solution space and look for better caregiver-vehicle-patient assignments, we proposed the caregiver swap heuristic algorithm. The proposed heuristic was inspired by the pheromone concept used in the Ant Colony Optimization (ACO) algorithm introduced by Colorni et al. (1991), in which pheromone is used to trace the most commonly visited paths to find the food source by ants.

In this heuristic, the pheromone density $\tau_{i,j}(t),\ i,j \in L$ is shared among all the caregivers at iteration $t$. Initially, the pheromone values are equal for all of the caregivers. Then the pheromone density between the caregivers in the same vehicle increases depending on their contributions to the solution. The higher the pheromone density among the caregivers, the more likely they are to be assigned to the same vehicle. In addition to the contribution to the solution, the pheromone density is also affected by the heuristic (visibility) value $\eta_{i,j}$ $i,j \in L$. Similar to Öztürkoğlu (2017), the pheromone density for all caregivers that are in the same vehicle is updated by $\tau_{i,j}(t) = (1-\rho) * \tau_{i,j}(t-1) + \rho * \left(\frac{\eta_{i,j}}{f_{best}(t-1)}\right),\ \forall(i,j) \in L$, where $\rho$ denotes the evaporation coefficient whose values lie between $(0,1)$, and $f_{best}$ is the best objective function value found until iteration $t-1$. Thus, the probability of assigning caregivers into the same vehicle is calculated by $\mathrm{P}_{i,j}(t) = \frac{\tau_{i,j}(t-1)}{\sum_{(k,l)\in L}\tau_{k,l}(t-1)}$, $(i,j) \in L$. Hence, the tournament selection procedure is performed to determine the other caregiver(s) who share the vehicle with the previously assigned caregivers. This process continues until all the caregivers are assigned to their respective vehicles according to the vehicle capacity.

For the proposed caregiver swap heuristic, we consider two different visibility values $\eta_{i,j}$ based on the common and unique number of patients that can or cannot be treated by caregivers $i$ and $j$. The idea behind common patients is that the possibility of a continuum of treating other patients by a caregiver increases after his/her colleague(s) is dropped off at a patient. Hence, this may efficiently use the DP policy by reducing the number of returns. On the other side, in the case of unique patients, the algorithm may cluster closer patients that have distinct requirements to each other. Hence, the closer distinct patients may increase the chance of using DP policy where a vehicle may go forth and back between them due to drop-off and pick-up. In section 5, we investigate if there is any difference between the common and unique visibility heuristics, as well as the effect of caregiver swap heuristic on the quality of the solution.

### 4.4 The repair function and termination criteria

Through the application of removal and insertion heuristics and DP local search heuristic algorithm, we only consider qualification and demand constraints. The total working time constraint is ignored to explore a high variety of solutions and to speed up the heuristics by avoiding recomputing the flow time after every insertion.

Therefore, a repair function is proposed to restore the feasibility of the solutions after all insertion and the DP heuristic are applied. Thus, a new feasible solution is being directed to the next iteration if accepted.

The proposed repair function given in Algorithm B.1 in Appendix B guarantees the feasibility of the solutions within two steps. In the first step, the algorithm removes the most time-consuming patient nodes from the routes to ensure the total working time limit of the vehicles. In the second step, the algorithm tries to assign the removed patients to the vehicles whose total working time is less than the max working time by applying the greedy heuristic.

After obtaining a new feasible solution, it is accepted as a current solution for the next iteration if the cost of the new solution is less than that of the current solution. Similar to the concept of the simulated annealing approach, the worse solution than the current solution may also be accepted with some probability to increase the exploration capability of the algorithm. This probability is calculated as $e^{-(f(x_{new})-f(x_{curr}))/T}$, where $f(x_{new})$ and $f(x_{curr})$ are the costs of the new and the current solutions, respectively. $T$ is the temperature having the cooling rate $c$ between $0 < c < 1$.

We also adopted an approach for updating the current solution to stay away from trapping into a local optimal solution and to increase the exploration capability of the algorithm. In our approach, if there is no improvement in the best solution in the last ω iterations, we apply random removal and Regret-3 insertion heuristics to the best-found solution so far and consider the resulting new solution as a current solution for the rest of the iteration. Last, the ALNS-VS algorithm is terminated when both the maximum number of iterations $\theta$ is reached and there is no improvement in the last $\underline{\theta}$ iterations. If the best solution is improved in the last $\underline{\theta}$ iterations, other $\underline{\theta}$ iterations are added to the search process until the condition is met.

## 5. Computational Experiments and Results

This section comprises computational experiments that were conducted to assess the performance of the proposed ALNS-VS algorithm, answer the research questions defined in section 3, and derive in-depth insights. The UBA and ALNS-VS algorithms described in the previous sections were implemented in C#. IBM ILOG CPLEX 12.6 optimization solver was used to solve the HHSRP-VS MILP model. CPLEX was run both with standard settings, the aim of which is to find a proven optimal solution, and with various settings that considered various MIP strategies. All of the experiments were conducted on a computer with a 2.50 GHz Intel Core i7-6500U CPU and 16 GB of RAM. Furthermore, the CPLEX solver was limited to 6 hours to obtain solutions.

### 5.1 Problem instances

A new set of problem instances are generated to evaluate the performance of the proposed algorithms and analyze the characteristics of the HHSRP-VS and the proposed policies. The features of the generated problem instances are described in Table 2. We considered 10 to 100 patients with 4 to 12 caregivers in a defined service area. The qualifications for the caregivers were obtained from Liu et al. (2017)'s data set. The patients were randomly located in a circular continuous area that is described by four different radiuses. The reason for considering areas of different sizes is to investigate the effect of area, or in other words, travel distance, on the effectiveness of proposed policies. In each instance class, the single HHC is located at the center of the area.

Table 2. Characteristics of the generated problem instances.

| **Feature** | **Description** |
|---|---|
| Number of patients and available caregivers (4 levels) | 10 patients with 4 caregivers; 30 patients with 4 caregivers; 50 patients with 6 caregivers; 100 patients with 12 caregivers |
| Service area radius (4 levels) | 10, 20, 30, and 40 minutes |
| Patients' Demand distributions (3 levels) | Level 0: 80/15/5: 80% basic, 15% moderate, 5% difficult.<br>Level 1: 60/30/10: 60% basic, 30% moderate, 10% difficult.<br>Level 2: 50/30/20: 50% basic, 30% moderate, 20% difficult. |
| Patients service requirement (illness) and corresponding service times (3 levels) | Basic: mean of 10 and standard deviation of 2.5 minutes<br>Moderate: mean of 20 and standard deviation of 5 minutes<br>Difficult: mean of 30 and standard deviation of 7.5 minutes |
| Capacity of a vehicle | 2 caregivers |

We defined three different types of care requirements concerning their difficulty level as basic, moderate, and difficult care. The reason for considering services with different difficulties is to investigate the effect of service time on the effectiveness of the proposed policies. The service time for each type of care was assumed to be normally distributed by three different means and standard deviations for care. Hence, we considered three different levels of patients' service demand distributions in the instance classes in which the first, second, and third numbers indicate the percentages of the patients that require basic, moderate, and difficult care, respectively. For example, the instance class h100_40_0 indicates that there is a total of 100 patients with 12 caregivers, the patients are randomly distributed in a circular area with a radius of 40 minutes, the demand distribution level is 0 indicating 80%, 15% and the remaining 5% of the patients require basic, moderate and difficult cares (80/15/5), respectively. Last, we generated five instances in each instance class by changing only the locations (coordinates) of the patients. Thus, an instance is described by the last index. For example, the last indices in h100_40_0_1 and h100_40_0_2 indicate that these are the first and the second instances in the instance class h100_40_0 such that only the locations (coordinates) of the patients are differentiated. Thus, there are 48 instance classes and 240 instances in total. Finally, each instance was run in five replications differentiated by five seeds used in a random number generator which resulted in 1200 runs.

### 5.2 Parameter tuning

We considered Ropke and Pisinger (2006a, 2006b)'s settings for many of the fundamental parameters used in a typical ALNS algorithm such as $\theta$, $\alpha$, $\beta$, $\mu$, $\upsilon$ and $c$ as 25000, 0.3, 0.1, 0.1, 0.1, and 0.99975, respectively. We took the additional number of iterations $(\underline{\theta})$ 250 as 10% of $\theta$ (Öztürkoğlu and Mağara, 2019). Furthermore, $\sigma$ and $\phi$ were assumed to be 0.5 and 0.8, respectively as explained in section 4.1. Last, we conducted a full factorial experimental design for the remaining parameters specific to our ALNS-VS algorithm which are update solution iteration ($\omega$), caregiver swap iteration ($\varphi$), maximum remove parameter ($\xi$) and evaporation rate ($\rho$).

After the preliminary experiments, 6 levels were defined for $\omega$ with ranging from 250 to 1500 with a step size of 250. $\varphi$ has 7 levels with ranging from 50 to 200 with a step size 25. Thus, we aimed to prevent the algorithm from being trapped in a local optimal solution due to the lack of proper caregiver assignment. $\xi$ has 5 levels such as $\xi \in \{0.4, 0.5, 0.6, 0.7, 0.8\}$ and finally $\rho$ has 5 levels as $\rho \in \{0.75, 0.8, 0.85, 0.9, 0.95\}$. In the literature, different values were used for evaporation rate which range from 0.75 to 0.95 (Fuellerer et al., 2009; Yu et al., 2009). In total, we had (6x7x5x5) 1.050 settings for parameters and performed 21.000 runs with 4 different tuning instances and 5 replications obtained by five seeds in a random number generator. To compare the solutions in the experiment, we normalized the best-found solutions for each run: $RPD_{i,j} = \left(\frac{f_{i,j} - f_{min,j}}{f_{min,j}}\right) * 100$, where $RPD_{i,j}$ is the normalized best-found solution of run $i$ for instance $j$; $f_{i,j}$ is the best-found solution by the algorithm in setting-replication pair $i$ for instance $j$, and $f_{min,j}$ is the best solution for instance $j$. These instances comprise of 30 patients, 4 caregivers, 2 vehicles with a capacity of 2 caregivers, area with a radius of 30 minutes. The experiment was conducted on Minitab 19 Statistical Software. The ANOVA and the Response Optimization tests were conducted to investigate the effects of parameters on the quality of the solutions with 95% confidence level. The tests' results showed that the optimal setting is ($\omega$, $\varphi$, $\xi$, $\rho$) = (250,100, 0.5, 0.95). See Table C.1 and Figure C.1 in Appendix C for the details of the test results. Hence, Table 3 summarizes the parameter settings used for the proposed ALNS-VS algorithm for the computational experiments.

Table 3. The parameter settings are used in the proposed ALNS-VS algorithm.

| **Parameters** | **Values** | **Parameters** | **Values** |
|---|---|---|---|
| Total number of iterations ($\theta$) | 25000 | First Shaw parameter ($\alpha$) | 0.3 |
| Additional iteration ($\underline{\theta}$) | 2500 | Second Shaw parameter ($\beta$) | 0.1 |
| Solution update iteration number ($\omega$) | 250 | Minimum dummy remove parameter | 0.5 |
| Number of caregiver swap | 100 | Maximum dummy remove parameter | 0.8 |
| Minimum remove parameter ($\upsilon$) | 0.1 | Evaporation coefficient ($\rho$) | 0.95 |
| Maximum remove parameter ($\xi$) | 0.5 | Noise parameter ($\mu$) | 0.1 |
| Cooling rate ($c$) | 0.99975 | | |

### 5.3 The effect of the variations of the caregiver swap heuristic

As previously described in Section 3, the first research question aims to investigate the effectiveness of the proposed variations of the caregiver swap heuristic algorithm. As highlighted in Section 4.3, this heuristic was designed to look for the best caregiver-vehicle assignment using the pheromone concept from the ACO algorithm with two different visibility heuristics that consider the common and unique number of patients. Thus, we proposed three ALNS-VS algorithms differentiated by the variations of caregiver swap heuristics: (1)

ALNS-VS_NoSwap does not include the caregiver swap heuristic, (2) ALNS-VS_Common consists of the heuristic with only common visibility heuristic, and (3) ALNS-VS_Unique considers only unique number of patients as a visibility heuristic.

After solving all of the 480 problem instances in 5 replications by each algorithm, we tested the following null hypothesis using a paired sample t-tests with a 99% confidence interval in Minitab 19. Whereas the following null hypotheses state that there is no difference between the means of the solutions obtained by the algorithms, the alternative hypotheses state that they are different. $\mu_{noSwap}$, $\mu_{Common}$ and $\mu_{Unique}$ indicate the averages of all of the solutions obtained by ALNS-VS_NoSwap, ALNS-VS_Common and ALNS-VS_Unique, respectively. For the sake of the flow of the manuscript, the solutions of the algorithms were provided in Tables D.1. through D.4 in Appendix D.

- $H_0^a: \mu_{Common} - \mu_{noSwap} = 0$, $H_1^a: \mu_{Common} - \mu_{noSwap} \neq 0$
- $H_0^b: \mu_{Unique} - \mu_{noSwap} = 0$, $H_1^b: \mu_{Unique} - \mu_{noSwap} \neq 0$
- $H_0^c: \mu_{Common} - \mu_{Unique} = 0$, $H_1^c: \mu_{Common} - \mu_{Unique} \neq 0$

Table 4 demonstrates the results of the paired t-tests for each hypothesis. As seen in the table, both ALNS-VS_Common and ALNS-VS_Unique are statistically different from ALNS-VS_NoSwap because p-values are less than 0.01. Additionally, ALNS-VS_Common and ALNS-VS_Unique present lower average total flow times than ALNS-VS_NoSwap with an average of 21 and 24 minutes. The analyzes also showed that there is no statistically significant evidence to reject the null hypothesis $H_0^c$ because the p-value (0.163) is greater than 0.01. Hence, we can conclude that ALNS-VS_Common and ALNS-VS_Unique provide statistically similar outputs. However, ALNS-VS_Common caused an average of 3 minutes more working time than ALNS-VS_Unique. Because of this small difference, we decided to use the ALNS-VS_Unique algorithm, hereafter called simply ALNS-VS again, and its solutions for further analyzes and comparisons.

Table 4. The result of the paired t-tests for the comparisons of the variants of the caregiver swap heuristics.

| | Mean | Std. Deviation | Std. Error Mean | Lower CI | Upper CI | *t* | *df* | *p* |
|---|---|---|---|---|---|---|---|---|
| $\mu_{Common} - \mu_{Unique}$ | 3.00 | 33.21 | 2.14 | -2.56 | 8.57 | 1.40 | 239 | 0.163 |
| $\mu_{Common} - \mu_{noSwap}$ | -21.12 | 43.12 | 2.78 | -28.34 | -13.89 | -7.59 | 239 | 0.000 |
| $\mu_{Unique} - \mu_{noSwap}$ | -24.12 | 44.38 | 2.86 | -31.56 | -16.68 | -8.42 | 239 | 0.000 |

### 5.4 The effectiveness of the proposed ALNS-VS algorithm

This section aims to provide answers to the second research question in which the effectiveness of the proposed algorithms is investigated in comparison to each other and CPLEX solutions. We limited the running time for processing the HHSRP-VS MILP model to 6 hours (21,600 seconds) because of the complexity of the problem. The quality of the solutions obtained by the CPLEX solver is defined as the discrepancy (GAP) between the best integer objective function value and the relaxed objective function value of the node remaining at the end of the time limit (Öztürkoğlu, 2020). Thus, if we did not obtain the global optimal solution within the time limit, we used the best-found solution so far with its gap for comparisons. We also calculated the computational time of the ALNS-VS and UBA algorithms in terms of seconds for accurate comparisons.

The CPLEX solver did not provide global optimal solutions for the HHSRP-VS problem within the time limit for any of the problem instances. In literature, many HHSRP studies also faced similar problems due to the complexity of the problem (Trautsamwieser and Hirsch, 2011; Trautsamwieser and Hirsch, 2014). We could obtain feasible integer solutions only for the instances with 10 patients except for four instances. For the other instances with more than 10 patients, we couldn't obtain any improved feasible integer solution despite the initial feasible solutions provided by UBA. Table A.1 in Appendix A demonstrates the solutions obtained by CPLEX and UBA for 10-patient instances. In the table, "NA" indicates that no integer feasible solution is available. Whereas the CPLEX provided 16.4% better solutions (see column % Imp.) than the given UBA solutions on average, the average GAP in CPLEX solutions is 40.7%. According to this result, it could be discussed that while the UBA presents a tighter upper bound in a short amount of time (0.05 sec. on average) the optimality GAP seems to be large due to poor lower bound, which is most likely caused by the fractional routing variables of vehicles and caregivers and subtours due to DP policy in linear-programming (LP)-relaxation.

In Table A.2 in Appendix A, we compared CPLEX solutions with ALNS-VS solutions only for 10-patient instances. For 10-patient instances, there are no unvisited patients in both CPLEX and ALNS-VS solutions. However, the ALNS-VS presented a maximum of 19.7% and an average of 6% lower total flow time than the CPLEX solutions only in 1.8 seconds on average.

Since the UBA does not consider the drop off and pick-up policy, its solutions can be considered weak benchmarks for evaluating the effectiveness of the ALNS-VS algorithm, especially in instances with more than 10 patients. Therefore, for an accurate comparison, we applied DP local search heuristic introduced in Section 4.2 to the solutions developed by UBA. The modified UBA by the DP heuristic is called UBA+DP. Table E.1 and Table E.2 in Appendix E demonstrate the best solutions obtained UBA and UBA+DP with their computational time in seconds, respectively. Additionally, these tables compare the quality of the solutions obtained by UBA, UBA+DP and ALNS-VS. In the columns of Table E.1, "UBA+DP-UBA" indicates the percentage improvement of the UBA+DP algorithm over UBA. Similarly, column "VS-UBA+DP" in Table E.2 indicates percentage improvement provided by the ALNS-VS algorithm over UBA+DP. Briefly, Table A.3 in Appendix A presents the aggregated best solutions, which are the averages of the best-found solutions of five instances in an instance class, of ALNS-VS, UBA, UBA+DP, the percentage improvement of UBA+DP over UBA in column "UBA+DP-UBA(%)", and the percentage improvement of ALNS-VS over UBA+DP in column "VS-UBA+DP".

When we applied the DP heuristic to the UBA solutions, we obtained 9.4, 14.6, 13.2 and 12.8 percentage improvement on average in the instances with 10, 30, 50 and 100 patients, respectively. It is obvious that these improvements were achieved by dropping and picking up caregivers on the route. Also, these improvements were achieved with milliseconds more computational effort to solve UBA+DP compared to UBA; where UBA+DP lasted 0.05, 0.2, 0.6, and 1.4 seconds on average in the 10-, 30-, 50-, and 100-patient instances, respectively.

On the other hand, the ALNS-VS solutions presented 13.1, 13.6, 19.3 and 15.9 percent lower total flow time than UBA+DP on average for the 10-, 30-, 50- and 100-patient instances, respectively. When they were compared with UBA solutions, as is expected the percentage improvements increase up to 30%, 35% and 34% for the instances with 30, 50 and 100 patients, respectively. Since ALNS-VS employs the DP policy throughout the iterations in contrast to UBA+DP, some portions of its savings on total flow time over UBA+DP seem to be achieved by additional drop-off and pick-ups. Whereas the caregivers were dropped off 4, 14, 23 and 41 times on average in 10-, 30-, 50- and 100-patient instances in the ALNS-VS solutions, they are 2, 10, 6, 30 in the UBA+DP solutions. It also seems that the number of drop off and pick-up increases as the instance size gets larger. Even though the ALNS-VS algorithm requires proportionally higher computational effort than UBA+DP, i.e. 23, 34, 119 seconds in the 30, 50- and 100-patient instances, respectively, we think that this could be negligible from the view of practitioners because a manual solution always takes a very long time and the expected planning time is also usually longer than 5 minutes in practice. Additionally, while there are several unvisited patients in UBA+DP solutions for 12 instances there are no unvisited patients in any of the ALNS-VS solutions. For example, there are averages of 0.4, 1.8, 0.6 and 1.4 unvisited patients in the UBA+DP solutions of h50_40_1, h50_40_2, h100_40_1 and h100_40_2 instance classes, respectively. For the sake of clarity, these unvisited patients were not shown in the tables. As a result, we can conclude that the proposed ALNS-VS algorithm seems to provide reasonably good solutions to the HHSRP-VS problems in a reasonable computational effort.

### 5.5 The effect of the proposed drop-off and pick-up (DP) policy

In the previous section, we highlighted that the DP policy seems to reduce the total flow time of caregivers when we compared ALNS-VS, UBA+DP and UBA solutions. Thus, this section aims to investigate the effectiveness of the DP policy in a detailed analysis and answer the third research question. To provide an accurate comparison, we introduced the HHSRP-M problem by removing only the DP policy in HHSRP-VS.

Hence, HHSRP-M only allows caregivers to share a vehicle without the possibility of drop-off and pick-up. The MILP model of HHSRP-M could be easily achieved by setting all $y_{i,k,l}$ decision variables to 0 and removing the set of dummy nodes $V_2$ in the HHSRP-VS MILP model.

**Proposition 1.** *The optimal total flow time of caregivers in HHSRP-VS ($f_{VS}^*$) is always less than or equal to that in HHSRP-M ($f_M^*$): $f_{VS}^* \leq f_M^*$.*

**Proof 1.** Suppose that $P_{VS}$ and $P_M$ are the optimal routes in HHSRP-VS and HHSRP-M, respectively. Since DP policy is the only difference between HHSRP-M and HHSRP-VS and it is not allowed in HHSRP-M, $P_M \subseteq P_{VS}$. Hence, it can be written that $f_M^* - \triangle_{DP} = f_{VS}^*$, where $\triangle_{DP}$ indicates savings in total flow time due to drop-off and pick-up. Hence, although the drop-off and pick-up require additional travel time if there exists at least one such a drop-off and pick-up option that reduces flow time of the caregivers by reducing wasted time of the caregivers who wait in the vehicle for the completion time of the occupied caregiver in HHSRP-M; if $\exists \triangle_{DP} > 0$ then $f_{VS}^* < f_M^*$; otherwise $f_{VS}^* = f_M^*$. ■

To compare HHSRP-VS solutions with HHSRP-M in an empirical analysis, we modified the ALNS-VS algorithm by removing its DP local search and dummy node removal heuristics, which were described in section 4. Hence, we called the modified algorithm ALNS-M to solve the HHSRP-M problem. After solving the problem instances with ALNS-M, we observed that there are no unvisited patients in any of the problem instances. We then calculated the percentage difference of total flow time between ALNS-M and ALNS-VS solutions as VS-M%=100*(ALNS-M - ALNS-VS)/ALNS-M to analyze the effect of DP policy on total flow time. Tables F.1 through F4 in Appendix F present the ALNS-M solutions and the percentage differences in details. Table A.3 in Appendix A also presents the aggregated best solutions of ALNS-M and their differences with ALNS-VS. It can be seen in the tables that the implementation of DP policy provides approximately 19, 25, 24 and 22% savings in caregivers' total working time on average for 10-, 30-, 50- and 100-patient instances.

Using the 240 solutions in Tables F.1-F.4 in Appendix F, we also performed a full factorial design of the experiment to investigate the effects of the problem features described in Table 2 on the contribution of DP policy at the 95% confidence level. Recall that there are 4 levels of a number of patients ($noP$), 4 levels of service area radiuses ($ra$) and 3 levels of patients' demand distributions ($dd$). The response (dependent variable) is the VS-M%. The results of the full factorial design of experiments (the ANOVA table) are given in Table F.5 in Appendix F. The main factors and their all-level interactions explain 94.27% of the total variation of the response ($R^2$). As seen in Table F.5, $noP$, $ra$ , and $dd$ are significant on the model. Moreover, $ra$ has the largest effect on the contribution of DP policy due to its high "Adj SS" value. This could also be seen in the main effects plot given in Figure 2.

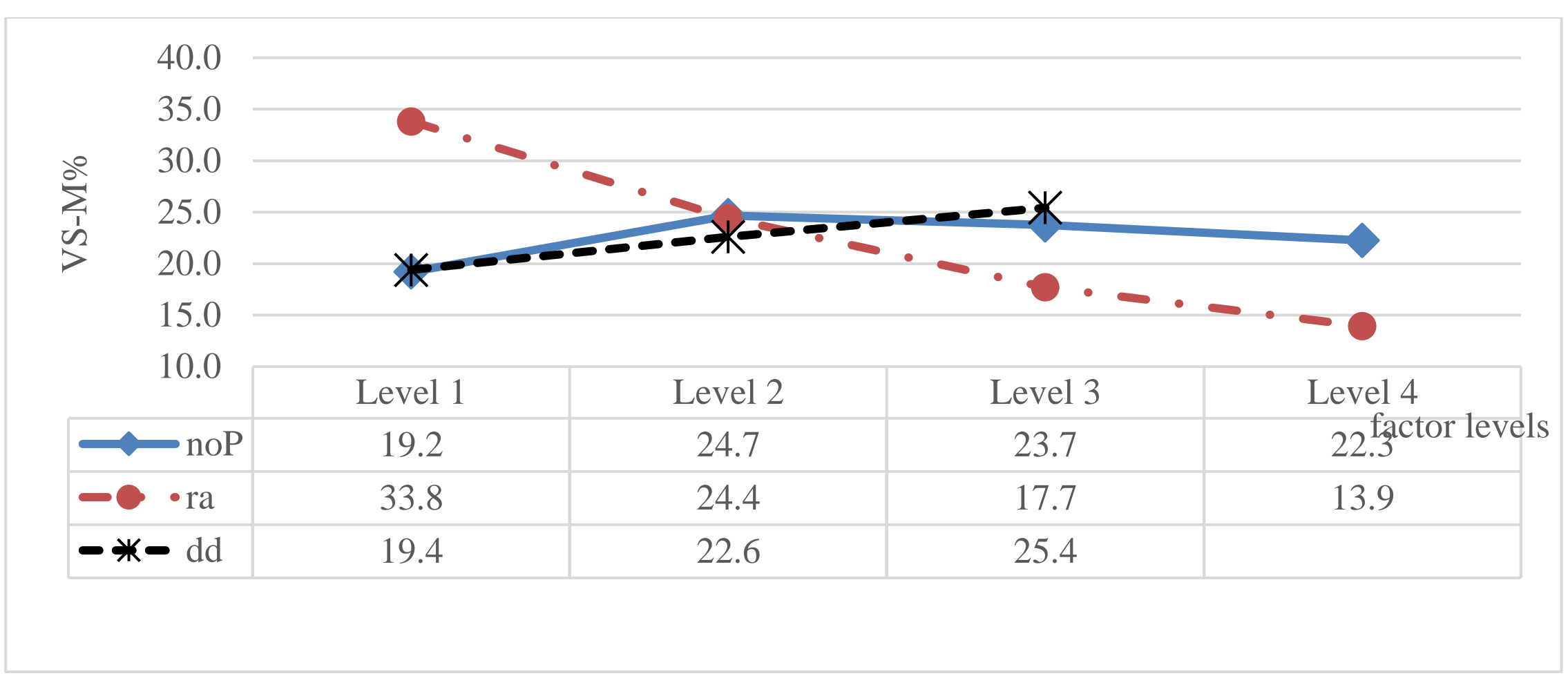

Figure 2. The main effects plot of the factors.

As the service area radius increases from 10 to 40 minutes, the contribution of DP policy steadily decreases from 34% to 14%. This shows that while the impact of DP policy on total flow time is very significant when patients are located in smaller areas such as urban or metropolitan areas, it also provides significant time savings for larger areas. Additionally, the contribution of DP policy steadily increases from 19% to 25% as the level of demand distribution increases. Note that while at the first $dd$ level only 20% of the patients were defined as requiring moderate and difficult care, this rate increases to 50% at the third level. Thus, this suggests that the higher the proportion of patients' difficult service requirements, the greater the contribution of the DP policy. The reason for this increasing contribution of DP policy with increasing demand for difficult services could be that a vehicle prefers to travel between patients rather than waiting in a patient due to high service time. Last, the contribution of the DP policy appeared to be the lowest when the number of patients is the smallest. Its contribution reaches its maximum when there is a moderate number of patients. In our experiments, the policy has shown its highest contribution in the 30-patient problem instances with an average of 25%. The reasons for decreasing contributions when there are few or many patients may be that (1) traveling back and forth due to the DP policy may not be very efficient because a small number of patients is highly likely to be dispersed far from each other, and (2) caregivers' may have longer waiting times at their patients due to the late arrival of the vehicle when there are too many patients to visit.

### 5.6 The effect of the vehicle sharing by multiple caregivers

As discussed in section 2, one of the common assumptions in existing HHSRP literature is that one vehicle carries only one caregiver. On the contrary, the proposed HHSRP-VS allows multiple caregivers to share a single vehicle for their travels. Hence, it is obvious that sharing a vehicle reduces the necessity of vehicles. However, this could also increase the total flow time of workers due to waiting for each other or a returning vehicle at a patient node. Thus, this section aims to investigate the scenarios where HHSRP-VS may provide potential cost savings and answer the fourth research question. For this purpose and accurate comparison, similar to conventional HHSRP we defined the HHSRP-STD problem in which every single caregiver is assumed to travel with a single vehicle with or without a driver. Therefore, vehicle sharing, and DP policies are irrelevant in the HHSRP-STD. The MILP model of the HHSRP-STD can be easily developed by setting the capacity of all vehicles to 1, $c_k = 1$, setting all $y_{i,k,l}$ decision variables to 0, and removing the set of dummy nodes $V_2$ in HHSRP-VS MILP model.

**Proposition 2. (Best-case scenario)** *If the sets of patients of $c$ caregivers, who can travel in a single shared vehicle, are assigned to the same locations and the patients at the same locations require the same type of service, then the optimal flow time of HHSRP-VS ($f_{VS}^*$), HHSRP-M ($f_M^*$) and HHSRP-STD ($f_{STD}^*$) are equal to each other. $f_{VS}^* = f_M^* = f_{STD}^*$.*

**Proof 2.** Suppose that there are $c$ caregivers who travel with their own vehicle in HHSRP-STD and with a single shared vehicle in HHSRP-M and HHSRP-VS. Suppose that they are assigned to serve the same number of patients ($n$), each located at the same node, i.e. in a mall, apartment or business center: the location of patient $i$ of caregiver $l$ is $v_i^l = v_i$ and $v_i \neq v_j, \forall i \neq j = 1, \ldots, n,$ , and $\forall l = 1, \ldots, c$. Suppose that the patients located at the same node require the same treatment: the service time of patient $i$ of caregiver $l$ is $p(v_i^l) = p(v_i), \forall i = 1, \ldots, n$ and $\forall l = 1, \ldots, c$. Since each caregiver must visit each patient, and patient treatment times are the same at the same location, the optimal tour for all caregivers in HHSRP-STD can be easily computed by solving a Traveling Salesman Problem (TSP) for just one caregiver. Hence, suppose that $P = P_{STD}^l = \{v_0^l = 0, v_1^l, \ldots, v_n^l, v_{n+1}^l = 0\}, \forall l = 1, \ldots, c$ indicates the optimal route of caregivers in HHSRP-STD. If $P$ minimizes the total route length for one caregiver, it must also be the optimal tour of the single shared vehicle in HHSRP-M since all of the caregivers' patients are located at the same points and their service times are the same. Hence,

$$f_{STD}^l = \sum_{i \in P_{Std}^l} \left( t_{v_{i-1}^l, v_i^l} + p(v_i^l) \right) = \sum_{i \in P} \left( t_{v_{i-1}, v_i} + p(v_i) \right) = T + S, \forall l = 1, \ldots, c, \tag{37}$$

where $\sum_{i \in P} (t_{v_{i-1}, v_i}) = T$ and $\sum_{i \in P} p(v_i^l) = S$.

$$f_{STD}^* = \sum_{l=1}^{c} f_{STD}^l = c \cdot (T + S) \tag{38}$$

$$f_M^* = \sum_{l=1}^{c} \sum_{i \in P} \left( t_{v_{i-1}, v_i} + p(v_i) \right) = c \cdot (T + S) \tag{39}$$

Because all caregivers in the shared vehicle leave at every patient node $v_i$ and treat their patients simultaneously with the same amount of service time, there is no need to implement a DP policy. Thus, $f_{VS}^* = f_M^* = f_{STD}^*$.■

**Proposition 3. (Practical best-case scenario)** *$f_{STD}^*$ is always less than $f_M^*$ when caregivers' patients located at the same nodes require different types of services contrary to Proposition 2.*

**Proof 3. (Bases on Proposition 2)** Suppose that caregiver $l$'s patient treatment time at patient node $i$ is not necessarily equal to the treatment times of other caregivers at the same node due to different service requirements: $p(v_i^j) \neq p(v_i^k), \forall i = 1, \ldots, n$ and $\forall j, k = 1, \ldots, c,$ and $j \neq k$ in Proposition 2. The optimal sequence of patients in HHSRP-STD ($P$) can be still obtained by solving a TSP for one caregiver because service times are constant. $P$ also minimizes the total travel time of the single shared vehicle in HHSRP-M. Let $T$ be the total travel time of caregivers or vehicles in the optimal path: $\sum_{i\in P_{Std}^{l}}\left(t_{v_{i-1}^{l},v_{i}^{l}}\right)=\sum_{i\in P}\left(t_{v_{i-1},v_{i}}\right)=T$, $\forall l=1,\dots,c$. Let $\sum_{i\in P}p\left(v_{i}^{l}\right)=S^{l}$ be the total service times of caregiver $l$'s patients, which are known and constant. The optimal flow time of the caregivers in HHSRP-STD is

$$f_{STD}^{l}=\sum_{i\in P_{Std}^{l}}\left(t_{v_{i-1}^{l},v_{i}^{l}}+p\left(v_{i}^{l}\right)\right)=T+S^{l},\forall l=1,\dots,c \tag{40}$$

$$f_{STD}^{*}=\sum_{l=1}^{c}f_{STD}^{l}=c\cdot T+\sum_{l=1}^{c}S^{l}. \tag{41}$$

In HHSRP-M, when $c$ caregivers visit their patients located at the same nodes with a shared vehicle, all other caregivers wait for the caregiver whose patient require the highest treatment time. Hence, $P$ still provides the optimal tour in HHSRP-M, and the optimal flow time in HHSRP-M can be written as in equation (42).

$$\begin{aligned} f_{M}^{*}&=\sum_{l=1}^{c}\sum_{i\in P}t_{v_{i-1},v_{i}}+\sum_{l=1}^{c}\sum_{i\in P}\max\left(\{p(v_{i}^{1}),\dots,p(v_{i}^{c})\}\right)\\ &=c\cdot T+c\cdot\sum_{i\in P}\max\left(\{p(v_{i}^{1}),\dots,p(v_{i}^{c})\}\right). \end{aligned} \tag{42}$$

As a result, since $S^{l}<\sum_{i\in P}\max\left(\{p\left(v_{i}^{1}\right),\dots,p\left(v_{i}^{c}\right)\}\right)$, $\forall l=1,\dots,c$, $f_{STD}^{*}<f_{M}^{*}$. ■

As it is seen in Propositions 2 and 3, sharing a vehicle without DP policy certainly increases caregivers' total flow time except for the best-case scenario. We also know from the previous sections that DP policy provides savings of the flow time when vehicle sharing is allowed. Therefore, to investigate the effect of vehicle sharing with DP and develop in-depth insights, we perform an empirical analysis. For this, we solved HHSRP-STD with the ALNS-STD algorithm, which was developed by removing DP local search, dummy node removal, and caregiver swap heuristics from ALNS-VS, for an accurate comparison.

As defined in Table 2, whereas 2 vehicles are assumed to be needed to serve 10 and 30 patients, 3 and 6 vehicles are required for 50 and 100 patients respectively in our problem instances in HHSRP-VS. However, the numbers of vehicles needed are 4, 4, 6, and 12 in HHSRP-STD because every caregiver needs a separate vehicle. Hence, the additional vehicle needs are 2, 2, 3, and 6 (doubled) in HHSRP-STD in those problems. After solving the same problem instances with the new number of vehicles using the ALNS-STD algorithm, we obtained the best-found solutions as given in Tables G.1 through G.4 in Appendix G in details. Table A.3 in Appendix A also presents the aggregated best solutions of ALNS-STD and their percentage improvement over ALNS-VS in column "STD-VS(%)". Because HHSRP-STD is less complex than HHSRP-VS and ALNS-STD requires a few local search heuristics, solving ALNS-STD requires a shorter amount of time: 1.7, 3.2, 5.1, and 19.5 seconds for 10, 30, 50, and 100 patients, respectively, which are much shorter than ALNS-VS that solved the same problems in 1.9, 22.9, 34.1, and 118.6 seconds. Furthermore, as expected the HHSRP-VS causes more total flow time than the HHSRP-STD.

- For 10-patient instances, the caregivers spent about 33% less time, on average 158 minutes, in HHSRP-STD than in HHSRP-VS.
- For 30, 50 and 100 patients, caregivers complete their tour in about 26%, 25% and 25% less time in HHSRP-STD than they are in HHSRP-VS on average, respectively. This leads to totals of 263, 400 and 858 minutes of savings on average for the same problem sets, respectively.

The abovementioned results showed that HHSRP-STD provides a considerable amount of savings in total flow time of caregivers' working time with a cost of additional vehicles, which may be special vehicles equipped with healthcare equipment. Because of this trade-off, we take our analysis further and compare HHSRP-VS and HHSRP-STD in light of the total cost of providing care services to find out deeper insights. For this purpose, we performed a break-even analysis.

Suppose that $TC_{STD}$ and $TC_{VS}$ are the total daily monetary cost of managing home health care services in HHSRP-STD and HHSRP-VS, respectively. Let $C_V$ be the hourly cost of vehicle ownership or usage that may consist of the rental or payment cost per hour of a vehicle, the hourly wage of a driver, the cost of fuel consumption for an hour, and all other costs related to the usage of the vehicle. Similarly, let $C_L$ be the average hourly cost of caregivers that may include their salaries, insurances, bonuses, and lunch payments. Last, $f^*_{STD}$ and $f_{VS}$ indicate the best objective function values (total flow time of caregivers in hours) of the HHSRP-STD and HHSRP-VS problem instances solved by ALNS-STD and ALNS-VS algorithms, respectively. Recall that $c$ is the capacity of the vehicles in HHSRP-VS. Thus, $TC_{STD}$ and $TC_{VS}$ can be simply written $TC^*_{STD} = f^*_{STD} * C_V + f^*_{STD} * C_L$ and $TC^*_{VS} = \frac{f^*_{VS}}{c} * C_V + f^*_{VS} * C_L$. The first terms in these equations indicate the total cost of vehicle ownership and the second terms identify the total cost of labor. In $TC^*_{VS}$, $f^*_{VS}$ is divided by $c$ to calculate the flow time of vehicles. Hence, the breakeven rate ($BER$) can be calculated by equation (36). Proposition 4 shows that the denominator is always positive.

$$BER = \frac{C^*_V}{C^*_L} = \frac{(f^*_{VS} - f^*_{STD})}{\left(f^*_{STD} - \frac{f^*_{VS}}{c}\right)}, \quad (36)$$

**Proposition 4.** *In the optimal solutions of HHSRP-STD* ($f^*_{STD}$)*, HHSPR-M* ($f^*_M$)*, and HHSRP-VS* ($f^*_{VS}$), $\frac{f^*_{VS}}{c} \leq \frac{f^*_M}{c} < f^*_{STD}$.

**Proof 4.** Suppose that $P_{STD} = \{P^l_{Std}, \forall l = 1 \dots c\}$ is the set of optimal assignments and routes of $c$ caregivers/vehicles in the optimal solution of a HHSRP-STD, where $P^l_{STD} = \{v^l_0 = 0, v^l_1, \dots, v^l_m, v^l_{m+1} = 2n + 1\}$ indicates the optimal route of caregiver $l$. Let $e^l_0 = \{v^l_0, v^l_1\}$ and $e^l_1 = \{v^l_m, v^l_{m+1}\}$ be the first and the last edges that are traversed in the route of caregiver $l$, respectively. Let $T^l$ be the total travel time of caregiver $l$, hence, similar to equation (40),

$$f_{STD}^{l} = \sum_{i \in P_{Std}^{l}} \left( t_{v_{i-1}^{l}, v_{i}^{l}} + p\left(v_{i}^{l}\right) \right) = T^{l} + S^{l} \tag{43}$$

$$f_{STD}^{*} = \sum_{l=1}^{c} f_{STD}^{l} = \sum_{l=1}^{c} T^{l} + \sum_{l=1}^{c} S^{l} \tag{44}$$

Suppose that caregivers have no common skills or there is no patient who can be treated by more than one caregiver. Suppose that each patient has a different service time and the locations of patients treated by each caregiver are placed apart from each other like in different regions or zones. See Figure 3 for an example representation of two caregivers' patients and paths. Contrary to Proposition 3, these assumptions define the worst case of the distribution and the assignments of patients. Because $P_{STD}^{l}$ identifies the optimal path for each caregiver, in the optimal solution of HHSRP-M the vehicle must follow through each caregivers' path with an elimination of return to the HHC after a caregiver's service completed. The caregivers who completed serving their patients must travel through the other caregivers' paths and wait in the shared vehicle until all caregivers completed their service. Hence, when we combine $P_M = P_{STD}^{1} \cup P_{STD}^{2} \dots \cup P_{STD}^{c} \cup \{\Delta^a\}/\{\Delta^s\}$ where $\Delta^a$ and $\Delta^s$ indicate the additional and the removed paths to complete a single circuit. Let us consider the example in Figure 3. We can develop optimal tour of HHSRP-M by combining two caregivers' routes $P_{STD}^{1}$ and $P_{STD}^{2}$ into one route of a vehicle $P_M$. Suppose that the vehicle visits patients in $P_{STD}^{1}$ then in $P_{STD}^{2}$ for minimum flow. Hence, $P_M = P_{STD}^{1}/\{e_1^1\} \cup \{e_{1,2}\} \cup P_{STD}^{2}/\{e_0^2\}$, where $e_{1,2} = \{v_m^1, v_1^2\}$ is the connection edge. For the example in Figure 3, $P_{STD}^{1} = \{0, 1, 2, 3, 0\}$, $P_{STD}^{2} = \{0, 4, 5, 6, 7, 0\}$ and $P_M = \{0, 1, 2, 3, 4, 5, 6, 7, 0\}$ where $e_1^1 = \{3,0\}$, $e_0^2 = \{0,4\}$ were removed from $P_{STD}^{1}$ and $P_{STD}^{2}$, respectively and connection edge $e_{1,2} = \{3,4\}$ was added.

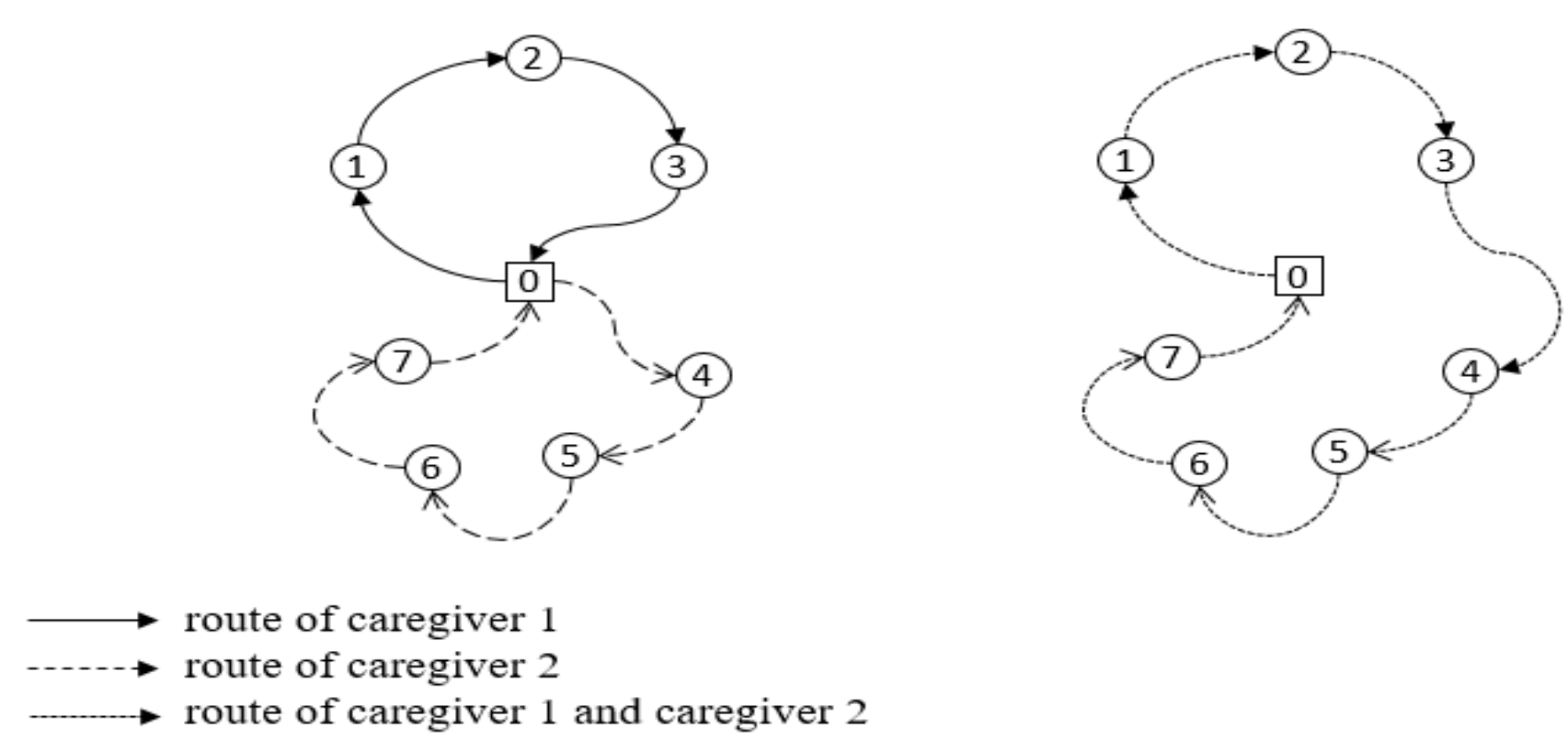

Figure 3. The representation of the optimal paths of two caregivers in the worst case.

Let $\mathcal{H}$ be the set of caregiver pairs in the shared vehicle in the optimal route of HHSRP-M. So, we can write the optimal flow time

$$f_{M}^{*} = c \cdot \left[ \sum_{l \in L} \sum_{i \in P_{Std}^{l}} t_{v_{i-1}^{l}, v_{i}^{l}} + \sum_{(i,i+1) \in \mathcal{H}, i \in \frac{L}{\{c\}}} \left( t_{e_{i,i+1}} - t_{e_1^i} - t_{e_0^{i+1}} \right) \right] + c \cdot \sum_{l \in L} \sum_{i \in P_{Std}^{l}} p\left(v_{i}^{l}\right) \tag{45}$$

In equation (45), the first and the second terms indicate the portion of total travel and service times in the total flow time of caregivers. With the triangle inequality assumption, the term $\sum_{(i,i+1)\in\mathcal{H},i\in\frac{L}{\{c\}}}\left(t_{e_{i,i+1}}-t_{e_1^i}-t_{e_0^{i+1}}\right)$ in total travel time is always non-positive. This can be seen in the example given in Figure 3 $e_{1,2}<e_1^1+e_0^2$. Hence, $f_M^*$ can be written as in equation (46).

$$f_M^*=c\cdot\sum_{l=1}^{c}T^l+\ c\cdot\sum_{l=1}^{c}S^l+c\cdot\Delta, \tag{46}$$

$$\frac{f_M^*}{c}=\sum_{l=1}^{c}T^l+\sum_{l=1}^{c}S^l+\Delta\le f_{STD}^* \tag{47}$$

Finally, with the help of Proof 1, $\frac{f_{VS}^*}{c}\le\frac{f_M}{c}<f_{STD}^*$.■

Tables G.1 – G.4 in Appendix G presents the $BER$ values calculated for each solution in detail. Table A.3 in Appendix A demonstrates the average $BER$ values for each instance class. Hence, if $\exists\frac{C_V}{C_L}>BER$, HHSRP-VS may be preferable to HHSRP-STD due to lower total cost of service; otherwise, HHSRP-STD is superior to HHSRP-VS in terms of the total cost of service. When we conducted a full factorial design of experiment with factors $noP$, $ra$, and $dd$ and a response $BER$, the analysis showed that every main factor and only $noP*ra$ two-way interaction is significant with a model of 76.75% $R^2$. The ANOVA table for this analysis is given in Table 5. Additionally, the number of patients ($noP$) and service area radius ($ra$) have the largest effect on $BER$ due to their high Adj SS values.

Table 5. The ANOVA table for break-even ratios.

| *Source* | *DF* | *Adj SS* | *Adj MS* | *F-Value* | *p-Value* |
|---|---|---|---|---|---|
| $\boldsymbol{noP}$ | 3 | 33.95 | 11.32 | 63.05 | 0.000 |
| $\boldsymbol{ra}$ | 3 | 57.57 | 19.19 | 106.93 | 0.000 |
| $\boldsymbol{dd}$ | 2 | 7.49 | 3.75 | 20.88 | 0.000 |
| $\boldsymbol{noP*ra}$ | 9 | 8.41 | 0.93 | 5.21 | 0.000 |
| $\boldsymbol{noP*\ dd}$ | 6 | 3.09 | 0.52 | 2.87 | 0.011 |
| $\boldsymbol{ra*dd}$ | 6 | 1.10 | 0.18 | 1.02 | 0.413 |
| $\boldsymbol{no*ra*dd}$ | 18 | 2.11 | 0.12 | 0.65 | 0.854 |
| *Error* | 192 | 34.46 | 0.18 | | |
| *Total* | 239 | 148.19 | | | |

Further analysis was also conducted to gain more insights into the effects of the main factors on $BER$. First, the Bonferroni t-test was used to examine the statistical significance of the different levels of $noP$, $ra$ and $dd$ with a 95% confidence level. If there is no statistically significant difference between the levels, then they are grouped and shown symbolically as demonstrated in Table 6.

Table 6. Multiple comparison test results for $BER$ according to the problem features.

| $noP$ | Mean $BER$ | | $ra$ | Mean $BER$ | | $dd$ | Mean $BER$ | |
|---|---|---|---|---|---|---|---|---|
| 10 | 1.99 | A | 40 | 1.91 | A | 0 | 1.57 | A |
| 30 | 1.30 | B | 30 | 1.69 | A | 1 | 1.37 | B |
| 100 | 1.08 | B C | 20 | 1.19 | B | 2 | 1.13 | C |
| 50 | 1.06 | C | 10 | 0.64 | C | | | |

As can be seen in Table 6, whereas the problems with 10 patients are statistically different from the others, it is interesting that there is no statistical difference between the problem instances of 30 and 100 patients and between 50 and 100 patients. The $BER$ values for the 10-patient problem instances are approximately twice as high as those for the other 50- and 100-patient problem instances. This result may be consistent with the observation made in section 5.5, where the contribution of DP policy to flow time savings was lowest when the number of patients is smallest. The analysis also revealed that while there was no statistical difference between service areas with a radius of 30 and 40 minutes, there was a difference among these and other service areas. It can also be seen that the $BER$ decreases as the service area gets smaller. This result is also consistent with the observation made in section 5.5 that the reduction in flow time is greatest when the service area is smallest. For example, in a 10-minute service area, HHSRP-VS has a lower total cost than HHSRP-STD as long as the hourly vehicle cost to hourly labor cost ratio is greater than 0.64. In other words, if the hourly labor cost is 100 units and the hourly vehicle cost is more than 64 units, car sharing with a drop-off policy may be preferred compared to the case where everyone uses their own vehicle to reduce the total cost in a 10-minute service area. Since labor costs may be higher than vehicle usage costs in developed countries, especially in the health sector, BER values less than 1 may indicate that the chance of using a shared vehicle with the DP policy is higher. However, the opposite might also be true for developing countries, where ownership or using cost of proper vehicles for home health care services might be more expensive than cost of labor. According to the results, we can say that vehicle sharing with DP policy provides cost savings mostly when the hourly vehicle cost is higher than the labor cost since $BER$ is mostly higher than 1 in many of the cases as can be seen in Table 6. Additionally, as shown in Proposition 2, $BER$ is equal to 0 in the best-case scenario where $f_{VS}^{*} = f_{STD}^{*}$. Hence, HHSRP-VS always costs less than HHSRP-STD, no matter how high the hourly labor cost in the best-case scenario. Last, the average $BER$ is statistically different at each level of the patient's demand distribution. The average $BER$ decreases as the percentage of difficult care requirement increases.

Figure 4 also demonstrates that the average $BER$ generally decreases as the number of patients increases, the service area decreases, and the level of service patients' demand distribution increases. If the practical $\frac{C_V}{C_L}$ can be assumed to be 1, where the hourly costs of vehicle ownership labor cost are equal, then we can say that sharing vehicles with DP policy provides savings in total service cost,

- when the service area is 10 minutes away from the HHC regardless of the number of patients and the difficulty of the service requirement.
- when the service area is 20 minutes away from the HHC and the number of patients in the area is more than 30.

● when the patients' demand distribution is 50/30/20, where the difficult and moderate care requirements are high, and there are more than 30 patients.

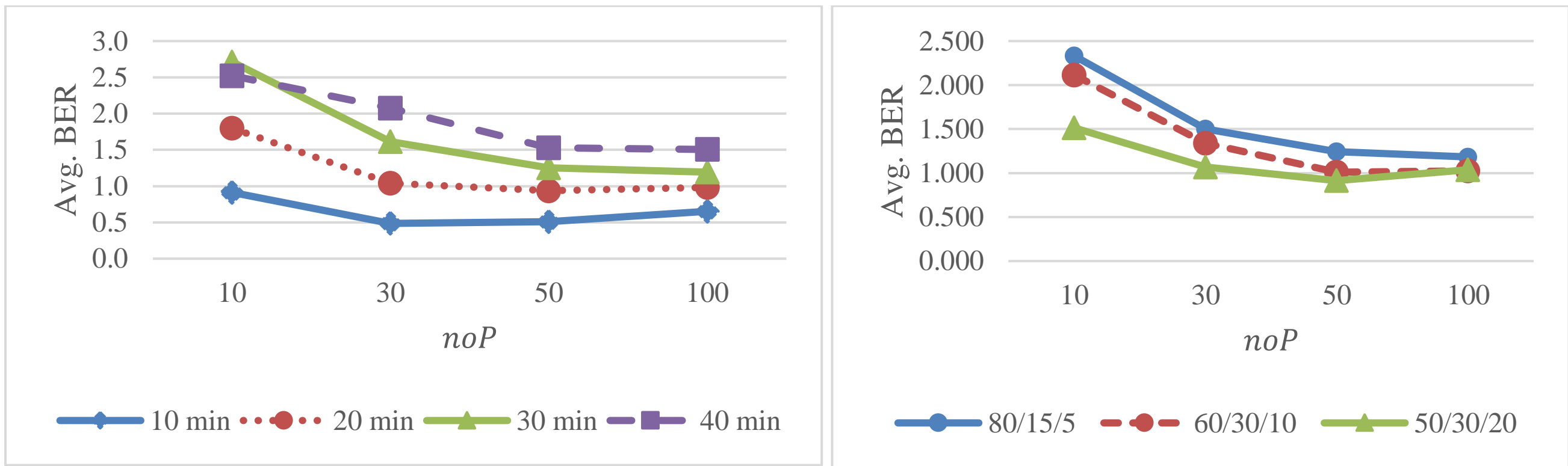

Figure 4. The changes on average $BER$ over service area (left) and the patients' demand distributions (right).

Finally, we performed further analysis to explore the effect of patient density in an area in terms of "number of patients located per unit service area ($PperA$) in terms of minute$^2$ (or kilometer$^2$ where a vehicle travels 60 km/h on average). In this analysis, $PperA = noP/(\pi * ra^2)$, where $\pi$ was taken 3.14. As seen in Figure 5, the average $BER$ mostly decreases as $PperA$ decreases. For example, the average BER is 0.98 when there are 0.08 patients in a unit service area, while it decreases to 0.48 when there are 0.096 patients in the same area. We can conclude that it is highly likely that sharing vehicles with DP policy will result in less total cost than HHSRP-STD when $PperA$ is greater than 0.075. Hence, this result also supports our previous observations, such that the denser the patients in an area the superior the HHSRP-VS model.

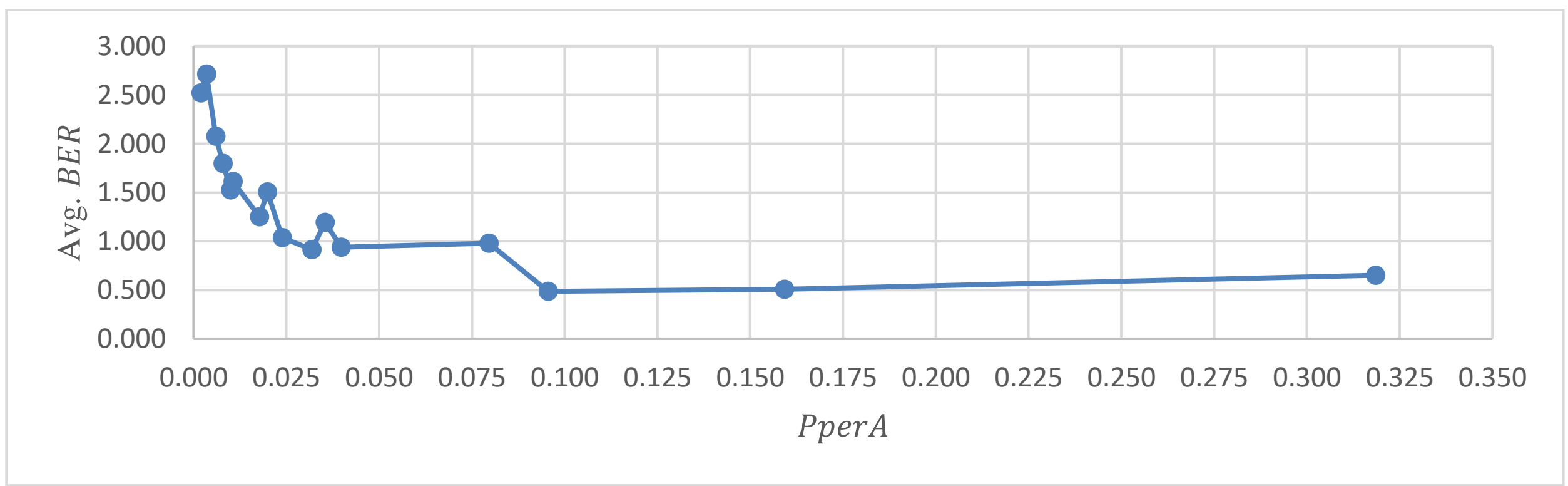

Figure 5. The effect of $PperA$ on average $BER$ values.

## 6. Conclusion

The main contribution of this study is to present a new generic problem to the literature of Workforce Scheduling and Routing Problem (WSRP). This problem introduces two distinct features. First, multiple independent workers can travel in a single shared vehicle. Second, a worker in a vehicle can be dropped off at a customer's location and then picked up by the same vehicle. Although the generic WSRP we introduced in this study can be applied in any field such as telecom, public utilities, or maintenance, we have defined it specifically in the context of the home healthcare industry. Hence, the problem is called Home Healthcare Scheduling and Routing Problem (HHSRP) with Vehicle Sharing (VS) and drop-off and pick-up (DP) policy. The objective of this HHSRP-VS is to minimize caregivers' total flow time and the penalty cost of unvisited patients.

We developed the mixed-integer linear programming (MILP) model of this problem using a two-layer approach to easily adapt the DP policy and avoid sub-tour elimination constraints. Since the complexity of the HHSRP-VS can be considered NP-Hard, we proposed a constructive matheuristic upper-bound algorithm (UBA) and an Adaptive Large Neighborhood Search (ALNS) algorithm with problem-specific local search heuristics to solve HHSRP-VS. We generated various problem instances based on some problem features such as the radius of the service area, the number of patients in an area, the patients' demand distribution of the difficulty of care. We then studied on four research questions.

i. We proposed two variations of the caregiver swap heuristic for the ALNS-VS algorithm, called the "common" and "unique" visibility heuristics. Statistical analysis showed no significant difference between these visibility heuristics.

ii. We analyzed the effectiveness of the proposed UBA and ALNS-VS algorithms. The CPLEX solver could only provide integer solutions for 10-patient instances with an average optimality gap of 40.7% in six hours. For the same instances, UBA developed 16.4% worse solutions than CPLEX in less than 1 seconds. However, the ALNS-VS presented a maximum of 19.7% and an average of 6% lower total flow time than the CPLEX solutions only in 1.8 seconds on average. Because of the lack of CPLEX solutions for the problems with more than 10 patients, we compared ALNS-VS solutions with UBA+DP solutions obtained by applying the proposed DP local search heuristic to UBA solutions. The ALNS-VS solutions presented 13.1%, 13.6% and 19.3% and 15.9% lower total flow time than UBA+DP in 1.9, 23, 34 and 119 seconds on average for 10-, 30-, 50- and 100-patient instances, respectively. While ALNS-VS did not result in any unvisited patients, there were 12 instances in UBA+DP where an average of 2 patients were not visited. We concluded that the proposed ALNS-VS algorithm offers both effective and efficient solutions for HHSRP-VS due to its solution qualities and short computation time.

iii. We investigated the effect of DP policy on the total flow time. For this purpose, we presented the HHSRP-M problem that allows vehicle sharing but DP. We first proved that the optimal solutions of the HHSRP-VS are always better than or equal to those in HHSRP-M. Next, in an empirical analysis, we also revealed that the DP policy saves up to 25% in total flow time. We also showed statistically that savings increase as service area gets smaller and patients need more difficult service.

iv. The effects of vehicle sharing with DP policy on total flow time and total service cost were analyzed. For this purpose, we presented the HHSRP-STD problem, which requires each caregiver to travel with their own vehicle, as in the conventional HHSRP. We proved that the optimal flow time of HHSRP-STD is always shorter than that of HHSRP-M, except in the best-case scenario. Next, we conducted an empirical break-even analysis to investigate under what conditions HHSRP-VS could reduce the total cost of service, including hourly vehicle ownership and labor costs. We explored that the denser the area, the higher the chance to reduce cost with the DP policy. Moreover, the possibility of reducing the cost of service by HHSRP-VS increases when the demand for difficult care increases.

These results and insights were obtained under various assumptions. First, we assume that the number of caregivers assigned to a vehicle is fixed and 2. In practice, however, more than two or varying numbers of caregivers can be assigned to vehicles. Because our assumption is so restrictive, this kind of flexibility can increase the likelihood of reducing total flow time with the vehicle sharing and DP policies. Therefore, we believe that relaxing this assumption could create new challenging problems and opportunities in HHSRP-VS. Second, the current model mandates that the caregiver be picked up by the same vehicle after being dropped off. Once this assumption is relaxed and caregivers are allowed to travel in any vehicle, the likelihood of lower total flow times may be very high. Third, researchers can also include multiple HHCs in the problem to develop more centralized decisions, reduce flow time, and increase patient satisfaction. Hence, the new generic problem introduced and insights developed in this study seem to have the potential to open up new discussions and challenging problems not only in the WSRP literature but also in the vehicle routing problems.

**Acknowledgement**

This research was supported by the TUBITAK (The Scientific and Technological Research Council of Turkey) under Grant 217M555.

## Appendix A The pseudocode of the proposed algorithms

**Algorithm A.1.** The pseudocode of the first stage of the proposed UBA

***input:*** *Service time of patients* $p_{is}$, *sets of patients* $V_1$ *and caregivers* $L$, *coordinates of patients* $p_i$
***output:*** $H_l$ *and* $c_{hl}$ *that indicate caregiver cluster* $l \in L$ *and the centroid of* $H_l$, *respectively.*

*1* **Start Stage 1: Initialize caregiver clusters:**
**forall** *caregivers* $l \in L$
  **Assign** *patient* $i \in \widehat{V_1}$ *where* $\widehat{V_1} \leftarrow V_1$ *to cluster* $H_l$ *if*
  - *patient* $i$ *can be treated by caregiver* $l$ *and the distance between* $p_i$ *and HHC is the maximum*

  **Update** $\widehat{V_1} \leftarrow \widehat{V_1}/\{i\}$, $H_l = \{i\}$, $ch_l = p_i$.
***If*** $H_{l'} = \emptyset$, $l' \in L$ *then* **forall** *caregivers* $l'$
  **Find** *a proper patient* $i$ *from the created clusters* $H_l$, **Remove** *it from that cluster and* **Assign** *it to* $H_{l'}$.
  **Assign** *the furthest patient* $i \in \widehat{V_1}$ *that can be treated by caregiver* $l$ *to* $H_l$
  **Update** $H_{l'} = \{i\}$, $ch_{l'} = p_i$, $H_l = \{i'\}$, $ch_l = p_{i'}$, $\widehat{V_1} \leftarrow \widehat{V_1}/\{i'\}$

*2* **Complete and Improve clusters (K-means algorithm with qualification constraint)**
**Repeat forall** *patient* $i \in V_1$
  **forall** *caregiver clusters* $l \in L$ *such that* $H_l \cap \{i\} = \emptyset$
    **Find** *the nearest cluster* $H_l$ *where caregiver* $l$ *can treat patient* $i$. *If there is no such a cluster,* **Move** *to the next patient. Otherwise;*
    **Remove** *patient* $i$ *from its clusters* $H_{l'}$ *and* **Add** *it to* $H_l$.
    **Update** $H_{l'} \leftarrow H_{l'}/\{i\}$, $H_l \leftarrow H_l \cup \{i\}$, $ch_l = \sum_{i \in H_l} p_i /|H_l|$, $l \in L$

**until** *the maximum number of iterations is reached or there is no further improvement*

3 **Recluster patients to balance total service workload**

**Compute** *the maximum total service time allowed per worker:* $\overline{ts} = \sum_{i \in V_1} \sum_{s \in S} p_{is}/l + \max_{i \in V_1, s \in S} p_{is}$

**forall** *caregiver clusters* $H_l, l \in L$

**If** *total workload in cluster* $l$ *exceeds the maximum allowance:* $ts_l > \overline{ts}$ *such that* $ts_l = \sum_{i \in H_l} \sum_{s \in S} p_{is}/|H_l|$.

**Remove** *the furthest patient* $i'$ *in cluster* $H_l$ **until** $ts_l \leq \overline{ts}$. **Assign** *patient* $i'$ *to a candidate list* $CL$.

**Update** $H_l \leftarrow H_l/\{i'\}$, $CL \leftarrow CL \cup \{i'\}$, $ch_l = \sum_{i \in H_l} p_i / |H_l|$, $l \in L$

**forall** *patient* $i \in CL$

**Find** *the nearest cluster* $H_l$ *where caregiver* $l$ *can treat patient* $i$ *and* $ts_l + p_{is} \leq \overline{ts}$. *Then* **Assign** *patient* $i$ *to* $H_l$.

**Update** $H_l \leftarrow H_l \cup \{i\}$, $CL \leftarrow CL/\{i\}$, $ch_l = \sum_{i \in H_l} p_i / |H_l|$, $l \in L$, $ts_l = ts_l + p_{is}$

**If** $CL \neq \emptyset$, **Repeat**

**Assign** *patient* $i \in CL$ *to the nearest* $H_l$ *where caregiver* $l$ *can treat patient* $i$ *even if total workload exceeds* $\overline{ts}$

**Update** $H_l \leftarrow H_l \cup \{i\}$, $CL \leftarrow CL/\{i\}$, $ch_l = \sum_{i \in H_l} p_i / |H_l|$, $l \in L$, $ts_l = ts_l + p_{is}$

**Until** $CL = \emptyset$

**STOP**

**Algorithm A.2.** The pseudocode of the second and third stages of the proposed UBA

***input:*** *Caregiver clusters* $H_l$ *and its centroid* $ch_l$ *from Algorithm 1. Set of vehicles* $K$, *maximum daily working time* $wTime$, *penalty cost of unvisited patient* $unv$, *capacity of vehicle* $c$.

***output:*** $A_k$ *and* $cv_k$ *that indicate the vehicle cluster and its centroid, respectively.* $H_l$ *is the visited patient list by caregiver* $l$, $u$ *is the unvisited patient list,* $z_k$ *is the tour length of vehicle* $k$, $\mu$ *is the total fitness value of the solution,* $\pi_k$ *is the route of vehicle* $k$,.

5 **Start Stage 2: Create vehicle clusters:**

**Assign** *the furthest caregiver cluster* $H_l$ *from the HHC to the first vehicle cluster.*

**Update** $A_1 \leftarrow A_1 + \{H_{l'}\}$, $cv_1 = ch_{l'}$, $H_l$, $L' \leftarrow L/\{l'\}$.

**forall** *vehicle cluster* $k \in K/\{1\}$

**Assign** *the furthest caregiver cluster* $H_l$, $l \in L'$ *from the centroid of the previously initialized vehicle clusters* $j = 1, \ldots, k-1$ *to the vehicle cluster.*

**Update** $A_k \leftarrow A_k + \{H_{l'}\}$, $cv_k = ch_{l'}$, $L' \leftarrow L'/\{l'\}$

**forall** *caregiver clusters* $H_l, l \in L'$

**Assign** *the nearest* $H_l$ *to vehicle cluster* $A_k$ *such that the capacity of vehicle* $k$ *is not exceeded such that* $|A_k| \leq c$

**Update** $A_k \leftarrow A_k + \{H_l\}$, $cv_k = \sum_{i \in H_l \in A_k} p_i / \sum_{H_l \in A_k} |H_l|$, $L' = L'/\{l\}$

**STOP**

6 **Start Stage 3: Construct the optimal routes of the vehicles**

**forall** *vehicle cluster* $k \in K$

**Solve** *the respective TSP using IBM ILOG CPLEX 12.6 to obtain the optimal route,*$\pi_k$, *of* $A_k$

*If* $z_k > wTime$

**Repeat forall** *patient* $i \in A_k$

**Compute** *the costliest patient* $i$ *considering its contribution to* $z_k$ *as similar to Clarke and Wright's savings algorithm. Let the cost of patient* $i$ *be* $d_i$.

**Remove** *patient* $i$ *and*

**If** $k \neq |K|$

**Assign** *it to the* $A_{k+1}$, *patient list of vehicle* $k+1$.

**Else**

**Assign** *it to the unvisited patient list* $u$.

**Update** $A_k \leftarrow A_k/\{i\}$, $u \leftarrow u \cup \{i\}$, $z_k = z_k - d_i - p_i$, $cv_k = \sum_{i \in A_k} p_i / |A_k|$,

**Until** $z_k \leq wTime$

**STOP**

**Algorithm A.3.** The pseudocode of the feedback loop and repair function of the proposed UBA

***input:*** *Set of vehicles* $K$, *Route of vehicles* $\pi_k = \{v_0, v_1, \dots, v_i, \dots, v_{2n+1}\}$, , $t_{i,j}$ *travel time between node* $i$ *and* $j$, $\delta_{i,j}^{k_1,k_2}$ *relocate value of assigning patient* $v_i$ *from vehicle* $k_1$ *to position* $j$ *of vehicle* $k_2$, $u$ *is the unvisited patient list, penalty cost of unvisited patient* $unv$

***output*** $z_k$ *is the tour length of vehicle* $k$, $\pi_k$ *is the route of vehicle* $k$, $\mu$ *is the total fitness value of the solution.*

*7* **Feedback Loop:**

**Do**

**forall** *vehicle* $k_1 \in K$

**forall** *position* $i \in \pi_{k_1}$ *such that* $v_i$ *is the patient of position* $i$

**forall** *vehicle* $k_2 \in K/\{k_1\}$

**forall** *feasible positions* $j \in \pi_{k_2}$

**Compute** *the relocate value* $\delta_{i,j}^{k_1,k_2}$ *such that,*

$$\delta_{i,j}^{k_1,k_2} = t_{v_{i-1},v_i} + t_{v_i,v_{i+1}} + t_{v_{j-1},v_j} - \left(t_{v_{j-1},v_i} + t_{v_{j-1},v_i}\right)$$

**end for**

**end for**

**end for**

**end for**

**Determine** $\delta_{i^*,j^*}^{k_1^*,k_2^*} = \max_{i,j,k_1,k_2} \left\{\delta_{i,j}^{k_1,k_2}\right\}$

**If** $\delta_{i^*,j^*}^{k_1^*,k_2^*} > 0$

**Remove** *the patient* $v_{i^*}$ *of position i from the vehicle* $k_1^*$ *and* **Assign** *it to the position* $j^*$ *of the vehicle* $k_2^*$

**Update** $z_{k_1^*}$ *and* $z_{k_2^*}$, *tour length of vehicles* $k_1^*$ *and* $k_2^*$, *respectively*

**While** $\delta_{i^*,j^*}^{k_1^*,k_2^*} > 0$

**STOP**

*8* **Repair Function:**

**While** $u \neq \emptyset$ **and** *there is any feasible assignment*

**forall** *unvisited patient* $v_i \in u$

**forall** *route of vehicle* $\pi_k \in K$

**forall** *feasible position* $j \in \pi_k$

**Compute** *the insertion cost of patient* $v_i$ *into position* $j$ *of vehicle* $\pi_k$

**end for**

**end for**

**end for**

**Insert** *the patient into the determined position of the vehicle that has minimum insertion cost*

**Update** $\pi_k \in K, z_k \in K, u$

**end while**

**Compute** *total service and travel time of the visited patients and the penalty cost for unvisited patients* $\mu$:

$\mu = \sum_k z_k + \sum_{l \in L} \sum_{i \in H_l} \sum_{s \in S} p_{is} + \sum_{i \in u_i} unv$

Table A.1. The best-found CPLEX solutions for 10-patient instances and their comparisons with the UBA.

| | **CPLEX** | | **UBA** | | | | **CPLEX** | | **UBA** | | |
|---|---|---|---|---|---|---|---|---|---|---|---|
| **Instance** | **Best-found Integer*** | **GAP (%)** | **UB**** | **CPU** | **% Imp.** | **Instance** | **Best-found Integer*** | **GAP (%)** | **UB**** | **CPU** | **% Imp.** |
| h10_10_0_0 | 258.5 | 39 | 335.1 | 0.11 | 22.9 | h10_30_0_0 | 506.5 | 49.2 | 557.4 | 0.05 | 9.1 |
| h10_10_0_1 | 256.3 | 35.5 | 341.4 | 0.05 | 24.9 | h10_30_0_1 | 531.9 | 46.9 | 614.9 | 0.03 | 13.5 |
| h10_10_0_2 | 248.5 | 30.5 | 338.3 | 0.06 | 26.5 | h10_30_0_2 | 503.5 | 34.6 | 533.9 | 0.03 | 5.7 |
| h10_10_0_3 | 255.8 | 38.6 | 343.9 | 0.05 | 25.6 | h10_30_0_3 | 522.2 | 52.4 | 580.4 | 0.03 | 10.0 |
| h10_10_0_4 | 233.8 | 29.1 | 318.3 | 0.03 | 26.5 | h10_30_0_4 | *NA* | *NA* | 514.2 | 0.03 | *NA* |
| h10_10_1_0 | 258.1 | 30.1 | 379.1 | 0.04 | 31.9 | h10_30_1_0 | 550.5 | 50.6 | 601.4 | 0.02 | 8.5 |
| h10_10_1_1 | 288.8 | 35.7 | 385.4 | 0.05 | 25.1 | h10_30_1_1 | 554.4 | 44.4 | 658.9 | 0.02 | 15.9 |
| h10_10_1_2 | 267.7 | 27 | 382.3 | 0.04 | 30.0 | h10_30_1_2 | 533.2 | 34.8 | 577.9 | 0.03 | 7.7 |
| h10_10_1_3 | 278.2 | 35.6 | 387.9 | 0.04 | 28.3 | h10_30_1_3 | 522.7 | 48.7 | 624.4 | 0.07 | 16.3 |
| h10_10_1_4 | 255.4 | 26.1 | 362.3 | 0.03 | 29.5 | h10_30_1_4 | 483.6 | 38.9 | 558.2 | 0.03 | 13.4 |
| h10_10_2_0 | 299.3 | 29.5 | 431.1 | 0.03 | 30.6 | h10_30_2_0 | 628.9 | 52 | 638.6 | 0.04 | 1.5 |
| h10_10_2_1 | 307.8 | 29.2 | 447.4 | 0.03 | 31.2 | h10_30_2_1 | 629.6 | 48.2 | 720.9 | 0.09 | 12.7 |
| h10_10_2_2 | 281.8 | 19.6 | 444.3 | 0.03 | 36.6 | h10_30_2_2 | 580 | 38.4 | 639.9 | 0.05 | 9.4 |
| h10_10_2_3 | 308 | 29.9 | 449.9 | 0.03 | 31.5 | h10_30_2_3 | 576.2 | 48 | 686.4 | 0.10 | 16.1 |
| h10_10_2_4 | 295.1 | 25.9 | 424.3 | 0.03 | 30.4 | h10_30_2_4 | *NA* | *NA* | 620.2 | 0.06 | *NA* |
| h10_20_0_0 | 399.9 | 47.9 | 446.8 | 0.03 | 10.5 | h10_40_0_0 | 567.4 | 45.4 | 668.6 | 0.02 | 15.1 |
| h10_20_0_1 | 403 | 45.2 | 469.1 | 0.03 | 14.1 | h10_40_0_1 | 644.7 | 43.7 | 728.2 | 0.08 | 11.5 |
| h10_20_0_2 | 378.4 | 38.1 | 452.7 | 0.06 | 16.4 | h10_40_0_2 | 614.7 | 39.7 | 637.0 | 0.06 | 3.5 |
| h10_20_0_3 | 390.4 | 49 | 457.7 | 0.22 | 14.7 | h10_40_0_3 | 646.7 | 49.7 | 698.6 | 0.14 | 7.4 |
| h10_20_0_4 | 418.6 | 48.1 | 418.7 | 0.03 | 0.0 | h10_40_0_4 | *NA* | *NA* | 608.7 | 0.03 | *NA* |
| h10_20_1_0 | 404.4 | 42.2 | 490.8 | 0.03 | 17.6 | h10_40_1_0 | 621 | 47.7 | 712.6 | 0.03 | 12.9 |
| h10_20_1_1 | 412.1 | 41.9 | 513.1 | 0.03 | 19.7 | h10_40_1_1 | 684.4 | 47.2 | 772.2 | 0.02 | 11.4 |
| h10_20_1_2 | 395 | 44 | 476.0 | 0.03 | 17.0 | h10_40_1_2 | 676.3 | 40.3 | 681.0 | 0.05 | 0.7 |
| h10_20_1_3 | 398.1 | 39.9 | 501.7 | 0.04 | 20.6 | h10_40_1_3 | 660.8 | 49.4 | 742.6 | 0.10 | 11.0 |
| h10_20_1_4 | *NA* | *NA* | 462.7 | 0.03 | *NA* | h10_40_1_4 | 652.6 | 47.2 | 652.7 | 0.05 | 0.0 |
| h10_20_2_0 | 465.6 | 45.2 | 535.6 | 0.03 | 13.1 | h10_40_2_0 | 708.7 | 50.3 | 741.5 | 0.03 | 4.4 |
| h10_20_2_1 | 466.4 | 43.2 | 575.1 | 0.02 | 18.9 | h10_40_2_1 | 745.2 | 47 | 834.2 | 0.02 | 10.7 |
| h10_20_2_2 | 445.8 | 37.1 | 558.7 | 0.02 | 20.2 | h10_40_2_2 | 708.6 | 39.6 | 743.0 | 0.03 | 4.6 |
| h10_20_2_3 | 451 | 44.1 | 563.7 | 0.04 | 20.0 | h10_40_2_3 | 687.7 | 48.1 | 804.6 | 0.07 | 14.5 |
| h10_20_2_4 | 384.1 | 29.7 | 524.7 | 0.03 | 26.8 | h10_40_2_4 | 637.8 | 41.2 | 714.7 | 0.02 | 10.8 |

*Best-found integer: The best objective function values of the best-found integer solutions by CPLEX.

**UB: the objective function values of the solutions obtained by the proposed UBA.

Table A.2. The ALNS-VS solutions and their comparisons with CPLEX solutions for 10-patient instances.

| **Instance** | **Best-found** | **Avg.** | **#DP** | **CPU** | **% Imp.** | **Instance** | **Best-found** | **Avg.** | **#DP** | **CPU** | **% Imp.** |
|---|---|---|---|---|---|---|---|---|---|---|---|
| h10_10_0_0 | 240.5 | 246.3 | 5.4 | 1.7 | 6.9 | h10_30_0_0 | 452.5 | 468.5 | 2.0 | 2.0 | 10.7 |

| h10_10_0_1 | 254.3 | 256.4 | 5.2 | 1.8 | 0.8 | h10_30_0_1 | 527.3 | 527.3 | 2.0 | 1.8 | 0.9 |
|---|---|---|---|---|---|---|---|---|---|---|---|
| h10_10_0_2 | 231.0 | 231.8 | 6.0 | 2.0 | 7.0 | h10_30_0_2 | 500.3 | 500.3 | 2.0 | 1.7 | 0.6 |
| h10_10_0_3 | 255.8 | 261.0 | 4.4 | 1.8 | 0.0 | h10_30_0_3 | 504.6 | 504.6 | 3.0 | 1.7 | 3.4 |
| h10_10_0_4 | 217.2 | 221.4 | 4.4 | 1.6 | 7.1 | h10_30_0_4 | 449.0 | 451.3 | 2.4 | 1.6 | *NA* |
| h10_10_1_0 | 252.4 | 257.5 | 6.2 | 1.9 | 2.2 | h10_30_1_0 | 496.3 | 515.3 | 3.8 | 2.0 | 9.9 |
| h10_10_1_1 | 282.0 | 283.3 | 5.4 | 1.7 | 2.4 | h10_30_1_1 | 544.6 | 544.6 | 4.0 | 1.7 | 1.8 |
| h10_10_1_2 | 243.0 | 245.4 | 7.0 | 1.9 | 9.2 | h10_30_1_2 | 535.1 | 549.2 | 4.4 | 1.8 | -0.4 |
| h10_10_1_3 | 272.0 | 275.9 | 6.2 | 1.8 | 2.2 | h10_30_1_3 | 527.8 | 527.8 | 3.0 | 1.8 | -1.0 |
| h10_10_1_4 | 235.8 | 238.8 | 6.0 | 1.7 | 7.7 | h10_30_1_4 | 484.2 | 484.2 | 4.2 | 1.7 | -0.1 |
| h10_10_2_0 | 278.0 | 287.9 | 6.8 | 1.9 | 7.1 | h10_30_2_0 | 504.8 | 531.8 | 3.6 | 1.9 | 19.7 |
| h10_10_2_1 | 305.3 | 308.0 | 6.8 | 1.8 | 0.8 | h10_30_2_1 | 548.5 | 598.4 | 3.4 | 1.7 | 12.9 |
| h10_10_2_2 | 276.6 | 277.3 | 6.0 | 1.8 | 1.9 | h10_30_2_2 | 577.6 | 578.9 | 3.2 | 1.7 | 0.4 |
| h10_10_2_3 | 303.2 | 306.3 | 4.8 | 1.8 | 1.6 | h10_30_2_3 | 548.7 | 551.4 | 4.0 | 2.0 | 4.8 |
| h10_10_2_4 | 279.2 | 282.9 | 5.0 | 1.6 | 5.4 | h10_30_2_4 | 496.1 | 496.1 | 4.0 | 1.6 | *NA* |
| h10_20_0_0 | 361.2 | 378.7 | 2.8 | 1.9 | 9.7 | h10_40_0_0 | 542.6 | 562.3 | 2.0 | 1.9 | 4.4 |
| h10_20_0_1 | 401.0 | 401.0 | 4.0 | 1.8 | 0.5 | h10_40_0_1 | 588.4 | 633.4 | 2.0 | 1.9 | 8.7 |
| h10_20_0_2 | 346.7 | 372.3 | 2.6 | 1.7 | 8.4 | h10_40_0_2 | 539.1 | 591.4 | 1.0 | 1.7 | 12.3 |
| h10_20_0_3 | 384.5 | 384.5 | 3.2 | 1.8 | 1.5 | h10_40_0_3 | 619.2 | 619.2 | 3.0 | 1.8 | 4.2 |
| h10_20_0_4 | 339.5 | 339.5 | 4.0 | 1.7 | 18.9 | h10_40_0_4 | 551.8 | 551.8 | 2.0 | 1.6 | *NA* |
| h10_20_1_0 | 357.1 | 377.2 | 5.4 | 2.0 | 11.7 | h10_40_1_0 | 593.3 | 620.0 | 3.0 | 1.9 | 4.5 |
| h10_20_1_1 | 413.1 | 413.2 | 3.6 | 1.8 | -0.2 | h10_40_1_1 | 665.4 | 668.6 | 4.4 | 1.7 | 2.8 |
| h10_20_1_2 | 357.4 | 385.4 | 6.0 | 1.9 | 9.5 | h10_40_1_2 | 606.2 | 649.8 | 1.8 | 1.7 | 10.4 |
| h10_20_1_3 | 397.7 | 398.0 | 4.2 | 1.7 | 0.1 | h10_40_1_3 | 650.3 | 650.3 | 3.0 | 1.9 | 1.6 |
| h10_20_1_4 | 356.8 | 356.8 | 5.6 | 1.6 | *NA* | h10_40_1_4 | 600.5 | 600.5 | 2.8 | 1.6 | 8.0 |
| h10_20_2_0 | 409.0 | 438.0 | 4.4 | 1.9 | 12.1 | h10_40_2_0 | 601.5 | 632.5 | 3.0 | 2.0 | 15.1 |
| h10_20_2_1 | 428.7 | 460.9 | 4.8 | 1.7 | 8.1 | h10_40_2_1 | 663.9 | 731.0 | 3.0 | 1.8 | 10.9 |
| h10_20_2_2 | 374.0 | 424.4 | 3.0 | 1.8 | 16.1 | h10_40_2_2 | 608.2 | 676.7 | 2.0 | 1.8 | 14.2 |
| h10_20_2_3 | 433.7 | 435.7 | 4.0 | 2.0 | 3.8 | h10_40_2_3 | 666.2 | 668.4 | 3.6 | 2.1 | 3.1 |
| h10_20_2_4 | 380.6 | 387.3 | 4.0 | 1.7 | 0.9 | h10_40_2_4 | 600.7 | 604.7 | 4.2 | 1.7 | 5.8 |

Table A.3. The aggregated results of ALNS-VS, UBA, ALNS_M, ALNS-STD and their comparisons.

| Instance | UBA | UBA+DP | ALNS-VS | ALNS-M | ALNS-STD | UBA+DP-UBA (%) | VS-UBA+DP (%) | VS-M (%) | STD-VS (%) | *BER* |
|---|---|---|---|---|---|---|---|---|---|---|
| h10_10_0 | 335.4 | 296.5 | 239.98 | 329.18 | 176.67 | 11.6 | 18.9 | 27.2 | 26.0 | 1.2 |
| h10_10_1 | 379.4 | 305.3 | 258.16 | 373.18 | 198.52 | 19.6 | 15.8 | 31.2 | 22.4 | 0.9 |
| h10_10_2 | 439.4 | 371.6 | 289.65 | 435.18 | 229.52 | 15.5 | 21.9 | 33.7 | 20.3 | 0.7 |
| h10_20_0 | 449.0 | 414.5 | 373.16 | 434.10 | 239.80 | 7.7 | 11.4 | 15.6 | 34.3 | 2.4 |
| h10_20_1 | 488.9 | 438.3 | 387.68 | 478.10 | 261.80 | 10.4 | 14.0 | 21.3 | 30.2 | 1.7 |
| h10_20_2 | 551.6 | 479.7 | 431.68 | 540.10 | 293.14 | 13.0 | 15.5 | 25.0 | 27.4 | 1.3 |
| h10_30_0 | 560.2 | 534.9 | 490.83 | 540.08 | 304.79 | 4.6 | 8.8 | 9.9 | 37.2 | 3.2 |
| h10_30_1 | 604.2 | 567.5 | 526.23 | 584.08 | 326.79 | 6.1 | 8.5 | 11.4 | 36.8 | 2.9 |
| h10_30_2 | 661.2 | 595.7 | 553.69 | 646.08 | 357.79 | 9.9 | 10.0 | 17.2 | 32.9 | 2.1 |
| h10_40_0 | 668.2 | 645.6 | 599.49 | 645.62 | 368.88 | 3.4 | 11.8 | 11.9 | 34.9 | 2.6 |
| h10_40_1 | 712.2 | 689.0 | 641.80 | 689.62 | 391.43 | 3.3 | 9.3 | 9.6 | 37.1 | 3.0 |
| h10_40_2 | 767.6 | 711.5 | 670.93 | 751.62 | 421.88 | 7.4 | 11.5 | 16.4 | 32.7 | 2.0 |
| h30_10_0 | 877.0 | 683.1 | 560.65 | 859.66 | 459.52 | 22.1 | 17.8 | 34.8 | 18.0 | 0.6 |
| h30_10_1 | 960.5 | 730.8 | 603.58 | 941.66 | 501.36 | 23.9 | 17.3 | 35.9 | 16.9 | 0.5 |
| h30_10_2 | 1104.3 | 860.0 | 665.52 | 1091.79 | 573.90 | 22.2 | 22.3 | 39.0 | 13.8 | 0.4 |
| h30_20_0 | 1057.3 | 898.5 | 785.82 | 1030.83 | 569.02 | 15.0 | 12.5 | 23.8 | 27.6 | 1.2 |
| h30_20_1 | 1131.1 | 947.4 | 817.17 | 1119.01 | 609.38 | 16.2 | 13.7 | 27.0 | 25.4 | 1.0 |
| h30_20_2 | 1283.3 | 1053.3 | 885.24 | 1265.90 | 684.17 | 17.9 | 15.9 | 30.1 | 22.7 | 0.8 |
| h30_30_0 | 1221.8 | 1106.4 | 1001.70 | 1202.96 | 676.04 | 9.5 | 9.5 | 16.7 | 32.5 | 1.9 |
| h30_30_1 | 1307.3 | 1169.7 | 1040.87 | 1287.13 | 719.03 | 10.5 | 10.9 | 19.1 | 30.9 | 1.6 |
| h30_30_2 | 1466.3 | 1272.1 | 1101.78 | 1434.96 | 789.51 | 13.3 | 13.3 | 23.2 | 28.3 | 1.3 |
| h30_40_0 | 1411.7 | 1323.2 | 1211.49 | 1380.25 | 789.16 | 6.3 | 8.4 | 12.2 | 34.9 | 2.3 |
| h30_40_1 | 1512.8 | 1402.3 | 1259.11 | 1466.82 | 829.19 | 7.4 | 9.9 | 14.2 | 34.1 | 2.2 |
| h30_40_2 | 1682.4 | 1492.1 | 1322.80 | 1613.96 | 904.68 | 11.4 | 11.2 | 18.0 | 31.6 | 1.7 |
| h50_10_0 | 1397.1 | 1123.8 | 917.19 | 1362.34 | 741.10 | 19.6 | 18.3 | 32.7 | 19.2 | 0.6 |
| h50_10_1 | 1673.3 | 1339.2 | 1049.48 | 1634.19 | 878.71 | 20.0 | 21.5 | 35.8 | 16.3 | 0.5 |
| h50_10_2 | 1976.9 | 1538.8 | 1207.55 | 1941.84 | 1030.20 | 22.2 | 21.4 | 37.8 | 14.7 | 0.4 |
| h50_20_0 | 1686.0 | 1490.6 | 1265.13 | 1617.41 | 928.82 | 11.6 | 15.1 | 21.8 | 26.6 | 1.1 |
| h50_20_1 | 1958.2 | 1711.2 | 1394.07 | 1890.92 | 1065.76 | 12.6 | 18.5 | 26.3 | 23.6 | 0.9 |
| h50_20_2 | 2266.3 | 1911.4 | 1558.52 | 2196.37 | 1218.73 | 15.7 | 18.3 | 29.0 | 21.8 | 0.8 |
| h50_30_0 | 1966.6 | 1817.2 | 1594.62 | 1875.04 | 1119.60 | 7.7 | 12.1 | 15.0 | 29.8 | 1.5 |
| h50_30_1 | 2251.2 | 2031.2 | 1729.75 | 2154.47 | 1255.32 | 9.8 | 14.6 | 19.7 | 27.4 | 1.2 |
| h50_30_2 | 2562.1 | 2231.5 | 1890.13 | 2464.78 | 1407.72 | 12.9 | 15.2 | 23.3 | 25.5 | 1.1 |
| h50_40_0 | 2279.4 | 2139.7 | 1916.01 | 2138.98 | 1310.78 | 6.1 | 10.4 | 10.4 | 31.6 | 1.7 |
| h50_40_1 | 2925.2 | 2756.1 | 2051.53 | 2416.16 | 1446.51 | 6.0 | 20.9 | 15.1 | 29.5 | 1.4 |
| h50_40_2 | 5162.6 | 4357.0 | 2256.00 | 2742.02 | 1598.29 | 14.2 | 45.8 | 17.7 | 29.2 | 1.4 |
| h100_10_0 | 2928.8 | 2433.3 | 2010.13 | 2870.71 | 1580.91 | 16.9 | 17.3 | 30.0 | 21.4 | 0.8 |
| h100_10_1 | 3600.9 | 2879.8 | 2360.71 | 3553.15 | 1920.09 | 20.0 | 18.0 | 33.6 | 18.7 | 0.6 |
| h100_10_2 | 3759.8 | 2954.9 | 2464.20 | 3719.20 | 1999.56 | 21.4 | 16.6 | 33.7 | 18.9 | 0.6 |
| h100_20_0 | 3499.7 | 3112.4 | 2659.69 | 3389.20 | 1967.69 | 11.1 | 14.5 | 21.5 | 26.0 | 1.1 |
| h100_20_1 | 4206.4 | 3632.8 | 3046.90 | 4082.18 | 2308.09 | 13.6 | 16.1 | 25.4 | 24.2 | 0.9 |
| h100_20_2 | 4341.8 | 3722.0 | 3141.26 | 4230.52 | 2391.76 | 14.3 | 15.6 | 25.7 | 23.9 | 0.9 |
| h100_30_0 | 4042.1 | 3751.3 | 3300.48 | 3909.16 | 2366.60 | 7.2 | 12.0 | 15.6 | 28.3 | 1.3 |
| h100_30_1 | 4710.9 | 4183.6 | 3677.09 | 4604.94 | 2702.44 | 11.2 | 12.1 | 20.1 | 26.5 | 1.1 |
| h100_30_2 | 4863.1 | 4352.5 | 3783.61 | 4783.56 | 2782.05 | 10.5 | 13.0 | 20.9 | 26.5 | 1.1 |
| h100_40_0 | 4613.2 | 4391.1 | 3968.25 | 4429.90 | 2755.97 | 4.8 | 9.6 | 10.4 | 30.5 | 1.6 |
| h100_40_1 | 6248.5 | 5493.3 | 4400.96 | 5192.64 | 3106.26 | 11.6 | 19.1 | 15.2 | 29.4 | 1.4 |
| h100_40_2 | 7098.9 | 6349.7 | 4535.37 | 5336.92 | 3173.82 | 10.6 | 27.4 | 15.0 | 30.0 | 1.5 |

## Appendix B Some details about the heuristics used in the proposed ALNS-VS algorithm

The removal and insertion heuristics used in the ALNS-VS algorithm are explained briefly below. Additionally, the pseudocode of the drop-off and pick-up local search heuristic algorithm and the repair function are given in Algorithms B.1 and B.2.

**Random Removal:** This heuristic algorithm randomly removes $q$ patients from the current solution $x_{curr}$ and adding them to the request bank R.

**Worst Removal:** This heuristic algorithm selects $q$ costliest patients in terms of distance from the current solution. The heuristic removes the selected patient $i \in x_{curr}$ from the current solution $x_{curr}$ and adds them to $R$. After removing patient $i$, the cost of the $x_{curr}$ is calculated as $f_{-i}$, whereas the cost of $i$ can be calculated as $\Delta f_i = f(x_{curr}) - f_{-i}$.

**Shaw Removal:** The main objective of this heuristic algorithm is to remove the most similar patients in terms of their locations and service times. The heuristic starts with selecting a random patient $i \in x_{curr}$ and adding it to the request bank $R$. The similarity measures $(d_{ij})$ between the selected patient $i$ and the rest of the patients $j \in \frac{x_{curr}}{\{i\}}$ in the solution $x_{curr}$ are calculated by $d_{ij} = \alpha * t_{i,j} + \beta * (|p_i - p_j|)$. In our problem, the lower the $d_{ij}$ is the higher the similarity. The most similar patient $j^*$ is selected and added to $R$ such that $j^* = argmin_{j \in x_{curr}} d_{ij}$, where $\alpha$ and $\beta$ are the shaw parameters, $p_i$ and $p_j$ are the service times of patients $i$ and $j$, and $t_{i,j}$ is the travel time between patient nodes $i$ and $j$. This heuristic algorithm is iteratively applied $q$ times to determine the removed patients such that the patient has the maximum similarity measure with the last removed patient.

**Route Removal:** This heuristic algorithm randomly selects a route of a vehicle $\underline{v}$ from $v$ (a set of routes of vehicles in $x_{curr}$), removes all the patients from it, and adds them to the $R$. The idea of route removal is to redesign the route to minimize the travel time by diversifying the search.

**Greedy Insertion:** All of the patients from $R$ are assigned to all possible positions of the routes $v$ of caregivers and an insertion cost is calculated for each position through $\Delta^l_{i,k,j} = t_{i,k} + t_{k,j} - t_{i,j}$ for $i, j = 1, \dots, n$ and $i \neq j$. In this process, only feasible assignments are considered. After insertion cost is calculated for all patients, the patient with the least insertion cost is assigned to determine the position of the route of the vehicle. This process continues until all patients are assigned to a route or no more insertion is possible. Since at each iteration only one route of a vehicle is changed, the insertion cost for the other routes does not need to be recalculated. This idea improves the computation time for all of the insertion heuristics.

**Greedy Insertion with Noise:** The idea of adding noise to the insertion cost is to provide randomization to the search process. This is done by considering the degree of freedom in determining the best location for a node. The steps of greedy insertion heuristic remain the same while the new insertion cost is calculated by $\Delta^l_{i,k,j} = t_{i,k} + t_{k,j} - t_{i,j} + t_{max} * \mu * \varepsilon$, where $t_{max}$ is the maximum time between patients, $\mu$ is the noise parameter which is used for the diversification and set to 0.1, and $\varepsilon$ is a random number between [-1,1].

**Regret-k Insertion**: Regret-k heuristics are proposed by Potvin and Rousseau (1993). Contrary to the greedy insertion, this heuristic considers the $k$ best positions (depending on choice) instead of the best one. Patients are assigned to positions to maximize the regret cost $(cost_i^k)$ which is computed as the difference between $k$ best position costs $\Delta_{i,m,j}^{l}$ i.e. change in objective value by inserting patient $m$ between patients $i$ and $j$ in route $v$. In this respect, the greedy heuristic can be seen as a regret-1 heuristic. The proposed algorithm considers regret-2 and regret-3 insertions.

**Regret-k Insertion with Noise:** The steps of this insertion heuristic are similar to the regret-k insertion heuristics but use the same cost function as discussed in the greedy insertion with noise.

**Algorithm B.1.** The framework of the drop-off and pick-up local search heuristic algorithm.

---

**input:** Route of vehicle $\pi_k, k \in K$ in the $x_{curr}$, and the saving of dropping the caregiver $l$ off at the patient $i$ and picking up after visiting the node $j$ by vehicle $k$ i.e. $dp_{i,j,k}^{l}$

**output:** A new feasible solution $x_{new}$

---

**forall** route of vehicle in $\pi_k, k \in K$

  **do**

    **forall** *caregivers* $l \in \pi_k^l$ *in vehicle* $k$

      **forall** *patients* $i \in \pi_k$

        **forall** patients $j \in \pi_k$ that are being visited after patient $i$

          *drop caregiver* $l$ *off at patient* $i$, *then add patient* $i$*'s dummy node after patient* $j$, *and calculate* $dp_{i,j,\pi_k}^{l}$ *using equation (35) in section 4.2.*

        **end for**

      **end for**

    **end for**

    **Update** $\pi_k$ *with the drop-off and picking-up decision where the maximum positive* $dp_{i,j,\pi_k}^{l}$ *occurs if it exists. Then, update the current solution.*

  **while** $dp_{i,j,\pi_k}^{l} > 0$

**end for**

**return** A new improved feasible solution $x_{new} \leftarrow x_{curr}$

---

**Algorithm B.2.** The pseudocode of the repair function to ensure feasibility.

---

**input:** Routes of vehicles $\pi_k$, $k \in K$ in the $x_{curr}$, where, $\pi_k = \{0, \ldots v_{i-1}, v_i, v_{i+1}, \ldots, 2n+1\}$, travel time between patient $i$ and $j$, $t_{i,j}$, time of arrival at node $i$, $av_{i,k}$, service time of patient $i, p_{i,s}$, maximum working time $wTime$ and request bank $R$.

**output:** Feasible solution $x_{curr}$

---

**forall** *vehicle routes* $\pi_k, k \in K$ *in the* $x_{curr}$,

  **while** $av_{2n+1,k} > wTime$

    **forall** *patient nodes* $v_i$ *in vehicle* $k$

      $cost_{v_i,k} = t_{v_{i-1},v_{i+1}} - t_{v_{i-1},v_i} - t_{v_i,v_{i+1}} + p_{v_i,s}$

    **end for**

    **Remove** *patient* $v_{i^*}$ *from vehicle* $k$, $v_{i^*} = \underset{v_i,k}{argmax}\{cost_{v_i,k}\}$ *and* **add** *to R*

  **end while**

**end for**

**Apply** *greedy insertion to all vehicle routes* $\pi_k$ *with patients that are in request bank* $R$. *Consider the remained patients in request bank* $R$ *as the unvisited patients.*
**return** route of vehicles $\pi_k,\ k \in K$

## Appendix C Details of the parameter tuning tests

The ANOVA results in Table C.1 indicate that all of the parameters are statistically significant since their p-values are less than 0.05. In addition, update solution iteration ($\omega$) has the greatest effect on the algorithm since it has the largest adjusted sum of square (Adj SS). In addition to the main effects, the two-way interaction of caregiver swap iteration ($\varphi$) and maximum remove parameter ($\xi$) is the only statistically significant interaction that affects the algorithm's output. Hence, we do not only consider the main effects but also $\varphi*\xi$ two-way interaction while determining the optimum setting for the parameters. For this purpose, we analyzed the main effects plot and used Response Optimizer module of Minitab 19. As seen in Figure C.1, the best setting for ($\omega$, $\varphi$, $\xi$, $\rho$) that minimizes the output is (250,150, 0.5, 0.95) when only main effects are considered. However, the result of the Response Optimization suggests a change on the value of $\varphi$ from 150 to 100 resulting that the optimal setting is ($\omega$, $\varphi$, $\xi$, $\rho$) = (250,100, 0.5, 0.95) with a 95% confidence interval of (3,185; 4,935).

Table C.1. ANOVA results for parameter tuning of the ALNS-VS.

| *Source* | *df* | *Adj SS* | *Adj MS* | *F-Value* | *p-Value* |
|---|---|---|---|---|---|
| Model | 1049 | 2955.5 | 2.8174 | 0.66 | 1.000 |
| Linear | 19 | 779.1 | 41.0042 | 9.63 | 0.000 |
| $\omega$ | 5 | 399.3 | 79.8577 | 18.75 | 0.000 |
| $\varphi$ | 6 | 81.3 | 13.5511 | 3.18 | 0.004 |
| $\xi$ | 4 | 74.5 | 18.6208 | 4.37 | 0.002 |
| $\rho$ | 4 | 224.0 | 56.0005 | 13.15 | 0.000 |
| 2-Way Interactions | 134 | 685.1 | 5.1127 | 1.20 | 0.057 |
| $\omega*\varphi$ | 30 | 132.5 | 4.4159 | 1.04 | 0.410 |
| $\omega*\xi$ | 20 | 39.3 | 1.9659 | 0.46 | 0.980 |
| $\omega*\rho$ | 20 | 9.9 | 0.4960 | 0.12 | 1.000 |
| $\varphi*\xi$ | 24 | 382.7 | 15.9467 | 3.74 | 0.000 |
| $\varphi*\rho$ | 24 | 51.2 | 2.1332 | 0.50 | 0.980 |
| $\xi*\rho$ | 16 | 69.5 | 4.3415 | 1.02 | 0.431 |
| 3-Way Interactions | 416 | 877.2 | 2.1086 | 0.50 | 1.000 |
| $\omega*\varphi*\xi$ | 120 | 402.2 | 3.3517 | 0.79 | 0.959 |
| $\omega*\varphi*\rho$ | 120 | 179.6 | 1.4966 | 0.35 | 1.000 |
| $\omega*\xi*\rho$ | 80 | 115.3 | 1.4413 | 0.34 | 1.000 |
| $\varphi*\xi*\rho$ | 96 | 180.1 | 1.8757 | 0.44 | 1.000 |
| 4-Way Interactions | 480 | 614.1 | 1.2794 | 0.30 | 1.000 |
| $\omega*\varphi*\xi*\rho$ | 480 | 614.1 | 1.2794 | 0.30 | 1.000 |
| Error | 19950 | 84952.0 | 4.2582 | | |
| Total | 20999 | 87907.4 | | | |

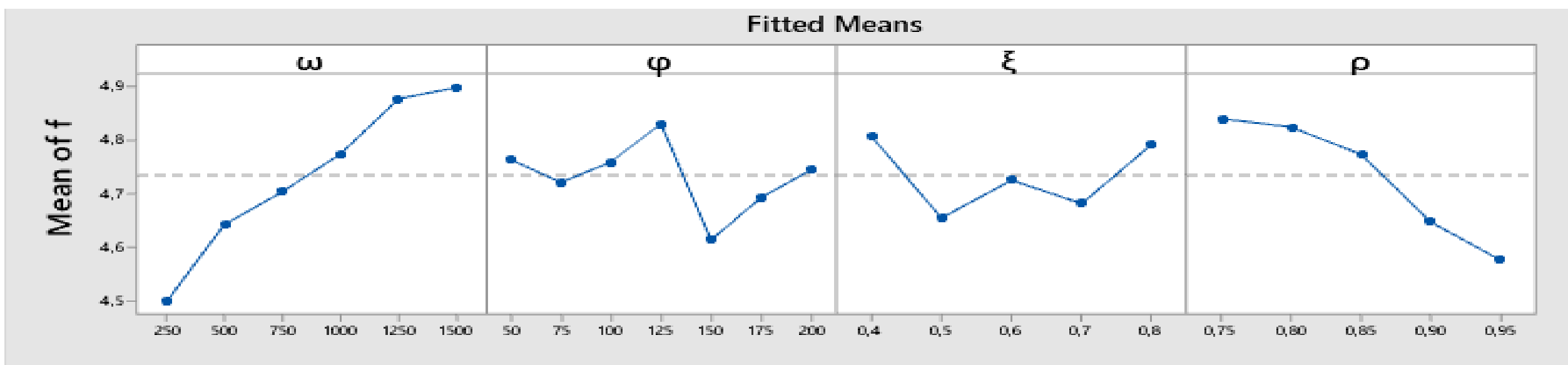

Figure C.1. The main effects plot for parameters.

## Appendix D The ALNS-VS solutions: variants of the caregiver swap heuristic

Tables D.1 through D.4 consists of the best-found solutions of the HHSRP problem instances by the variations of the ALNS-VS algorithm for 10, 30, 50 and 100 patients, respectively. These solutions are used in the analyzes in sections 5.3, 5.4, 5.5 and 5.6. In the following tables, the “Best-found” and “Avg.” columns indicate the objective values of the best-found and the averages of the best solutions found in five replications, respectively. The column “#DP” shows how many times the caregivers were dropped off. Last, “CPU” presents the computational time of the algorithm in seconds.

Table D.1. The best-found solutions for the instances with 10 patients by the ALNS-VS algorithms.

| | ALNS-VS_Unique | | | | ALNS-VS_Common | | | | ALNS-VS_No-Swap | | | |
|---|---|---|---|---|---|---|---|---|---|---|---|---|
| Instance | Best-found | Avg. | # DP | CPU | Best-found | Avg. | # DP | CPU | Best-found | Avg. | # DP | CPU |
| h10_10_0_0 | 240.54 | 246.304 | 5.2 | 1.94 | 240.54 | 246.304 | 5.2 | 1.85 | 247.48 | 250.056 | 5.00 | 1.77 |
| h10_10_0_1 | 254.28 | 256.42 | 5.2 | 1.78 | 254.28 | 256.42 | 5.2 | 1.77 | 254.28 | 256.328 | 5.2 | 1.70 |
| h10_10_0_2 | 232.1 | 233.628 | 5.8 | 1.87 | 232.1 | 233.628 | 5.8 | 1.89 | 231 | 233.22 | 5.2 | 1.82 |
| h10_10_0_3 | 255.76 | 257.088 | 5.2 | 1.80 | 255.76 | 257.088 | 5.2 | 1.78 | 255.76 | 258.844 | 4.8 | 1.70 |
| h10_10_0_4 | 217.24 | 220.508 | 4.8 | 2.09 | 217.24 | 220.508 | 4.8 | 1.58 | 217.24 | 220.508 | 4.8 | 1.54 |
| h10_10_1_0 | 256.26 | 258.22 | 6 | 1.92 | 256.26 | 258.22 | 6 | 1.85 | 256.26 | 259.148 | 5.8 | 2.86 |
| h10_10_1_1 | 281.98 | 283.932 | 5.4 | 1.77 | 281.98 | 283.932 | 5.4 | 1.74 | 281.98 | 283.848 | 5.6 | 3.32 |
| h10_10_1_2 | 243.04 | 244.848 | 7 | 1.81 | 243.04 | 244.848 | 7 | 1.78 | 245.3 | 245.3 | 7.0 | 3.58 |
| h10_10_1_3 | 271.96 | 275.88 | 6.2 | 1.78 | 271.96 | 275.88 | 6.2 | 1.75 | 276.48 | 277.276 | 5.2 | 1.96 |
| h10_10_1_4 | 237.54 | 241.524 | 6.4 | 1.63 | 237.54 | 241.524 | 6.4 | 1.57 | 235.84 | 240.14 | 5.2 | 1.58 |
| h10_10_2_0 | 278.02 | 290.968 | 7 | 1.74 | 278.02 | 290.968 | 7 | 1.76 | 280.82 | 289.9 | 6.6 | 1.71 |
| h10_10_2_1 | 308.08 | 309.436 | 6.8 | 1.81 | 308.08 | 309.436 | 6.8 | 1.96 | 305.34 | 307.8 | 7.0 | 1.77 |
| h10_10_2_2 | 276.56 | 276.56 | 6 | 1.80 | 276.56 | 276.56 | 6 | 1.87 | 276.56 | 276.56 | 5.6 | 1.69 |
| h10_10_2_3 | 306.36 | 308.12 | 5.2 | 2.00 | 306.36 | 308.12 | 5.2 | 1.88 | 304.76 | 308 | 4.8 | 1.71 |
| h10_10_2_4 | 279.24 | 281.272 | 5.4 | 1.60 | 279.24 | 281.272 | 5.4 | 1.62 | 279.24 | 280.252 | 5.8 | 1.54 |
| h10_20_0_0 | 371.66 | 377.892 | 3.6 | 1.99 | 371.66 | 377.892 | 3.6 | 2.18 | 371.66 | 379.332 | 3.6 | 1.86 |
| h10_20_0_1 | 401.04 | 401.04 | 4 | 1.79 | 401.04 | 401.04 | 4 | 1.74 | 401.04 | 401.416 | 3.8 | 1.71 |
| h10_20_0_2 | 369.14 | 378.356 | 2.8 | 1.82 | 369.14 | 378.356 | 2.8 | 1.99 | 366.94 | 377.916 | 3.2 | 1.72 |
| h10_20_0_3 | 384.52 | 384.52 | 3 | 1.90 | 384.52 | 384.52 | 3 | 1.93 | 384.52 | 384.52 | 3.2 | 1.77 |
| h10_20_0_4 | 339.46 | 339.46 | 4 | 1.72 | 339.46 | 339.46 | 4 | 1.85 | 339.46 | 339.46 | 4.2 | 1.60 |
| h10_20_1_0 | 379.22 | 382.368 | 5.8 | 2.01 | 379.22 | 382.368 | 5.8 | 2.20 | 379.22 | 384.612 | 5.8 | 3.63 |
| h10_20_1_1 | 413.12 | 413.18 | 4.8 | 1.78 | 413.12 | 413.18 | 4.8 | 1.80 | 413.12 | 413.18 | 4.8 | 5.56 |
| h10_20_1_2 | 391.6 | 391.6 | 6 | 1.92 | 391.6 | 391.6 | 6 | 2.12 | 391.6 | 391.6 | 6.0 | 6.35 |
| h10_20_1_3 | 397.7 | 398.096 | 4.6 | 1.85 | 397.7 | 398.096 | 4.6 | 1.92 | 397.7 | 397.964 | 3.8 | 5.06 |
| h10_20_1_4 | 356.78 | 356.78 | 5.2 | 1.66 | 356.78 | 356.78 | 5.2 | 1.80 | 356.78 | 356.78 | 5.2 | 4.76 |
| h10_20_2_0 | 440.16 | 441.372 | 4.8 | 1.92 | 440.16 | 441.372 | 4.8 | 2.18 | 441.22 | 441.948 | 4.6 | 4.77 |
| h10_20_2_1 | 466.62 | 471.716 | 4.8 | 1.72 | 466.62 | 471.716 | 4.8 | 2.07 | 466.62 | 467.172 | 4.6 | 1.74 |
| h10_20_2_2 | 437.02 | 437.048 | 3.2 | 1.79 | 437.02 | 437.048 | 3.2 | 2.34 | 437.02 | 437.076 | 3.4 | 2.00 |
| h10_20_2_3 | 433.98 | 435.724 | 4 | 2.04 | 433.98 | 435.724 | 4 | 2.33 | 433.7 | 437.108 | 3.4 | 2.43 |
| h10_20_2_4 | 380.64 | 386.36 | 5.2 | 1.72 | 380.64 | 386.36 | 5.2 | 1.96 | 380.64 | 385.668 | 5.0 | 2.41 |
| h10_30_0_0 | 472.46 | 472.46 | 2 | 1.92 | 472.46 | 472.46 | 2 | 2.07 | 472.46 | 472.46 | 2.0 | 2.74 |
| h10_30_0_1 | 527.26 | 527.26 | 2 | 1.81 | 527.26 | 527.26 | 2 | 1.95 | 527.26 | 527.26 | 2.0 | 2.62 |
| h10_30_0_2 | 500.28 | 500.28 | 2.2 | 1.77 | 500.28 | 500.28 | 2.2 | 1.83 | 500.28 | 500.28 | 2.0 | 1.95 |
| h10_30_0_3 | 504.56 | 504.56 | 3 | 1.99 | 504.56 | 504.56 | 3 | 2.02 | 504.56 | 504.56 | 3.0 | 2.17 |
| h10_30_0_4 | 449.6 | 452.08 | 2.2 | 1.63 | 449.6 | 452.08 | 2.2 | 1.66 | 449.6 | 452.08 | 2.2 | 1.83 |
| h10_30_1_0 | 519.86 | 519.86 | 4 | 2.08 | 519.86 | 519.86 | 4 | 2.12 | 518.48 | 519.584 | 4.2 | 2.32 |
| h10_30_1_1 | 544.58 | 544.58 | 3.8 | 1.76 | 544.58 | 544.58 | 3.8 | 1.74 | 544.58 | 544.58 | 4.0 | 1.95 |
| h10_30_1_2 | 554.74 | 554.74 | 4 | 1.95 | 554.74 | 554.74 | 4 | 1.93 | 549.12 | 552.892 | 4.4 | 2.02 |
| h10_30_1_3 | 527.82 | 527.82 | 3 | 1.86 | 527.82 | 527.82 | 3 | 1.87 | 527.82 | 527.82 | 3.0 | 2.09 |
| h10_30_1_4 | 484.16 | 484.16 | 4.2 | 1.69 | 484.16 | 484.16 | 4.2 | 1.65 | 484.16 | 484.728 | 3.8 | 1.91 |
| h10_30_2_0 | 538.44 | 538.44 | 4 | 1.90 | 538.44 | 538.44 | 4 | 1.92 | 538.44 | 541.812 | 3.4 | 2.23 |
| h10_30_2_1 | 607.64 | 610.08 | 3.8 | 1.84 | 607.64 | 610.08 | 3.8 | 1.73 | 607.64 | 609.136 | 4.0 | 1.90 |
| h10_30_2_2 | 577.62 | 577.62 | 3 | 1.84 | 577.62 | 577.62 | 3 | 1.86 | 577.62 | 578.624 | 3.2 | 1.98 |
| h10_30_2_3 | 548.68 | 551.352 | 3.8 | 1.97 | 548.68 | 551.352 | 3.8 | 1.97 | 548.68 | 551.352 | 3.8 | 2.21 |

Table D.1. The best-found solutions for the instances with 10 patients by the ALNS-VS algorithms (cont.).

| | **ALNS-VS_Unique** | | | | **ALNS-VS_Common** | | | | **ALNS-VS_No-Swap** | | | |
|---|---|---|---|---|---|---|---|---|---|---|---|---|
| **Instance** | **Best-found** | **Avg.** | **# DP** | **CPU** | **Best-found** | **Avg.** | **# DP** | **CPU** | **Best-found** | **Avg.** | **# DP** | **CPU** |
| h10_30_2_4 | 496.08 | 496.08 | 4 | 1.64 | 496.08 | 496.08 | 4 | 1.64 | 496.08 | 496.08 | 3.8 | 1.79 |
| h10_40_0_0 | 567.26 | 567.26 | 2 | 1.79 | 567.26 | 567.26 | 2 | 1.80 | 567.26 | 567.26 | 2.0 | 2.06 |
| h10_40_0_1 | 644.64 | 644.64 | 2 | 2.40 | 644.64 | 644.64 | 2 | 1.80 | 644.64 | 644.64 | 2.0 | 1.92 |
| h10_40_0_2 | 614.5 | 614.5 | 1 | 1.74 | 614.5 | 614.5 | 1 | 1.75 | 614.5 | 614.5 | 1.0 | 1.78 |
| h10_40_0_3 | 619.24 | 619.24 | 3 | 2.00 | 619.24 | 619.24 | 3 | 1.94 | 619.24 | 619.24 | 3.0 | 2.08 |
| h10_40_0_4 | 551.8 | 551.8 | 2 | 1.62 | 551.8 | 551.8 | 2 | 1.60 | 551.8 | 551.8 | 2.0 | 1.68 |
| h10_40_1_0 | 614.74 | 626.04 | 2.8 | 2.61 | 614.74 | 626.04 | 2.8 | 2.01 | 614.74 | 627.808 | 2.6 | 2.12 |
| h10_40_1_1 | 665.44 | 666.496 | 4.8 | 2.02 | 665.44 | 666.496 | 4.8 | 1.75 | 665.44 | 666.496 | 4.8 | 1.83 |
| h10_40_1_2 | 678.02 | 678.02 | 2 | 1.85 | 678.02 | 678.02 | 2 | 1.78 | 678.02 | 678.02 | 2.0 | 1.86 |
| h10_40_1_3 | 650.3 | 650.3 | 3 | 1.91 | 650.3 | 650.3 | 3 | 1.83 | 650.3 | 650.3 | 3.0 | 2.00 |
| h10_40_1_4 | 600.48 | 600.48 | 3 | 1.72 | 600.48 | 600.48 | 3 | 1.57 | 600.48 | 600.48 | 2.8 | 1.70 |
| h10_40_2_0 | 640.18 | 640.18 | 3 | 1.93 | 640.18 | 640.18 | 3 | 1.97 | 640.18 | 640.18 | 3.0 | 2.08 |
| h10_40_2_1 | 743.28 | 743.352 | 3.4 | 1.80 | 743.28 | 743.352 | 3.4 | 1.81 | 743.28 | 743.376 | 3.2 | 2.05 |
| h10_40_2_2 | 704.28 | 705.2 | 2.8 | 1.81 | 704.28 | 705.2 | 2.8 | 1.90 | 704.28 | 704.28 | 3.0 | 1.91 |
| h10_40_2_3 | 666.2 | 668.412 | 4 | 2.10 | 666.2 | 668.412 | 4 | 2.00 | 666.2 | 666.2 | 4.0 | 1.91 |
| h10_40_2_4 | 600.7 | 604.672 | 4.2 | 1.61 | 600.7 | 604.672 | 4.2 | 1.71 | 600.7 | 608.636 | 3.2 | 2.16 |

Table D.2. The best-found solutions for the instances with 30 patients by the ALNS-VS algorithms

| | **ALNS-VS_Unique** | | | | **ALNS-VS_Common** | | | | **ALNS-VS_No-Swap** | | | |
|---|---|---|---|---|---|---|---|---|---|---|---|---|
| **Instance** | **Best-found** | **Avg.** | **# DP** | **CPU** | **Best-found** | **Avg.** | **# DP** | **CPU** | **Best-found** | **Avg.** | **# DP** | **CPU** |
| h30_10_0_0 | 548.5 | 563.568 | 17.4 | 24 | 538.76 | 555.344 | 17.6 | 18 | 573.98 | 579.24 | 17.6 | 24 |
| h30_10_0_1 | 571.6 | 579.116 | 15.4 | 19 | 568.22 | 577 | 14.6 | 18 | 569.52 | 578.56 | 15.4 | 24 |
| h30_10_0_2 | 552.34 | 562.172 | 20 | 13 | 561.66 | 566.736 | 20.6 | 15 | 559.76 | 568.376 | 20 | 20 |
| h30_10_0_3 | 571.22 | 577.948 | 15.8 | 15 | 576.48 | 577.88 | 16.6 | 21 | 556.06 | 573.336 | 16 | 22 |
| h30_10_0_4 | 559.58 | 564.284 | 16.2 | 20 | 559.84 | 566.768 | 16.2 | 29 | 558.4 | 573.14 | 15.4 | 21 |
| h30_10_1_0 | 604.16 | 622.404 | 17.8 | 15 | 604.16 | 628.3 | 16.8 | 27 | 638.6 | 638.6 | 16.6 | 23 |
| h30_10_1_1 | 614.22 | 623.92 | 18.8 | 16 | 614.22 | 622.272 | 18.6 | 34 | 619.96 | 632.488 | 18.6 | 22 |
| h30_10_1_2 | 604.38 | 613.864 | 21.6 | 17 | 611.5 | 616.04 | 20.4 | 25 | 610.78 | 614.156 | 20.2 | 22 |
| h30_10_1_3 | 597.14 | 600.38 | 17.6 | 22 | 597.14 | 602.984 | 18 | 29 | 595.46 | 604.856 | 18.8 | 22 |
| h30_10_1_4 | 597.98 | 604.148 | 19 | 23 | 599.12 | 607.832 | 18.8 | 28 | 601.34 | 609.864 | 19.2 | 21 |
| h30_10_2_0 | 664.76 | 678.128 | 16.2 | 29 | 659.7 | 676.34 | 15.8 | 28 | 670.74 | 680.964 | 17 | 21 |
| h30_10_2_1 | 665.44 | 690.324 | 17.2 | 28 | 663.06 | 681.62 | 18.6 | 29 | 666.94 | 684.94 | 18 | 27 |
| h30_10_2_2 | 671.58 | 682.828 | 17.4 | 27 | 669.14 | 678.1 | 20.2 | 29 | 675.68 | 680.876 | 17.6 | 21 |
| h30_10_2_3 | 652.6 | 664.488 | 17.4 | 29 | 651.54 | 664.096 | 17.8 | 30 | 661.56 | 676.724 | 16.8 | 21 |
| h30_10_2_4 | 673.22 | 679.768 | 15.6 | 21 | 663.02 | 677.964 | 15.8 | 22 | 670.54 | 678.972 | 15.2 | 18 |
| h30_20_0_0 | 769.9 | 788.924 | 12.4 | 30 | 770.96 | 786.868 | 13.8 | 28 | 765.02 | 774.152 | 13 | 23 |
| h30_20_0_1 | 795.58 | 810.9 | 14.8 | 32 | 795.58 | 815.54 | 16 | 32 | 800.84 | 803.496 | 14.4 | 25 |
| h30_20_0_2 | 818.36 | 836.032 | 14.8 | 28 | 814.02 | 828.084 | 16.4 | 26 | 824.42 | 826.392 | 16.2 | 20 |
| h30_20_0_3 | 767.94 | 777.92 | 16 | 30 | 773.48 | 779.252 | 16 | 27 | 768.48 | 778.844 | 15.8 | 20 |
| h30_20_0_4 | 777.32 | 779.608 | 16.6 | 27 | 768.92 | 777.264 | 15.8 | 27 | 777.34 | 779.472 | 15.2 | 22 |
| h30_20_1_0 | 816.72 | 831.776 | 15.4 | 30 | 818.06 | 832.088 | 16.4 | 34 | 824.16 | 833.692 | 14.6 | 25 |
| h30_20_1_1 | 866.12 | 880.496 | 14.4 | 30 | 855.12 | 875.128 | 13.4 | 21 | 873.28 | 883.736 | 13 | 25 |
| h30_20_1_2 | 808.28 | 829.98 | 16 | 25 | 809.8 | 824.556 | 14.4 | 15 | 825.18 | 833.808 | 15.4 | 20 |
| h30_20_1_3 | 785.9 | 798.828 | 17.8 | 30 | 796.66 | 802.68 | 16.4 | 18 | 799.96 | 812.512 | 17 | 20 |
| h30_20_1_4 | 808.82 | 822.028 | 13.4 | 20 | 808.82 | 820.572 | 12.8 | 11 | 823.44 | 824.344 | 13.6 | 15 |
| h30_20_2_0 | 874.98 | 890.644 | 14.6 | 37 | 845.38 | 878.92 | 15.8 | 20 | 854.42 | 875.464 | 16.4 | 30 |
| h30_20_2_1 | 923.62 | 940.1 | 15.4 | 28 | 905.16 | 936.428 | 14.8 | 17 | 930.2 | 948.36 | 16.2 | 27 |
| h30_20_2_2 | 894.72 | 904.356 | 15.8 | 26 | 894.72 | 901.516 | 16 | 15 | 894.72 | 904.348 | 15.6 | 20 |
| h30_20_2_3 | 849.02 | 866.828 | 16.6 | 25 | 832.84 | 854.684 | 17 | 16 | 861.08 | 874.564 | 16.2 | 21 |
| h30_20_2_4 | 883.88 | 891.828 | 13.4 | 20 | 865.94 | 888 | 14.2 | 10 | 878.08 | 893.408 | 14.4 | 18 |
| h30_30_0_0 | 959.54 | 972.352 | 10.2 | 24 | 974.96 | 977.052 | 10 | 15 | 958.58 | 964.776 | 10.2 | 23 |
| h30_30_0_1 | 1039.02 | 1044.46 | 8.6 | 15 | 1036.04 | 1044.176 | 10.2 | 15 | 1042.08 | 1044.824 | 8.6 | 20 |

Table D.2. The best-found solutions for the instances with 30 patients by the ALNS-VS algorithms (cont.)

| | **ALNS-VS_Unique** | | | | **ALNS-VS_Common** | | | | **ALNS-VS_No-Swap** | | | |
|---|---|---|---|---|---|---|---|---|---|---|---|---|
| **Instance** | **Best-found** | **Avg.** | **# DP** | **CPU** | **Best-found** | **Avg.** | **# DP** | **CPU** | **Best-found** | **Avg.** | **# DP** | **CPU** |
| h30_30_0_2 | 1053.54 | 1069.328 | 12.8 | 15 | 1054.2 | 1068.404 | 13.2 | 15 | 1050.98 | 1067.116 | 13.6 | 18 |
| h30_30_0_3 | 989.12 | 992.924 | 12.6 | 16 | 969.52 | 986.4 | 13.2 | 15 | 991.5 | 999.876 | 15 | 22 |
| h30_30_0_4 | 967.3 | 983.896 | 12.6 | 15 | 986.98 | 989.408 | 12.4 | 16 | 986.98 | 993.932 | 11.2 | 20 |
| h30_30_1_0 | 1045.36 | 1053.992 | 16 | 16 | 1045.8 | 1055.556 | 16 | 15 | 1046.62 | 1058.228 | 14.4 | 28 |
| h30_30_1_1 | 1106.12 | 1112.152 | 11 | 15 | 1081.98 | 1106.804 | 11.2 | 15 | 1088.9 | 1103.64 | 11.6 | 20 |
| h30_30_1_2 | 1047.8 | 1064.38 | 15.8 | 16 | 1031.38 | 1056.944 | 15 | 17 | 1051.7 | 1058.096 | 15.8 | 20 |
| h30_30_1_3 | 996.64 | 1031.664 | 13.4 | 22 | 998.02 | 1035.208 | 13.2 | 16 | 1037.76 | 1049.976 | 14 | 22 |
| h30_30_1_4 | 1008.44 | 1011.236 | 12 | 21 | 1014 | 1021.616 | 11 | 10 | 1004.88 | 1020.872 | 10.6 | 17 |
| h30_30_2_0 | 1069.3 | 1087.996 | 15.6 | 30 | 1089.06 | 1107.484 | 12.6 | 19 | 1093.58 | 1102.92 | 14.8 | 24 |
| h30_30_2_1 | 1163.3 | 1172.428 | 14 | 26 | 1159.6 | 1174.088 | 11.6 | 15 | 1163.86 | 1184.056 | 14.2 | 20 |
| h30_30_2_2 | 1144.22 | 1149.812 | 14 | 25 | 1140.68 | 1143.916 | 15 | 16 | 1143.66 | 1147.144 | 14.2 | 22 |
| h30_30_2_3 | 1057.34 | 1083.984 | 13.6 | 28 | 1062.86 | 1085.592 | 14.6 | 15 | 1096.34 | 1108.792 | 13 | 20 |
| h30_30_2_4 | 1074.72 | 1080.016 | 11.6 | 21 | 1074.72 | 1081.74 | 13 | 11 | 1064.38 | 1078.572 | 12.8 | 16 |
| h30_40_0_0 | 1163.52 | 1172.112 | 10.6 | 26 | 1163.52 | 1171.192 | 9.6 | 15 | 1164.2 | 1174.66 | 9.6 | 21 |
| h30_40_0_1 | 1249.28 | 1249.28 | 7 | 25 | 1246.38 | 1247.368 | 9.2 | 15 | 1246.44 | 1248.508 | 7.6 | 20 |
| h30_40_0_2 | 1274.94 | 1283.996 | 7.8 | 25 | 1281.48 | 1287.564 | 8.2 | 15 | 1274.94 | 1281.036 | 8 | 20 |
| h30_40_0_3 | 1179.64 | 1202.1 | 10.6 | 24 | 1186.42 | 1199.2 | 10.2 | 15 | 1207.42 | 1220.948 | 9.4 | 20 |
| h30_40_0_4 | 1190.08 | 1196.688 | 9.2 | 20 | 1190.08 | 1194.176 | 9.2 | 11 | 1190.08 | 1195.7 | 9 | 16 |
| h30_40_1_0 | 1253.14 | 1261.06 | 10.8 | 25 | 1252.72 | 1266.272 | 9.6 | 14 | 1246.96 | 1261.52 | 11.4 | 23 |
| h30_40_1_1 | 1328.04 | 1336.316 | 9.2 | 29 | 1336.52 | 1341 | 9.2 | 15 | 1325.7 | 1332.864 | 8.8 | 25 |
| h30_40_1_2 | 1273.12 | 1280.696 | 11.8 | 26 | 1271.72 | 1282.864 | 13.2 | 15 | 1281.36 | 1283.688 | 12.6 | 21 |
| h30_40_1_3 | 1224.14 | 1239.648 | 11.6 | 25 | 1225.26 | 1250.016 | 7.8 | 15 | 1264.06 | 1273.464 | 9.8 | 20 |
| h30_40_1_4 | 1217.12 | 1221.356 | 8.4 | 21 | 1220.02 | 1223.52 | 8.4 | 10 | 1210.7 | 1224.636 | 9 | 20 |
| h30_40_2_0 | 1298.9 | 1315.564 | 11.8 | 25 | 1301.42 | 1314.672 | 12 | 15 | 1298.9 | 1303.552 | 13.8 | 21 |
| h30_40_2_1 | 1413.4 | 1423.804 | 10.6 | 19 | 1416 | 1436.756 | 10.2 | 15 | 1422.44 | 1425.116 | 9.6 | 21 |
| h30_40_2_2 | 1348.32 | 1363.456 | 10.6 | 15 | 1360.74 | 1366.648 | 11 | 15 | 1363.52 | 1370.188 | 11 | 20 |
| h30_40_2_3 | 1306.24 | 1319.656 | 11.8 | 15 | 1292.98 | 1315.408 | 10.4 | 15 | 1305.46 | 1319.32 | 11.6 | 21 |
| h30_40_2_4 | 1247.14 | 1256.468 | 11.4 | 10 | 1245.02 | 1256.672 | 10.2 | 10 | 1247.14 | 1257.02 | 11.4 | 15 |

Table D.3. The best-found solutions for the instances with 50 patients by the ALNS-VS algorithms

| **ALNS-VS_Unique** | **ALNS-VS_Common** | **ALNS-VS_No-Swap** |
|---|---|---|

| Instance | Best-found | Avg. | # DP | CPU | Best-found | Avg. | # DP | CPU | Best-found | Avg. | # DP | CPU |
|---|---|---|---|---|---|---|---|---|---|---|---|---|
| h50_10_0_0 | 909.82 | 938.08 | 27.6 | 36 | 925.44 | 943.028 | 28 | 39 | 930.86 | 952.352 | 29 | 36 |
| h50_10_0_1 | 930.92 | 945.58 | 30 | 32 | 931.62 | 943.516 | 33 | 35.4 | 937.54 | 949.216 | 31 | 36.6 |
| h50_10_0_2 | 942.04 | 951.084 | 30.2 | 36 | 940.8 | 951.136 | 30.6 | 39.6 | 932.86 | 954.168 | 30 | 36.8 |
| h50_10_0_3 | 889.4 | 909.504 | 28.2 | 32 | 860.06 | 921.512 | 29.2 | 40.2 | 940.48 | 953.92 | 28.8 | 41.6 |
| h50_10_0_4 | 913.76 | 925.072 | 28 | 35 | 927.3 | 931.408 | 27.6 | 34.6 | 910.2 | 919.164 | 28.4 | 43.6 |
| h50_10_1_0 | 1070.26 | 1080.828 | 28.2 | 32 | 1063.7 | 1072.544 | 28.2 | 39 | 1113.02 | 1124.7 | 24.8 | 34.8 |
| h50_10_1_1 | 1072.84 | 1086.144 | 27.8 | 31 | 1089.38 | 1102.112 | 25.2 | 34.2 | 1082.86 | 1094.712 | 25 | 30 |
| h50_10_1_2 | 1053.3 | 1078.44 | 29 | 33 | 1066.1 | 1098.616 | 29.4 | 33.4 | 1063.92 | 1084.536 | 30.8 | 37.2 |
| h50_10_1_3 | 1021 | 1064.352 | 28.2 | 33 | 1055.6 | 1075.012 | 28.8 | 32 | 1059.56 | 1080.78 | 28 | 38.8 |
| h50_10_1_4 | 1030.02 | 1046.452 | 30.8 | 42 | 1019.32 | 1043.08 | 32 | 40.8 | 1027.76 | 1042.816 | 31 | 42.8 |
| h50_10_2_0 | 1211.42 | 1240.408 | 28.4 | 47 | 1219.9 | 1234.612 | 28.2 | 42.4 | 1246.84 | 1265.812 | 25.6 | 37.4 |
| h50_10_2_1 | 1219.42 | 1242.992 | 28.4 | 36 | 1216.34 | 1234.952 | 26 | 37.8 | 1228.12 | 1253.668 | 27 | 40.2 |
| h50_10_2_2 | 1216.24 | 1244.288 | 31.6 | 36 | 1246.3 | 1257.284 | 28.6 | 34 | 1238.1 | 1258.672 | 29.2 | 40.2 |
| h50_10_2_3 | 1183 | 1219.128 | 27.8 | 31 | 1224.58 | 1233.184 | 27.6 | 38 | 1195.12 | 1228.564 | 27.8 | 38.4 |
| h50_10_2_4 | 1207.66 | 1221.068 | 28.8 | 32 | 1177.94 | 1201.628 | 28.2 | 41.6 | 1203.4 | 1227.228 | 28.4 | 34.6 |
| h50_20_0_0 | 1267.5 | 1290.608 | 24.6 | 39 | 1291.12 | 1295.832 | 25.8 | 34.2 | 1312.28 | 1338.592 | 24.4 | 38.6 |
| h50_20_0_1 | 1299.14 | 1313.872 | 23.8 | 30 | 1314.48 | 1338.432 | 23.8 | 31.4 | 1274.34 | 1306 | 25 | 31.2 |
| h50_20_0_2 | 1301.56 | 1324.108 | 24.8 | 34 | 1294.1 | 1322.084 | 25.6 | 38.2 | 1316.54 | 1333.508 | 27.6 | 32.4 |
| h50_20_0_3 | 1176.6 | 1194.74 | 28 | 38 | 1188.76 | 1192.504 | 26.6 | 35 | 1187.82 | 1194.772 | 27.4 | 31.8 |

Table D.3. The best-found solutions for the instances with 50 patients by the ALNS-VS algorithms (cont.)

| | ALNS-VS_Unique | | | | ALNS-VS_Common | | | | ALNS-VS_No-Swap | | | |
|---|---|---|---|---|---|---|---|---|---|---|---|---|
| **Instance** | **Best-found** | **Avg.** | **# DP** | **CPU** | **Best-found** | **Avg.** | **# DP** | **CPU** | **Best-found** | **Avg.** | **# DP** | **CPU** |
| h50_20_0_4 | 1280.86 | 1310.496 | 23.6 | 32 | 1263.54 | 1284.024 | 23.4 | 35.6 | 1281.58 | 1306.288 | 24 | 31.8 |
| h50_20_1_0 | 1406.52 | 1432.5 | 25.2 | 37 | 1404.26 | 1415 | 25.6 | 33.6 | 1478.94 | 1492.788 | 23.4 | 31.2 |
| h50_20_1_1 | 1382.12 | 1430.876 | 23.4 | 31 | 1398.38 | 1419.384 | 26 | 35.4 | 1440.68 | 1470.408 | 22.4 | 42.2 |
| h50_20_1_2 | 1465.96 | 1479.756 | 23.2 | 36 | 1468.94 | 1488.236 | 24.8 | 33 | 1446.44 | 1458.032 | 26 | 34.2 |
| h50_20_1_3 | 1331.9 | 1347.188 | 25.6 | 30 | 1322.08 | 1352.208 | 26 | 36.2 | 1374.6 | 1386.48 | 24.2 | 34.4 |
| h50_20_1_4 | 1383.84 | 1389.088 | 28.8 | 36 | 1388.82 | 1402.252 | 27.8 | 37.6 | 1373.54 | 1401.568 | 27.2 | 33 |
| h50_20_2_0 | 1591.36 | 1609.98 | 26.6 | 34 | 1546.5 | 1597.96 | 28.4 | 41.4 | 1624.22 | 1642.432 | 23.2 | 41.8 |
| h50_20_2_1 | 1588.36 | 1607.168 | 27.4 | 29 | 1557.82 | 1590.504 | 28.2 | 46 | 1585.92 | 1608.26 | 26 | 35.6 |
| h50_20_2_2 | 1602.74 | 1638.724 | 27 | 36 | 1575.06 | 1614.196 | 26.4 | 35.2 | 1594.08 | 1620.212 | 28.4 | 40.8 |
| h50_20_2_3 | 1465.66 | 1509.78 | 21.4 | 32 | 1463.98 | 1499.7 | 25 | 33.2 | 1522.36 | 1548.98 | 27 | 33.2 |
| h50_20_2_4 | 1544.46 | 1589.408 | 26.8 | 42 | 1592.3 | 1612 | 24.6 | 31 | 1573.76 | 1615.104 | 25.6 | 30.8 |
| h50_30_0_0 | 1569.1 | 1599.296 | 19.6 | 35 | 1608.34 | 1624.424 | 21.2 | 35.6 | 1665.62 | 1676.012 | 20.2 | 32.4 |
| h50_30_0_1 | 1598.98 | 1635.804 | 20.8 | 31 | 1639.72 | 1669.436 | 20 | 35.2 | 1652.24 | 1672.1 | 21.2 | 33.4 |
| h50_30_0_2 | 1679.3 | 1688.92 | 19.4 | 30 | 1672.28 | 1692.956 | 21.8 | 42.6 | 1677.04 | 1696.328 | 21.6 | 34.6 |
| h50_30_0_3 | 1504.66 | 1514.452 | 21.4 | 33 | 1449.36 | 1484.812 | 25.2 | 43 | 1496.68 | 1509.08 | 23.4 | 33.4 |
| h50_30_0_4 | 1621.08 | 1648.012 | 20.8 | 36 | 1637.52 | 1655.068 | 18.6 | 31.4 | 1645.18 | 1657.992 | 20 | 31.6 |
| h50_30_1_0 | 1723.9 | 1734.768 | 22.2 | 31 | 1696.5 | 1720.348 | 20.4 | 34.4 | 1813.5 | 1844.34 | 17.6 | 28.6 |
| h50_30_1_1 | 1746.08 | 1794.456 | 18.4 | 28 | 1753.18 | 1801.108 | 17.8 | 30.6 | 1803.38 | 1816.192 | 18.2 | 35.4 |
| h50_30_1_2 | 1808.54 | 1835.792 | 21 | 27 | 1800.54 | 1826.184 | 20.6 | 30.6 | 1839.04 | 1859.068 | 17.2 | 30 |
| h50_30_1_3 | 1625.76 | 1653.1 | 22.6 | 37 | 1644.98 | 1662.32 | 22 | 33.4 | 1682.7 | 1696.252 | 20 | 35 |
| h50_30_1_4 | 1744.46 | 1798.956 | 20 | 33 | 1730.98 | 1790.524 | 19 | 31.2 | 1771.5 | 1783.4 | 22.4 | 31 |
| h50_30_2_0 | 1836.42 | 1922.96 | 25.2 | 42 | 1923.7 | 1976.232 | 22.4 | 33.4 | 1967.3 | 2025.196 | 22 | 35 |
| h50_30_2_1 | 1938.74 | 1968.944 | 20 | 30 | 1957.26 | 1963.848 | 22 | 34.6 | 1947.74 | 1961.46 | 24.4 | 31.8 |
| h50_30_2_2 | 1980.06 | 2001.136 | 20.2 | 27 | 1965.12 | 2008.84 | 21.6 | 30.4 | 2023.9 | 2035.716 | 19.8 | 32.4 |
| h50_30_2_3 | 1764.08 | 1803.304 | 23.6 | 38 | 1752.46 | 1805.9 | 22.2 | 35.8 | 1804.64 | 1834.144 | 22 | 33.6 |
| h50_30_2_4 | 1931.36 | 1940.228 | 19.4 | 30 | 1931.16 | 1943.368 | 20.4 | 32.4 | 1912.78 | 1936.14 | 21.8 | 30.8 |
| h50_40_0_0 | 1871.68 | 1915.668 | 19 | 51 | 1896.68 | 1931.132 | 17.2 | 30.6 | 2042.96 | 2050.812 | 14.6 | 36.8 |
| h50_40_0_1 | 1971.42 | 2004.052 | 14.2 | 39 | 1958.68 | 2007.612 | 15.4 | 29 | 2032.58 | 2039.244 | 15.2 | 35 |
| h50_40_0_2 | 2012.02 | 2070.376 | 15.4 | 42 | 2078.08 | 2097.748 | 15.2 | 36.4 | 2099.42 | 2104.348 | 14 | 36.4 |
| h50_40_0_3 | 1774.16 | 1809.528 | 19.2 | 43 | 1773.56 | 1798.12 | 20 | 42.2 | 1801.58 | 1815.32 | 15.6 | 41.4 |
| h50_40_0_4 | 1950.78 | 1969.724 | 12.8 | 29 | 1962.04 | 1980.316 | 15.6 | 37.8 | 1977.02 | 1985.492 | 16.6 | 35.8 |
| h50_40_1_0 | 2047.46 | 2091.54 | 17.6 | 38 | 2021.92 | 2078.472 | 18.8 | 37.6 | 2124.92 | 2159.1 | 16.8 | 41.4 |
| h50_40_1_1 | 2071.16 | 2108.1 | 15 | 32 | 2072.68 | 2113.672 | 17.6 | 36.8 | 2129.42 | 2163.3 | 12.6 | 32.6 |
| h50_40_1_2 | 2144.36 | 2193.98 | 14.8 | 28 | 2130.7 | 2217.676 | 16 | 42.2 | 2239.06 | 2272.792 | 18 | 33.2 |
| h50_40_1_3 | 1914.2 | 1936.308 | 16.4 | 31 | 1875.3 | 1916.38 | 22.4 | 45.2 | 1982.58 | 2004.908 | 19.2 | 32.2 |
| h50_40_1_4 | 2080.46 | 2126.724 | 17.4 | 30 | 2071.92 | 2149.152 | 16.4 | 31.2 | 2120.08 | 2137.74 | 17.2 | 29.8 |
| h50_40_2_0 | 2287.3 | 2328.432 | 18 | 28 | 2242.92 | 2335.78 | 18.8 | 33.4 | 2393.82 | 2414.124 | 16.4 | 32.4 |
| h50_40_2_1 | 2292.64 | 2326.444 | 16.6 | 31 | 2313.12 | 2334.24 | 17.6 | 33.2 | 2335.02 | 2350.908 | 17.6 | 30.8 |
| h50_40_2_2 | 2352.22 | 2402.028 | 17.4 | 34 | 2353.68 | 2391.196 | 18 | 36.4 | 2404.32 | 2418.796 | 18.4 | 35.8 |
| h50_40_2_3 | 2056.96 | 2139.992 | 18.4 | 30 | 2062.88 | 2097.872 | 19 | 46.4 | 2096 | 2137.224 | 19 | 38.4 |
| h50_40_2_4 | 2290.88 | 2307.472 | 19.6 | 28 | 2284.68 | 2298.344 | 21.4 | 34.4 | 2291.94 | 2306.404 | 17.6 | 30.6 |

Table D.4. The best-found solutions for the instances with 100 patients by the ALNS-VS algorithms.

| | **ALNS-VS_Unique** | | | | **ALNS-VS_Common** | | | | **ALNS-VS_No-Swap** | | | |
|---|---|---|---|---|---|---|---|---|---|---|---|---|
| **Instance** | **Best-found** | **Avg.** | **# DP** | **CPU** | **Best-found** | **Avg.** | **# DP** | **CPU** | **Best-found** | **Avg.** | **# DP** | **CPU** |
| h100_10_0_0 | 1975.94 | 2033.988 | 49.4 | 149.08 | 2019.76 | 2066.468 | 50.4 | 115.4 | 2040.06 | 2064.648 | 50 | 138 |
| h100_10_0_1 | 2051.56 | 2070.744 | 48.4 | 113.68 | 1964.26 | 2012.772 | 52.4 | 135.6 | 2029.7 | 2067.368 | 51.2 | 146 |
| h100_10_0_2 | 2041.64 | 2061.932 | 48.8 | 99.75 | 1977.2 | 2024.948 | 52 | 137 | 1991.1 | 2023.512 | 50.8 | 168 |
| h100_10_0_3 | 2013.56 | 2035.148 | 53.8 | 101.77 | 1953.18 | 2035.676 | 49.8 | 117.6 | 2018.1 | 2027.964 | 53.2 | 202 |
| h100_10_0_4 | 1967.94 | 2021.876 | 50.6 | 110.50 | 1940.72 | 2018.312 | 52.2 | 142.4 | 2076.9 | 2097.88 | 51.6 | 152 |
| h100_10_1_0 | 2382.64 | 2436.08 | 47.4 | 138.95 | 2405.42 | 2453.616 | 49.2 | 126.2 | 2465.1 | 2496.02 | 49.2 | 146 |

Table D.4. The best-found solutions for the instances with 100 patients by the ALNS-VS algorithms (cont.).

| | **ALNS-VS_Unique** | | | | **ALNS-VS_Common** | | | | **ALNS-VS_No-Swap** | | | |
|---|---|---|---|---|---|---|---|---|---|---|---|---|
| **Instance** | **Best-found** | **Avg.** | **# DP** | **CPU** | **Best-found** | **Avg.** | **# DP** | **CPU** | **Best-found** | **Avg.** | **# DP** | **CPU** |
| h100_10_1_1 | 2326.82 | 2357.72 | 52 | 148.36 | 2352.7 | 2397.42 | 48.8 | 140 | 2328.68 | 2402.856 | 54.6 | 186 |
| h100_10_1_2 | 2388.9 | 2449.052 | 47 | 109.69 | 2387.84 | 2407.896 | 49.8 | 112 | 2368.82 | 2393.16 | 50 | 149 |
| h100_10_1_3 | 2327.8 | 2406.86 | 52.4 | 111.42 | 2393.56 | 2434.272 | 51.6 | 119.6 | 2379.28 | 2411.784 | 52.4 | 192 |
| h100_10_1_4 | 2377.38 | 2423.956 | 50 | 121.26 | 2403.4 | 2428.488 | 51.2 | 114 | 2470.76 | 2508.92 | 49.8 | 196 |
| h100_10_2_0 | 2492.14 | 2552.196 | 49.4 | 119.28 | 2481.96 | 2560.168 | 51.2 | 123.4 | 2555.28 | 2580.456 | 49.6 | 175 |
| h100_10_2_1 | 2413.36 | 2472.932 | 49 | 147.52 | 2455.66 | 2532.704 | 47 | 98.4 | 2464.08 | 2517.132 | 50.4 | 164 |
| h100_10_2_2 | 2480.8 | 2532.168 | 49.2 | 110.87 | 2498.76 | 2527.124 | 51.8 | 131.2 | 2501.24 | 2510.068 | 47.6 | 182 |
| h100_10_2_3 | 2457.34 | 2489.384 | 52.4 | 144.96 | 2391.6 | 2533.36 | 52.4 | 134.4 | 2516.92 | 2547.644 | 51.4 | 128 |
| h100_10_2_4 | 2477.34 | 2517.396 | 53.8 | 146.06 | 2470.08 | 2507.384 | 53 | 121.2 | 2534.56 | 2577.452 | 52.6 | 154 |
| h100_20_0_0 | 2657.8 | 2711.224 | 43 | 128.59 | 2614.92 | 2713.1 | 40.8 | 114.4 | 2691.18 | 2707.952 | 44 | 180 |
| h100_20_0_1 | 2681.96 | 2711.828 | 42.6 | 125.32 | 2662.06 | 2719.924 | 41.2 | 106.4 | 2733.38 | 2774.04 | 40 | 164 |
| h100_20_0_2 | 2611.18 | 2678.72 | 44.6 | 128.44 | 2690.94 | 2740.396 | 38.8 | 129.6 | 2704.62 | 2726.696 | 42 | 169 |
| h100_20_0_3 | 2629.6 | 2675.424 | 44 | 105.09 | 2574.78 | 2647.04 | 44.2 | 134.6 | 2667.82 | 2702.704 | 42.2 | 143 |
| h100_20_0_4 | 2717.9 | 2769.444 | 42.8 | 122.81 | 2687.62 | 2737.068 | 44.4 | 132.4 | 2850.54 | 2885.18 | 37.2 | 139 |
| h100_20_1_0 | 3024.92 | 3068.608 | 45 | 119.07 | 3058.46 | 3090.732 | 44.6 | 114.2 | 2990.66 | 3061.632 | 43.8 | 168 |
| h100_20_1_1 | 3003.9 | 3066.524 | 47.2 | 102.61 | 3004.4 | 3065.98 | 44.4 | 123.8 | 3044.9 | 3095.756 | 43.4 | 154 |
| h100_20_1_2 | 3036.4 | 3072.892 | 46.4 | 119.15 | 3053.54 | 3081.452 | 47.8 | 144.6 | 3082.14 | 3110.568 | 44.6 | 149 |
| h100_20_1_3 | 3078.32 | 3105.324 | 45.8 | 112.63 | 3064.24 | 3097.34 | 47.8 | 148.6 | 3017.06 | 3072.512 | 44 | 150 |
| h100_20_1_4 | 3090.96 | 3174.912 | 45.8 | 112.93 | 3125.54 | 3168.024 | 48.4 | 101.2 | 3206.98 | 3234.464 | 39.8 | 152 |
| h100_20_2_0 | 3175.98 | 3262.4 | 48.8 | 99.51 | 3173.98 | 3253.584 | 49.4 | 130.2 | 3167.02 | 3205.34 | 50.2 | 164 |
| h100_20_2_1 | 3134.16 | 3204.636 | 45.8 | 116.13 | 3149.68 | 3165.056 | 46.8 | 131.6 | 3134.74 | 3164.36 | 47 | 147 |
| h100_20_2_2 | 3139.24 | 3176.716 | 49.4 | 126.49 | 3205.72 | 3244.504 | 46.4 | 121.8 | 3187.52 | 3253.492 | 44 | 131 |
| h100_20_2_3 | 3082.96 | 3170.284 | 48.6 | 132.11 | 3124.34 | 3199.832 | 48 | 127 | 3150.58 | 3193.792 | 47.4 | 173 |
| h100_20_2_4 | 3173.98 | 3233.616 | 48.2 | 137.97 | 3246.96 | 3288.032 | 46 | 110.4 | 3339.36 | 3348.956 | 45.4 | 178 |
| h100_30_0_0 | 3273.16 | 3341.556 | 33.2 | 104.89 | 3393.88 | 3426.432 | 33.2 | 113 | 3406.66 | 3452.156 | 32.6 | 182 |
| h100_30_0_1 | 3398.24 | 3429.252 | 32.2 | 120.67 | 3276.74 | 3369.552 | 32.2 | 103.2 | 3463.1 | 3488.116 | 26.8 | 139 |
| h100_30_0_2 | 3241.28 | 3351.3 | 34.6 | 113.79 | 3364.72 | 3403.68 | 36 | 117.2 | 3364.26 | 3374.548 | 35.4 | 143 |

| h100_30_0_3 | 3191.18 | 3266.752 | 36.2 | 101.53 | 3230.68 | 3303.508 | 34.6 | 107 | 3165.96 | 3197.292 | 39 | 168 |
|---|---|---|---|---|---|---|---|---|---|---|---|---|
| h100_30_0_4 | 3398.52 | 3439.18 | 34.8 | 120.98 | 3383.56 | 3400.452 | 38 | 113.8 | 3504.04 | 3551.208 | 31 | 138 |
| h100_30_1_0 | 3702.12 | 3782.592 | 41.4 | 139.78 | 3696.76 | 3785.92 | 37.8 | 110.8 | 3808.16 | 3892.712 | 36.2 | 135 |
| h100_30_1_1 | 3713.72 | 3778.296 | 33.2 | 110.19 | 3704.4 | 3748.708 | 34 | 97.8 | 3639.32 | 3727.416 | 36.6 | 129 |
| h100_30_1_2 | 3647.28 | 3784.8 | 39 | 103.34 | 3630.7 | 3744.148 | 41.4 | 152 | 3808.5 | 3831.992 | 37 | 116 |
| h100_30_1_3 | 3586.44 | 3688.588 | 39.2 | 103.71 | 3606.88 | 3676.636 | 40 | 142.4 | 3572.16 | 3662.64 | 40 | 163 |
| h100_30_1_4 | 3735.9 | 3864.744 | 37.2 | 116.17 | 3820.04 | 3888.968 | 35.2 | 96.2 | 3912.46 | 3957.376 | 37.6 | 164 |
| h100_30_2_0 | 3875.58 | 3914.144 | 41 | 103.81 | 3825.66 | 3916.224 | 40.8 | 106 | 3908.42 | 3957.996 | 39.8 | 152 |
| h100_30_2_1 | 3829.38 | 3873.244 | 39.6 | 92.95 | 3798.62 | 3919.956 | 38.6 | 101.2 | 3851.7 | 3916.98 | 38.4 | 176 |
| h100_30_2_2 | 3692.94 | 3874.692 | 40 | 113.00 | 3895.78 | 3997.084 | 37.8 | 91.2 | 3835.2 | 3894.888 | 40.4 | 157 |
| h100_30_2_3 | 3736.7 | 3766.84 | 43.8 | 155.08 | 3718.26 | 3812.804 | 44.2 | 109.4 | 3750.54 | 3776.84 | 43.8 | 155 |
| h100_30_2_4 | 3783.46 | 3932.988 | 36.8 | 113.65 | 3874.5 | 3957.136 | 39.8 | 134.8 | 4029.02 | 4069.096 | 34 | 123 |
| h100_40_0_0 | 3872.08 | 4025.224 | 25.2 | 107.70 | 4037.28 | 4102.708 | 25.8 | 119.2 | 4065.76 | 4100.28 | 25 | 116 |
| h100_40_0_1 | 3937.38 | 4028.892 | 25.8 | 110.50 | 3924.94 | 4005.34 | 27.4 | 91 | 4082.98 | 4129.188 | 25.6 | 125 |
| h100_40_0_2 | 4027.06 | 4096.788 | 22.6 | 88.54 | 4038.82 | 4072.064 | 26.2 | 101.8 | 4070.9 | 4106.112 | 23 | 146 |
| h100_40_0_3 | 3882.64 | 3923.128 | 31.2 | 110.08 | 3861.82 | 3921.808 | 31.8 | 136.6 | 3919.56 | 3974.664 | 32.2 | 133 |
| h100_40_0_4 | 4122.08 | 4186.268 | 23.6 | 123.97 | 4042.88 | 4084.268 | 29 | 102.2 | 4143.36 | 4195.98 | 24 | 143 |
| h100_40_1_0 | 4447.4 | 4514.596 | 33.2 | 101.58 | 4349.46 | 4476.232 | 29.4 | 123.8 | 4503.38 | 4528.632 | 30.6 | 163 |
| h100_40_1_1 | 4386.1 | 4452.664 | 33.8 | 116.57 | 4380.64 | 4446.904 | 34.8 | 125 | 4439.68 | 4481.556 | 32 | 139 |
| h100_40_1_2 | 4396.34 | 4436.012 | 33 | 129.34 | 4396.52 | 4465.572 | 34.4 | 111.6 | 4437.78 | 4496.212 | 33.4 | 176 |
| h100_40_1_3 | 4295.88 | 4335.04 | 35.8 | 118.47 | 4275.06 | 4382.084 | 32.8 | 131.8 | 4339.76 | 4388.512 | 31.8 | 142 |
| h100_40_1_4 | 4479.1 | 4568.864 | 29.8 | 98.98 | 4544.14 | 4590.904 | 32.4 | 94.4 | 4527.8 | 4586.44 | 33.6 | 145 |
| h100_40_2_0 | 4545.96 | 4634.152 | 29.2 | 104.15 | 4603.64 | 4638.652 | 34.4 | 129.8 | 4608.8 | 4652.392 | 30.6 | 141 |
| h100_40_2_1 | 4480.66 | 4623.752 | 32 | 108.06 | 4491.28 | 4609.904 | 34.4 | 122.2 | 4561.18 | 4622.664 | 32.6 | 165 |
| h100_40_2_2 | 4612.04 | 4645.636 | 32 | 128.22 | 4601.72 | 4636.368 | 31.4 | 116.8 | 4576.02 | 4623.932 | 29.2 | 131 |
| h100_40_2_3 | 4369.82 | 4432.904 | 38.6 | 183.18 | 4377.48 | 4461.34 | 38 | 125.4 | 4439.96 | 4481.936 | 36.6 | 153 |
| h100_40_2_4 | 4668.38 | 4742.964 | 32.4 | 108.38 | 4712.68 | 4746.248 | 29.2 | 119.6 | 4640.92 | 4728.256 | 29.6 | 143 |

## Appendix E The solutions obtained by the UBA and their comparisons with the ALNS-VS solutions

As mentioned in the manuscript, ALNS-VS_Unique solutions in Tables D1.-D.4 are adopted for further analysis. The column "UB" presents the objective function values of the solutions obtained by the proposed UBA. The "VS-UBA(%)" column indicates how much improvement on objective is offered by ALNS-VS over the UBA.

Table E.1. UBA solutions and their comparisons with ALNS-VS solutions.

| **Instance** | **UB** | **CPU** | **VS-UBA(%)** | **Instance** | **UB** | **CPU** | **VS-UBA(%)** | **Instance** | **UB** | **CPU** | **VS-UBA(%)** |
|---|---|---|---|---|---|---|---|---|---|---|---|
| h30_10_0_0 | 911.7 | 0.57 | 39.8 | h50_10_0_0 | 1434.6 | 0.70 | 36.6 | h100_10_0_0 | 3004.5 | 1.61 | 34.2 |
| h30_10_0_1 | 892.2 | 0.15 | 35.9 | h50_10_0_1 | 1456.1 | 0.30 | 36.1 | h100_10_0_1 | 2979.1 | 1.07 | 31.1 |
| h30_10_0_2 | 901.4 | 0.21 | 38.7 | h50_10_0_2 | 1460.3 | 0.43 | 35.5 | h100_10_0_2 | 2999.6 | 0.66 | 31.9 |
| h30_10_0_3 | 907.7 | 0.26 | 37.1 | h50_10_0_3 | 1442.8 | 0.45 | 38.4 | h100_10_0_3 | 2998.3 | 0.98 | 32.8 |
| h30_10_0_4 | 884.7 | 0.07 | 36.8 | h50_10_0_4 | 1402.6 | 0.31 | 34.9 | h100_10_0_4 | 3032.2 | 1.15 | 35.1 |
| h30_10_1_0 | 1011.3 | 0.22 | 40.3 | h50_10_1_0 | 1714.7 | 0.51 | 37.6 | h100_10_1_0 | 3686.0 | 0.97 | 35.4 |
| h30_10_1_1 | 979.3 | 0.16 | 37.3 | h50_10_1_1 | 1702.2 | 0.18 | 37.0 | h100_10_1_1 | 3685.0 | 0.75 | 36.9 |
| h30_10_1_2 | 983.4 | 0.15 | 38.5 | h50_10_1_2 | 1719.6 | 0.24 | 38.7 | h100_10_1_2 | 3704.7 | 0.78 | 35.5 |
| h30_10_1_3 | 984.1 | 0.32 | 39.3 | h50_10_1_3 | 1716.8 | 0.41 | 40.5 | h100_10_1_3 | 3683.6 | 1.07 | 36.8 |
| h30_10_1_4 | 966.7 | 0.07 | 38.1 | h50_10_1_4 | 1676.5 | 0.32 | 38.6 | h100_10_1_4 | 3696.6 | 1.25 | 35.7 |
| h30_10_2_0 | 1100.9 | 0.10 | 39.6 | h50_10_2_0 | 2018.7 | 0.51 | 40.0 | h100_10_2_0 | 3863.7 | 1.52 | 35.5 |
| h30_10_2_1 | 1118.2 | 0.22 | 40.5 | h50_10_2_1 | 2032.5 | 0.31 | 40.0 | h100_10_2_1 | 3834.7 | 0.97 | 37.1 |
| h30_10_2_2 | 1127.4 | 0.14 | 40.4 | h50_10_2_2 | 2038.3 | 0.50 | 40.3 | h100_10_2_2 | 3848.0 | 1.45 | 35.5 |
| h30_10_2_3 | 1128.1 | 0.30 | 42.1 | h50_10_2_3 | 2020.8 | 0.38 | 41.5 | h100_10_2_3 | 3845.6 | 1.70 | 36.1 |
| h30_10_2_4 | 1110.7 | 0.06 | 39.4 | h50_10_2_4 | 1974.7 | 0.29 | 38.8 | h100_10_2_4 | 3859.2 | 2.40 | 35.8 |

| h30_20_0_0 | 1117.1 | 0.10 | 31.1 | h50_20_0_0 | 1789.9 | 0.46 | 29.2 | h100_20_0_0 | 3553.0 | 2.64 | 25.2 |
|---|---|---|---|---|---|---|---|---|---|---|---|
| h30_20_0_1 | 1092.5 | 0.20 | 27.2 | h50_20_0_1 | 1787.1 | 0.23 | 27.3 | h100_20_0_1 | 3706.7 | 1.29 | 27.6 |
| h30_20_0_2 | 1113.9 | 0.24 | 26.5 | h50_20_0_2 | 1766.2 | 0.27 | 26.3 | h100_20_0_2 | 3710.5 | 1.88 | 29.6 |
| h30_20_0_3 | 1135.7 | 0.18 | 32.4 | h50_20_0_3 | 1680.0 | 0.61 | 30.0 | h100_20_0_3 | 3668.2 | 1.61 | 28.3 |
| h30_20_0_4 | 1074.2 | 0.08 | 27.6 | h50_20_0_4 | 1818.4 | 0.23 | 29.6 | h100_20_0_4 | 3666.2 | 0.95 | 25.9 |
| h30_20_1_0 | 1197.9 | 0.48 | 31.8 | h50_20_1_0 | 2042.2 | 0.40 | 31.1 | h100_20_1_0 | 4274.2 | 2.05 | 29.2 |
| h30_20_1_1 | 1212.2 | 0.10 | 28.5 | h50_20_1_1 | 2043.0 | 0.17 | 32.3 | h100_20_1_1 | 4415.4 | 1.20 | 32.0 |
| h30_20_1_2 | 1194.9 | 0.18 | 32.4 | h50_20_1_2 | 2077.9 | 0.24 | 29.5 | h100_20_1_2 | 4425.0 | 3.63 | 31.4 |
| h30_20_1_3 | 1205.3 | 0.83 | 34.8 | h50_20_1_3 | 1889.1 | 0.51 | 29.5 | h100_20_1_3 | 4270.6 | 1.37 | 27.9 |
| h30_20_1_4 | 1161.3 | 0.13 | 30.4 | h50_20_1_4 | 2061.2 | 0.56 | 32.9 | h100_20_1_4 | 4356.7 | 0.78 | 29.1 |
| h30_20_2_0 | 1343.1 | 0.15 | 34.9 | h50_20_2_0 | 2346.2 | 0.42 | 32.2 | h100_20_2_0 | 4430.4 | 1.86 | 28.3 |
| h30_20_2_1 | 1318.5 | 0.31 | 30.0 | h50_20_2_1 | 2365.1 | 0.19 | 32.8 | h100_20_2_1 | 4501.0 | 1.27 | 30.4 |
| h30_20_2_2 | 1339.9 | 0.26 | 33.2 | h50_20_2_2 | 2344.2 | 0.25 | 31.6 | h100_20_2_2 | 4613.5 | 2.50 | 32.0 |
| h30_20_2_3 | 1361.2 | 1.77 | 37.6 | h50_20_2_3 | 2290.8 | 0.50 | 36.0 | h100_20_2_3 | 4443.8 | 2.16 | 30.6 |
| h30_20_2_4 | 1300.2 | 0.07 | 32.0 | h50_20_2_4 | 2293.8 | 0.65 | 32.7 | h100_20_2_4 | 4523.2 | 1.55 | 29.8 |
| h30_30_0_0 | 1426.2 | 0.22 | 32.7 | h50_30_0_0 | 2093.8 | 0.52 | 25.1 | h100_30_0_0 | 4307.1 | 2.82 | 24.0 |
| h30_30_0_1 | 1288.4 | 0.13 | 19.4 | h50_30_0_1 | 2161.8 | 0.25 | 26.0 | h100_30_0_1 | 4221.4 | 0.55 | 19.5 |
| h30_30_0_2 | 1299.6 | 0.13 | 18.9 | h50_30_0_2 | 2177.6 | 0.22 | 22.9 | h100_30_0_2 | 4213.4 | 1.14 | 23.1 |
| h30_30_0_3 | 1322.4 | 0.32 | 25.2 | h50_30_0_3 | 1886.8 | 2.34 | 20.3 | h100_30_0_3 | 4275.9 | 0.67 | 25.4 |
| h30_30_0_4 | 1246.7 | 0.16 | 22.4 | h50_30_0_4 | 2174.7 | 0.18 | 25.5 | h100_30_0_4 | 4402.6 | 0.56 | 22.8 |
| h30_30_1_0 | 1409.2 | 0.14 | 25.8 | h50_30_1_0 | 3315.6 | 0.54 | 48.0 | h100_30_1_0 | 4931.5 | 0.88 | 24.9 |
| h30_30_1_1 | 1426.2 | 0.07 | 22.4 | h50_30_1_1 | 2364.6 | 0.19 | 26.2 | h100_30_1_1 | 5076.5 | 0.54 | 26.8 |
| h30_30_1_2 | 1381.6 | 0.10 | 24.2 | h50_30_1_2 | 2490.6 | 0.25 | 27.4 | h100_30_1_2 | 5018.3 | 0.74 | 27.3 |
| h30_30_1_3 | 1388.5 | 0.36 | 28.2 | h50_30_1_3 | 2160.8 | 2.37 | 24.8 | h100_30_1_3 | 4855.2 | 2.20 | 26.1 |
| h30_30_1_4 | 1336.2 | 0.06 | 24.5 | h50_30_1_4 | 2398.7 | 0.66 | 27.3 | h100_30_1_4 | 4982.6 | 0.52 | 25.0 |
| h30_30_2_0 | 1561.0 | 0.11 | 31.5 | h50_30_2_0 | 3566.2 | 0.52 | 48.5 | h100_30_2_0 | 5050.6 | 2.70 | 23.3 |
| h30_30_2_1 | 1514.4 | 0.11 | 23.2 | h50_30_2_1 | 2732.7 | 0.32 | 29.1 | h100_30_2_1 | 5111.0 | 0.58 | 25.1 |
| h30_30_2_2 | 1472.1 | 0.14 | 22.3 | h50_30_2_2 | 2731.3 | 0.29 | 27.5 | h100_30_2_2 | 5199.9 | 1.41 | 29.0 |
| h30_30_2_3 | 1532.5 | 0.38 | 31.0 | h50_30_2_3 | 2464.8 | 2.91 | 28.4 | h100_30_2_3 | 5148.3 | 0.85 | 27.4 |
| h30_30_2_4 | 1472.7 | 0.09 | 27.0 | h50_30_2_4 | 2584.1 | 0.54 | 25.3 | h100_30_2_4 | 5171.9 | 0.57 | 26.8 |
| h30_40_0_0 | 1545.5 | 0.32 | 24.7 | h50_40_0_0 | 2424.6 | 0.92 | 22.8 | h100_40_0_0 | 4964.7 | 3.76 | 22.0 |
| h30_40_0_1 | 1487.8 | 0.16 | 16.0 | h50_40_0_1 | 2602.5 | 0.34 | 24.2 | h100_40_0_1 | 4851.8 | 0.60 | 18.8 |
| h30_40_0_2 | 1505.3 | 0.13 | 15.3 | h50_40_0_2 | 2427.0 | 0.19 | 17.1 | h100_40_0_2 | 4917.0 | 1.14 | 18.1 |

Table E.1. UBA solutions and their comparisons with ALNS-VS solutions (cont.).

| **Instance** | **UB** | **CPU** | **VS-UBA(%)** | **Instance** | **UB** | **CPU** | **VS-UBA(%)** | **Instance** | **UB** | **CPU** | **VS-UBA(%)** |
|---|---|---|---|---|---|---|---|---|---|---|---|
| h30_40_0_3 | 1521.0 | 0.18 | 22.4 | h50_40_0_3 | 2149.1 | 1.72 | 17.4 | h100_40_0_3 | 4911.8 | 0.49 | 21.0 |
| h30_40_0_4 | 1437.6 | 0.15 | 17.2 | h50_40_0_4 | 3492.6 | 0.65 | 44.1 | h100_40_0_4 | 5091.4 | 0.54 | 19.0 |
| h30_40_1_0 | 1703.2 | 0.22 | 26.4 | h50_40_1_0 | 5516.2 | 0.48 | 62.9 | h100_40_1_0 | 7646.1 | 0.88 | 41.8 |
| h30_40_1_1 | 1585.0 | 0.07 | 16.2 | h50_40_1_1 | 3776.5 | 0.48 | 45.2 | h100_40_1_1 | 9488.7 | 1.82 | 53.8 |
| h30_40_1_2 | 1597.5 | 0.22 | 20.3 | h50_40_1_2 | 4709.0 | 0.24 | 54.5 | h100_40_1_2 | 10561.2 | 0.79 | 58.4 |
| h30_40_1_3 | 1586.5 | 0.26 | 22.8 | h50_40_1_3 | 2423.1 | 2.10 | 21.0 | h100_40_1_3 | 12520.1 | 1.17 | 65.7 |
| h30_40_1_4 | 1530.5 | 0.08 | 20.5 | h50_40_1_4 | 2607.6 | 0.83 | 20.2 | h100_40_1_4 | 11516.5 | 0.92 | 61.1 |
| h30_40_2_0 | 1736.5 | 0.09 | 25.2 | h50_40_2_0 | 5791.2 | 0.53 | 60.5 | h100_40_2_0 | 11453.7 | 2.92 | 60.3 |
| h30_40_2_1 | 1713.8 | 0.10 | 17.5 | h50_40_2_1 | 8733.6 | 0.28 | 73.7 | h100_40_2_1 | 12504.8 | 1.13 | 64.2 |
| h30_40_2_2 | 1658.1 | 0.14 | 18.7 | h50_40_2_2 | 6680.1 | 0.28 | 64.8 | h100_40_2_2 | 9613.9 | 4.91 | 52.0 |
| h30_40_2_3 | 2650.4 | 0.28 | 50.7 | h50_40_2_3 | 2841.8 | 0.59 | 27.6 | h100_40_2_3 | 9537.6 | 1.79 | 54.2 |
| h30_40_2_4 | 1663.6 | 0.06 | 25.0 | h50_40_2_4 | 3849.0 | 0.39 | 40.5 | h100_40_2_4 | 10626.1 | 0.91 | 56.1 |

## Appendix F The effect of DP policy and the HHSRP-M solutions

In section 5.5, we discussed the effect of DP policy on total flow time by comparing the solutions of the HHSRP-M with the HHSRP-VS. In this appendix, Tables F.1 through F.4 demonstrates the solutions obtained by the ALNS-M and their comparisons with the ALNS-VS in details. As mentioned in the manuscript, ALNS-VS_Unique solutions in Tables D1.-D.4 are adopted for further analysis. In the following tables, the "Best-found" and "Avg." columns indicate the objective values of the best-found and the averages of the best solutions found in five replications by the ALNS-M, respectively. Additionally, the column "VS-M%" presents the percentage improvement on the objective offered by ALNS-VS over the ALNS-M.

Table F.1. ALNS-M solutions for 10-patient instances and their comparisons with ALNS-VS.

| **Instance** | **Best-found** | **Avg.** | **CPU** | **VS-M(%)** | **Instance** | **Best-found** | **Avg.** | **CPU** | **VS-M(%)** |
|---|---|---|---|---|---|---|---|---|---|
| h10_10_0_0 | 318.6 | 318.6 | 1.09 | 24.5 | h10_30_0_0 | 506.5 | 506.5 | 1.05 | 6.7 |
| h10_10_0_1 | 338.7 | 338.7 | 1.04 | 24.9 | h10_30_0_1 | 569.1 | 569.1 | 1.07 | 7.3 |
| h10_10_0_2 | 333.2 | 333.2 | 1.06 | 30.3 | h10_30_0_2 | 552.3 | 552.3 | 1.04 | 9.4 |
| h10_10_0_3 | 337.2 | 337.2 | 1.07 | 24.2 | h10_30_0_3 | 558.5 | 558.5 | 1.06 | 9.7 |
| h10_10_0_4 | 318.2 | 318.2 | 1.02 | 31.7 | h10_30_0_4 | 514.1 | 514.1 | 1.00 | 12.5 |
| h10_10_1_0 | 362.6 | 362.6 | 1.04 | 29.3 | h10_30_1_0 | 550.5 | 550.5 | 1.11 | 5.6 |
| h10_10_1_1 | 382.7 | 382.7 | 1.03 | 26.3 | h10_30_1_1 | 613.1 | 613.1 | 1.02 | 11.2 |
| h10_10_1_2 | 377.2 | 377.2 | 1.02 | 35.6 | h10_30_1_2 | 596.3 | 596.3 | 1.06 | 7.0 |
| h10_10_1_3 | 381.2 | 381.2 | 1.02 | 28.7 | h10_30_1_3 | 602.5 | 602.5 | 1.06 | 12.4 |
| h10_10_1_4 | 362.2 | 362.2 | 0.96 | 34.4 | h10_30_1_4 | 558.1 | 558.1 | 0.97 | 13.2 |
| h10_10_2_0 | 424.6 | 424.6 | 1.02 | 34.5 | h10_30_2_0 | 612.5 | 612.5 | 1.05 | 12.1 |
| h10_10_2_1 | 444.7 | 444.7 | 1.02 | 30.7 | h10_30_2_1 | 675.1 | 675.1 | 1.09 | 10.0 |
| h10_10_2_2 | 439.2 | 439.2 | 1.02 | 37.0 | h10_30_2_2 | 658.3 | 658.3 | 1.07 | 12.3 |
| h10_10_2_3 | 443.2 | 443.2 | 1.02 | 30.9 | h10_30_2_3 | 664.5 | 664.5 | 1.06 | 17.4 |
| h10_10_2_4 | 424.2 | 424.2 | 0.98 | 34.2 | h10_30_2_4 | 620.1 | 620.1 | 0.98 | 20.0 |
| h10_20_0_0 | 415.4 | 415.4 | 1.06 | 10.5 | h10_40_0_0 | 602.7 | 602.7 | 1.06 | 5.9 |
| h10_20_0_1 | 449.0 | 449.0 | 1.04 | 10.7 | h10_40_0_1 | 687.4 | 687.4 | 1.05 | 6.2 |
| h10_20_0_2 | 442.8 | 442.8 | 1.04 | 16.6 | h10_40_0_2 | 659.9 | 659.9 | 1.04 | 6.9 |
| h10_20_0_3 | 444.7 | 444.7 | 1.57 | 13.5 | h10_40_0_3 | 669.5 | 669.5 | 1.03 | 7.5 |

Table F.1. ALNS-M solutions for 10-patient instances and their comparisons with ALNS-VS (cont.).

| **Instance** | **Best-found** | **Avg.** | **CPU** | **VS-M(%)** | **Instance** | **Best-found** | **Avg.** | **CPU** | **VS-M(%)** |
|---|---|---|---|---|---|---|---|---|---|
| h10_20_0_4 | 418.6 | 418.6 | 1.01 | 18.9 | h10_40_0_4 | 608.6 | 608.6 | 1.12 | 9.3 |
| h10_20_1_0 | 459.4 | 459.4 | 1.06 | 17.5 | h10_40_1_0 | 646.7 | 646.7 | 2.85 | 4.9 |
| h10_20_1_1 | 493.0 | 493.0 | 1.06 | 16.2 | h10_40_1_1 | 731.4 | 731.4 | 1.51 | 9.0 |
| h10_20_1_2 | 486.8 | 486.8 | 1.05 | 19.6 | h10_40_1_2 | 703.9 | 703.9 | 1.00 | 3.7 |
| h10_20_1_3 | 488.7 | 488.7 | 1.10 | 18.6 | h10_40_1_3 | 713.5 | 713.5 | 1.00 | 8.9 |
| h10_20_1_4 | 462.6 | 462.6 | 0.97 | 22.9 | h10_40_1_4 | 652.6 | 652.6 | 0.93 | 8.0 |
| h10_20_2_0 | 521.4 | 521.4 | 1.04 | 15.6 | h10_40_2_0 | 708.7 | 708.7 | 1.01 | 9.7 |
| h10_20_2_1 | 555.0 | 555.0 | 1.04 | 15.9 | h10_40_2_1 | 793.4 | 793.4 | 0.99 | 6.3 |
| h10_20_2_2 | 548.8 | 548.8 | 1.08 | 20.4 | h10_40_2_2 | 765.9 | 765.9 | 0.99 | 8.0 |
| h10_20_2_3 | 550.7 | 550.7 | 1.06 | 21.2 | h10_40_2_3 | 775.5 | 775.5 | 1.00 | 14.1 |
| h10_20_2_4 | 524.6 | 524.6 | 0.97 | 27.4 | h10_40_2_4 | 714.6 | 714.6 | 1.03 | 15.9 |

Table F.2. ALNS-M solutions for 30-patient instances and their comparisons with ALNS-VS.

| Instance | Best-found | Avg. | CPU | VS-M(%) | Instance | Best-found | Avg. | CPU | VS-M(%) |
|---|---|---|---|---|---|---|---|---|---|
| h30_10_0_0 | 846.8 | 849.7 | 17.1 | 35.4 | h30_30_0_0 | 1182.9 | 1188.7 | 11.0 | 19.3 |
| h30_10_0_1 | 863.3 | 864.7 | 12.9 | 33.9 | h30_30_0_1 | 1226.4 | 1228.6 | 12.4 | 15.4 |
| h30_10_0_2 | 876.0 | 879.5 | 10.9 | 37.2 | h30_30_0_2 | 1230.5 | 1230.5 | 12.0 | 14.4 |
| h30_10_0_3 | 856.4 | 864.5 | 12.0 | 33.9 | h30_30_0_3 | 1203.3 | 1206.9 | 13.4 | 18.0 |
| h30_10_0_4 | 855.7 | 855.8 | 11.1 | 34.6 | h30_30_0_4 | 1171.6 | 1171.6 | 13.0 | 17.4 |
| h30_10_1_0 | 928.8 | 933.8 | 12.6 | 35.3 | h30_30_1_0 | 1272.2 | 1274.6 | 15.3 | 18.0 |
| h30_10_1_1 | 945.3 | 945.3 | 13.8 | 35.0 | h30_30_1_1 | 1312.0 | 1312.0 | 18.3 | 15.7 |
| h30_10_1_2 | 958.0 | 961.5 | 11.0 | 37.1 | h30_30_1_2 | 1312.5 | 1312.5 | 17.1 | 20.2 |
| h30_10_1_3 | 938.4 | 946.5 | 12.4 | 36.9 | h30_30_1_3 | 1285.3 | 1288.9 | 17.7 | 22.7 |
| h30_10_1_4 | 937.7 | 937.8 | 12.4 | 36.2 | h30_30_1_4 | 1253.6 | 1253.6 | 15.1 | 19.6 |
| h30_10_2_0 | 1082.8 | 1082.8 | 12.3 | 38.6 | h30_30_2_0 | 1422.2 | 1427.2 | 17.8 | 25.1 |
| h30_10_2_1 | 1098.3 | 1099.6 | 11.9 | 39.5 | h30_30_2_1 | 1458.2 | 1458.2 | 18.1 | 20.2 |
| h30_10_2_2 | 1106.3 | 1106.3 | 10.0 | 39.3 | h30_30_2_2 | 1467.5 | 1467.5 | 16.0 | 22.0 |
| h30_10_2_3 | 1089.6 | 1089.8 | 12.2 | 40.1 | h30_30_2_3 | 1429.3 | 1432.8 | 18.4 | 26.2 |
| h30_10_2_4 | 1081.9 | 1081.9 | 9.4 | 37.8 | h30_30_2_4 | 1397.6 | 1397.6 | 15.3 | 23.1 |
| h30_20_0_0 | 1019.0 | 1019.0 | 11.9 | 24.4 | h30_40_0_0 | 1350.8 | 1362.0 | 18.4 | 14.6 |
| h30_20_0_1 | 1047.8 | 1049.7 | 11.6 | 24.2 | h30_40_0_1 | 1413.7 | 1413.7 | 19.6 | 11.6 |
| h30_20_0_2 | 1045.1 | 1050.1 | 10.2 | 22.1 | h30_40_0_2 | 1426.5 | 1426.5 | 18.5 | 10.6 |
| h30_20_0_3 | 1017.1 | 1025.1 | 11.8 | 25.1 | h30_40_0_3 | 1369.7 | 1374.1 | 18.9 | 14.1 |
| h30_20_0_4 | 1025.2 | 1025.2 | 9.4 | 24.2 | h30_40_0_4 | 1340.6 | 1340.6 | 16.4 | 11.2 |
| h30_20_1_0 | 1112.9 | 1113.3 | 10.7 | 26.6 | h30_40_1_0 | 1432.8 | 1434.6 | 17.6 | 12.7 |
| h30_20_1_1 | 1131.6 | 1132.7 | 11.5 | 23.5 | h30_40_1_1 | 1498.7 | 1499.2 | 18.6 | 11.4 |
| h30_20_1_2 | 1133.4 | 1133.4 | 9.9 | 28.7 | h30_40_1_2 | 1508.5 | 1508.5 | 18.7 | 15.6 |
| h30_20_1_3 | 1110.0 | 1115.7 | 11.2 | 29.6 | h30_40_1_3 | 1471.6 | 1476.4 | 19.6 | 17.1 |
| h30_20_1_4 | 1107.2 | 1107.2 | 9.5 | 26.9 | h30_40_1_4 | 1422.6 | 1422.6 | 17.3 | 14.4 |
| h30_20_2_0 | 1256.9 | 1257.8 | 11.0 | 30.4 | h30_40_2_0 | 1579.7 | 1610.7 | 18.5 | 19.4 |
| h30_20_2_1 | 1280.7 | 1280.7 | 10.9 | 27.9 | h30_40_2_1 | 1654.0 | 1654.0 | 20.2 | 14.5 |
| h30_20_2_2 | 1282.3 | 1282.3 | 9.5 | 30.2 | h30_40_2_2 | 1657.8 | 1664.1 | 17.0 | 19.0 |
| h30_20_2_3 | 1258.4 | 1260.3 | 11.0 | 32.6 | h30_40_2_3 | 1611.7 | 1620.9 | 18.5 | 19.4 |
| h30_20_2_4 | 1251.2 | 1251.2 | 9.8 | 29.4 | h30_40_2_4 | 1566.6 | 1566.6 | 16.3 | 20.4 |

Table F.3. ALNS-M solutions for 50-patient instances and their comparisons with ALNS-VS.

| Instance | Best-found | Avg. | CPU | VS-M(%) | Instance | Best-found | Avg. | CPU | VS-M(%) |
|---|---|---|---|---|---|---|---|---|---|
| h50_10_0_0 | 1371.0 | 1374.0 | 33 | 33.6 | h50_30_0_0 | 1883.9 | 1892.9 | 38 | 16.7 |
| h50_10_0_1 | 1367.1 | 1372.0 | 31 | 31.9 | h50_30_0_1 | 1901.0 | 1915.3 | 38 | 15.9 |
| h50_10_0_2 | 1367.9 | 1372.9 | 34 | 31.1 | h50_30_0_2 | 1898.0 | 1938.3 | 38 | 11.5 |
| h50_10_0_3 | 1339.5 | 1343.4 | 35 | 33.6 | h50_30_0_3 | 1806.0 | 1820.7 | 39 | 16.7 |
| h50_10_0_4 | 1366.2 | 1371.5 | 41 | 33.1 | h50_30_0_4 | 1886.3 | 1900.2 | 38 | 14.1 |
| h50_10_1_0 | 1642.4 | 1647.6 | 40 | 34.8 | h50_30_1_0 | 2171.1 | 2182.5 | 35 | 20.6 |
| h50_10_1_1 | 1641.1 | 1645.0 | 35 | 34.6 | h50_30_1_1 | 2160.6 | 2178.7 | 28 | 19.2 |
| h50_10_1_2 | 1638.5 | 1644.8 | 42 | 35.7 | h50_30_1_2 | 2190.9 | 2201.9 | 20 | 17.5 |
| h50_10_1_3 | 1612.2 | 1619.5 | 39 | 36.7 | h50_30_1_3 | 2088.2 | 2097.1 | 20 | 22.1 |

| h50_10_1_4 | 1636.8 | 1643.2 | 42 | 37.1 | h50_30_1_4 | 2161.5 | 2183.0 | 20 | 19.3 |
|---|---|---|---|---|---|---|---|---|---|
| h50_10_2_0 | 1950.0 | 1952.8 | 41 | 37.9 | h50_30_2_0 | 2464.0 | 2498.7 | 20 | 25.5 |
| h50_10_2_1 | 1948.7 | 1951.2 | 37 | 37.4 | h50_30_2_1 | 2488.0 | 2498.8 | 20 | 22.1 |
| h50_10_2_2 | 1953.4 | 1955.6 | 40 | 37.7 | h50_30_2_2 | 2504.9 | 2523.6 | 20 | 21.0 |
| h50_10_2_3 | 1916.3 | 1918.6 | 41 | 38.3 | h50_30_2_3 | 2398.0 | 2403.9 | 20 | 26.4 |
| h50_10_2_4 | 1940.8 | 1950.9 | 35 | 37.8 | h50_30_2_4 | 2469.0 | 2495.5 | 20 | 21.8 |
| h50_20_0_0 | 1627.0 | 1628.9 | 44 | 22.1 | h50_40_0_0 | 2145.5 | 2182.5 | 20 | 12.8 |
| h50_20_0_1 | 1625.8 | 1636.9 | 36 | 20.1 | h50_40_0_1 | 2161.1 | 2191.3 | 20 | 8.8 |
| h50_20_0_2 | 1631.3 | 1638.1 | 42 | 20.2 | h50_40_0_2 | 2186.5 | 2207.8 | 20 | 8.0 |
| h50_20_0_3 | 1578.7 | 1580.7 | 24 | 25.5 | h50_40_0_3 | 2056.0 | 2063.1 | 20 | 13.7 |
| h50_20_0_4 | 1624.2 | 1627.6 | 27 | 21.1 | h50_40_0_4 | 2145.7 | 2153.8 | 20 | 9.1 |
| h50_20_1_0 | 1901.0 | 1909.5 | 26 | 26.0 | h50_40_1_0 | 2420.2 | 2471.9 | 20 | 15.4 |
| h50_20_1_1 | 1899.8 | 1912.6 | 25 | 27.2 | h50_40_1_1 | 2435.1 | 2463.1 | 20 | 14.9 |
| h50_20_1_2 | 1909.0 | 1922.2 | 33 | 23.2 | h50_40_1_2 | 2467.4 | 2485.9 | 20 | 13.1 |
| h50_20_1_3 | 1844.8 | 1851.4 | 43 | 27.8 | h50_40_1_3 | 2330.3 | 2343.8 | 20 | 17.9 |
| h50_20_1_4 | 1900.0 | 1903.6 | 45 | 27.2 | h50_40_1_4 | 2427.7 | 2461.9 | 20 | 14.3 |
| h50_20_2_0 | 2207.8 | 2211.4 | 42 | 27.9 | h50_40_2_0 | 2744.9 | 2806.3 | 20 | 16.7 |
| h50_20_2_1 | 2203.8 | 2216.8 | 38 | 27.9 | h50_40_2_1 | 2754.6 | 2799.3 | 20 | 16.8 |
| h50_20_2_2 | 2212.1 | 2219.1 | 39 | 27.5 | h50_40_2_2 | 2809.8 | 3002.9 | 20 | 16.3 |
| h50_20_2_3 | 2152.3 | 2156.7 | 35 | 31.9 | h50_40_2_3 | 2662.0 | 2662.0 | 20 | 22.7 |
| h50_20_2_4 | 2205.8 | 2220.2 | 36 | 30.0 | h50_40_2_4 | 2738.8 | 2774.0 | 20 | 16.4 |

Table F.4. ALNS-M solutions for 100-patient instances and their comparisons with ALNS-VS.

| **Instance** | **Best-found** | **Avg.** | **CPU** | **VS-M(%)** | **Instance** | **Best-found** | **Avg.** | **CPU** | **VS-M(%)** |
|---|---|---|---|---|---|---|---|---|---|
| h100_10_0_0 | 2867.7 | 2891.2 | 63 | 31.1 | h100_30_0_0 | 3927.5 | 3965.8 | 62 | 16.7 |
| h100_10_0_1 | 2861.3 | 2868.6 | 65 | 28.3 | h100_30_0_1 | 3883.6 | 3912.8 | 62 | 12.5 |
| h100_10_0_2 | 2876.0 | 2886.5 | 65 | 29.0 | h100_30_0_2 | 3930.7 | 3950.4 | 64 | 17.5 |
| h100_10_0_3 | 2859.4 | 2884.4 | 64 | 29.6 | h100_30_0_3 | 3864.3 | 3893.5 | 64 | 17.4 |
| h100_10_0_4 | 2889.1 | 2900.7 | 63 | 31.9 | h100_30_0_4 | 3939.7 | 3966.2 | 63 | 13.7 |
| h100_10_1_0 | 3564.8 | 3571.7 | 67 | 33.2 | h100_30_1_0 | 4610.1 | 4645.9 | 60 | 19.7 |

Table F.4. ALNS-M solutions for 100-patient instances and their comparisons with ALNS-VS (cont.).

| **Instance** | **Best-found** | **Avg.** | **CPU** | **VS-M(%)** | **Instance** | **Best-found** | **Avg.** | **CPU** | **VS-M(%)** |
|---|---|---|---|---|---|---|---|---|---|
| h100_10_1_1 | 3544.1 | 3551.1 | 65 | 34.3 | h100_30_1_1 | 4610.6 | 4620.2 | 60 | 19.5 |
| h100_10_1_2 | 3545.8 | 3559.2 | 65 | 32.6 | h100_30_1_2 | 4599.5 | 4627.1 | 61 | 20.7 |
| h100_10_1_3 | 3546.0 | 3561.1 | 66 | 34.4 | h100_30_1_3 | 4557.4 | 4580.9 | 60 | 21.3 |
| h100_10_1_4 | 3565.1 | 3578.5 | 65 | 33.3 | h100_30_1_4 | 4647.2 | 4671.9 | 60 | 19.6 |
| h100_10_2_0 | 3724.1 | 3738.9 | 65 | 33.1 | h100_30_2_0 | 4762.8 | 4814.4 | 60 | 18.6 |
| h100_10_2_1 | 3707.7 | 3712.7 | 65 | 34.9 | h100_30_2_1 | 4770.2 | 4770.2 | 60 | 19.7 |

| h100_10_2_2 | 3715.7 | 3723.4 | 65 | 33.2 | h100_30_2_2 | 4800.1 | 4810.5 | 60 | 23.1 |
|---|---|---|---|---|---|---|---|---|---|
| h100_10_2_3 | 3711.8 | 3718.9 | 65 | 33.8 | h100_30_2_3 | 4715.6 | 4731.1 | 60 | 20.8 |
| h100_10_2_4 | 3736.7 | 3748.2 | 66 | 33.7 | h100_30_2_4 | 4869.2 | 4885.9 | 63 | 22.3 |
| h100_20_0_0 | 3412.3 | 3412.3 | 63 | 22.1 | h100_40_0_0 | 4409.2 | 4475.7 | 61 | 12.2 |
| h100_20_0_1 | 3372.6 | 3400.2 | 65 | 20.5 | h100_40_0_1 | 4394.0 | 4491.5 | 62 | 10.4 |
| h100_20_0_2 | 3396.5 | 3402.0 | 65 | 23.1 | h100_40_0_2 | 4468.0 | 4497.0 | 63 | 9.9 |
| h100_20_0_3 | 3349.3 | 3379.4 | 64 | 21.5 | h100_40_0_3 | 4404.9 | 4427.3 | 62 | 11.9 |
| h100_20_0_4 | 3415.4 | 3458.0 | 64 | 20.4 | h100_40_0_4 | 4473.4 | 4540.0 | 60 | 7.9 |
| h100_20_1_0 | 4077.7 | 4091.4 | 61 | 25.8 | h100_40_1_0 | 5220.2 | 5225.9 | 61 | 14.8 |
| h100_20_1_1 | 4089.2 | 4101.6 | 63 | 26.5 | h100_40_1_1 | 5149.0 | 5205.4 | 62 | 14.8 |
| h100_20_1_2 | 4073.7 | 4076.1 | 66 | 25.5 | h100_40_1_2 | 5182.6 | 5186.0 | 63 | 15.2 |
| h100_20_1_3 | 4059.2 | 4066.6 | 67 | 24.2 | h100_40_1_3 | 5145.1 | 5186.9 | 60 | 16.5 |
| h100_20_1_4 | 4111.1 | 4133.8 | 65 | 24.8 | h100_40_1_4 | 5266.3 | 5280.1 | 64 | 14.9 |
| h100_20_2_0 | 4215.0 | 4255.3 | 62 | 24.6 | h100_40_2_0 | 5365.4 | 5409.0 | 67 | 15.3 |
| h100_20_2_1 | 4221.5 | 4240.4 | 66 | 25.8 | h100_40_2_1 | 5390.7 | 5435.7 | 62 | 16.9 |
| h100_20_2_2 | 4234.5 | 4242.4 | 66 | 25.9 | h100_40_2_2 | 5324.2 | 5344.4 | 63 | 13.4 |
| h100_20_2_3 | 4229.5 | 4231.1 | 64 | 27.1 | h100_40_2_3 | 5311.0 | 5323.8 | 61 | 17.7 |
| h100_20_2_4 | 4252.2 | 4290.1 | 67 | 25.4 | h100_40_2_4 | 5293.3 | 5404.9 | 64 | 11.8 |

Table F.5. ANOVA table for analyzing the contribution of DP.

| *Source* | *DF* | *Adj SS* | *Adj MS* | *F-Value* | *p-Value* |
|---|---|---|---|---|---|
| $\boldsymbol{noP}$ | 3 | 2171.55 | 723.85 | 138.08 | 0.000 |
| $\boldsymbol{ra}$ | 3 | 14833.40 | 4944.47 | 943.17 | 0.000 |
| $\boldsymbol{dd}$ | 2 | 1269.85 | 634.92 | 121.11 | 0.000 |
| $\boldsymbol{noP} * \boldsymbol{ra}$ | 9 | 164.41 | 18.27 | 3.48 | 0.001 |
| $\boldsymbol{noP} * \boldsymbol{dd}$ | 6 | 94.16 | 15.69 | 2.99 | 0.008 |
| $\boldsymbol{ra} * \boldsymbol{dd}$ | 6 | 24.24 | 4.04 | 0.77 | 0.594 |
| $\boldsymbol{no} * \boldsymbol{ra} * \boldsymbol{dd}$ | 18 | 54.51 | 3.03 | 0.58 | 0.913 |
| ***Error*** | 192 | 1006.54 | 5.24 | | |
| ***Total*** | 239 | 19618.65 | | | |

## Appendix G The effect of vehicle sharing with DP policy and the HHSRP-STD solutions

In section 5.6, we discussed the effect of the vehicle sharing with DP policy on total flow time by comparing the solutions of the HHSRP-STD with the HHSRP-VS. In this appendix, Tables G.1 through G.4 demonstrates the solutions obtained by the ALNS-STD. In the following tables, the “Best-found” and “Avg.” columns indicate the objective values of the best-found and the averages of the best solutions found in five replications by the ALNS-STD, respectively. The column “ALNS-VS” shows the best-found solution by the ALNS-VS.

The column “ADD” presents the increase in total working time of the caregivers caused by the vehicle sharing with DP policy, which is simple the different between the best-found solutions of ALNS-STD and ALNS-VS. Moreover, the column “$BER$" demonstrates the break-even ratios.

Table G.1 ALNS-STD solutions and break-even ratios for 10-patient instances.

| Instance | Best-found | Avg. | ALNS-VS | ADD | $BER$ | Instance | Best-found | Avg. | ALNS-VS | ADD | $BER$ |
|---|---|---|---|---|---|---|---|---|---|---|---|
| h10_10_0_0 | 174.62 | 174.62 | 240.54 | 65.92 | 1.2 | h10_30_0_0 | 296.95 | 299.102 | 472.46 | 175.51 | 2.9 |
| h10_10_0_1 | 178.48 | 178.48 | 254.28 | 75.8 | 1.5 | h10_30_0_1 | 311.45 | 311.45 | 527.26 | 215.81 | 4.5 |
| h10_10_0_2 | 177.72 | 177.72 | 232.1 | 54.38 | 0.9 | h10_30_0_2 | 307.78 | 308.232 | 500.28 | 192.5 | 3.3 |
| h10_10_0_3 | 176.36 | 176.36 | 255.76 | 79.4 | 1.6 | h10_30_0_3 | 301.45 | 301.45 | 504.56 | 203.11 | 4.1 |
| h10_10_0_4 | 176.15 | 177.454 | 217.24 | 41.09 | 0.6 | h10_30_0_4 | 306.31 | 314.118 | 449.6 | 143.29 | 1.8 |
| h10_10_1_0 | 195.87 | 196.798 | 256.26 | 60.39 | 0.9 | h10_30_1_0 | 318.95 | 321.102 | 519.86 | 200.91 | 3.4 |
| h10_10_1_1 | 200.48 | 200.48 | 281.98 | 81.5 | 1.4 | h10_30_1_1 | 333.45 | 333.45 | 544.58 | 211.13 | 3.5 |
| h10_10_1_2 | 199.72 | 199.72 | 243.04 | 43.32 | 0.6 | h10_30_1_2 | 329.78 | 330.684 | 554.74 | 224.96 | 4.3 |
| h10_10_1_3 | 198.36 | 198.36 | 271.96 | 73.6 | 1.2 | h10_30_1_3 | 323.45 | 323.45 | 527.82 | 204.37 | 3.4 |
| h10_10_1_4 | 198.15 | 198.15 | 237.54 | 39.39 | 0.5 | h10_30_1_4 | 328.31 | 331.506 | 484.16 | 155.85 | 1.8 |
| h10_10_2_0 | 226.87 | 227.17 | 278.02 | 51.15 | 0.6 | h10_30_2_0 | 349.95 | 352.102 | 538.44 | 188.49 | 2.3 |
| h10_10_2_1 | 231.48 | 231.48 | 308.08 | 76.6 | 1.0 | h10_30_2_1 | 364.45 | 364.45 | 607.64 | 243.19 | 4.0 |
| h10_10_2_2 | 230.72 | 230.72 | 276.56 | 45.84 | 0.5 | h10_30_2_2 | 360.78 | 360.78 | 577.62 | 216.84 | 3.0 |
| h10_10_2_3 | 229.36 | 229.36 | 306.36 | 77.0 | 1.0 | h10_30_2_3 | 354.45 | 354.45 | 548.68 | 194.23 | 2.4 |
| h10_10_2_4 | 229.15 | 230.454 | 279.24 | 50.09 | 0.6 | h10_30_2_4 | 359.31 | 367.118 | 496.08 | 136.77 | 1.2 |
| h10_20_0_0 | 236 | 237.986 | 371.66 | 135.66 | 2.7 | h10_40_0_0 | 358.74 | 360.948 | 567.26 | 208.52 | 2.8 |
| h10_20_0_1 | 243.81 | 243.81 | 401.04 | 157.23 | 3.6 | h10_40_0_1 | 380.51 | 380.51 | 644.64 | 264.13 | 4.5 |
| h10_20_0_2 | 242.07 | 242.07 | 369.14 | 127.07 | 2.2 | h10_40_0_2 | 371.65 | 371.818 | 614.5 | 242.85 | 3.8 |
| h10_20_0_3 | 236.34 | 236.34 | 384.52 | 148.18 | 3.4 | h10_40_0_3 | 364.31 | 364.31 | 619.24 | 254.93 | 4.7 |
| h10_20_0_4 | 240.78 | 242.842 | 339.46 | 98.68 | 1.4 | h10_40_0_4 | 369.18 | 369.18 | 551.8 | 182.62 | 2.0 |
| h10_20_1_0 | 258 | 259.344 | 379.22 | 121.22 | 1.8 | h10_40_1_0 | 383.5 | 383.5 | 614.74 | 231.24 | 3.0 |
| h10_20_1_1 | 265.81 | 265.81 | 413.12 | 147.31 | 2.5 | h10_40_1_1 | 402.51 | 402.51 | 665.44 | 262.93 | 3.8 |
| h10_20_1_2 | 264.07 | 264.07 | 391.6 | 127.53 | 1.9 | h10_40_1_2 | 393.65 | 393.734 | 678.02 | 284.37 | 5.2 |
| h10_20_1_3 | 258.34 | 258.34 | 397.7 | 139.36 | 2.3 | h10_40_1_3 | 386.31 | 386.31 | 650.3 | 263.99 | 4.3 |
| h10_20_1_4 | 262.78 | 266.904 | 356.78 | 94.0 | 1.1 | h10_40_1_4 | 391.18 | 391.18 | 600.48 | 209.3 | 2.3 |
| h10_20_2_0 | 290.68 | 291.322 | 440.16 | 149.48 | 2.1 | h10_40_2_0 | 411.74 | 413.396 | 640.18 | 228.44 | 2.5 |
| h10_20_2_1 | 296.81 | 296.81 | 466.62 | 169.81 | 2.7 | h10_40_2_1 | 433.51 | 433.51 | 743.28 | 309.77 | 5.0 |
| h10_20_2_2 | 295.07 | 295.07 | 437.02 | 141.95 | 1.9 | h10_40_2_2 | 424.65 | 424.734 | 704.28 | 279.63 | 3.9 |
| h10_20_2_3 | 289.34 | 289.34 | 433.98 | 144.64 | 2.0 | h10_40_2_3 | 417.31 | 417.31 | 666.2 | 248.89 | 3.0 |
| h10_20_2_4 | 293.78 | 298.466 | 380.64 | 86.86 | 0.8 | h10_40_2_4 | 422.18 | 422.18 | 600.7 | 178.52 | 1.5 |

Table G.2 ALNS-STD solutions and break-even ratios for 30-patient instances.

| Instance | Best-found | Avg. | ALNS-VS | ADD | $BER$ | Instance | Best-found | Avg. | ALNS-VS | ADD | $BER$ |
|---|---|---|---|---|---|---|---|---|---|---|---|
| h30_10_0_0 | 453 | 458.258 | 548.5 | 95.5 | 0.5 | h30_30_0_0 | 660.92 | 674.008 | 959.54 | 298.62 | 1.6 |
| h30_10_0_1 | 463.61 | 465.666 | 571.6 | 107.99 | 0.6 | h30_30_0_1 | 695.92 | 701.784 | 1039.02 | 343.1 | 1.9 |

| h30_10_0_2 | 462.48 | 462.48 | 552.34 | 89.86 | 0.5 | h30_30_0_2 | 682.78 | 682.78 | 1053.54 | 370.76 | 2.4 |
|---|---|---|---|---|---|---|---|---|---|---|---|
| h30_10_0_3 | 459.45 | 459.45 | 571.22 | 111.77 | 0.6 | h30_30_0_3 | 670.36 | 675.106 | 989.12 | 318.76 | 1.8 |
| h30_10_0_4 | 459.04 | 460.506 | 559.58 | 100.54 | 0.6 | h30_30_0_4 | 670.21 | 673.722 | 967.3 | 297.09 | 1.6 |
| h30_10_1_0 | 495.53 | 498.028 | 604.16 | 108.63 | 0.6 | h30_30_1_0 | 701.5 | 711.156 | 1045.36 | 343.86 | 1.9 |
| h30_10_1_1 | 507.18 | 507.18 | 614.22 | 107.04 | 0.5 | h30_30_1_1 | 739.49 | 742.474 | 1106.12 | 366.63 | 2.0 |
| h30_10_1_2 | 503.48 | 503.48 | 604.38 | 100.9 | 0.5 | h30_30_1_2 | 723.78 | 723.78 | 1047.8 | 324.02 | 1.6 |
| h30_10_1_3 | 500.45 | 500.45 | 597.14 | 96.69 | 0.5 | h30_30_1_3 | 714.99 | 717.03 | 996.64 | 281.65 | 1.3 |
| h30_10_1_4 | 500.18 | 503.552 | 597.98 | 97.8 | 0.5 | h30_30_1_4 | 715.39 | 715.96 | 1008.44 | 293.05 | 1.4 |
| h30_10_2_0 | 570.34 | 570.618 | 664.76 | 94.42 | 0.4 | h30_30_2_0 | 777.77 | 782.312 | 1069.3 | 291.53 | 1.2 |
| h30_10_2_1 | 579.18 | 579.18 | 665.44 | 86.26 | 0.3 | h30_30_2_1 | 808.21 | 814.642 | 1163.3 | 355.09 | 1.6 |
| h30_10_2_2 | 575.48 | 575.48 | 671.58 | 96.1 | 0.4 | h30_30_2_2 | 795.78 | 795.78 | 1144.22 | 348.44 | 1.6 |
| h30_10_2_3 | 572.45 | 572.45 | 652.6 | 80.15 | 0.3 | h30_30_2_3 | 782.59 | 788.15 | 1057.34 | 274.75 | 1.1 |
| h30_10_2_4 | 572.04 | 574.7 | 673.22 | 101.18 | 0.4 | h30_30_2_4 | 783.21 | 790.366 | 1074.72 | 291.51 | 1.2 |
| h30_20_0_0 | 558.61 | 568.012 | 769.9 | 211.29 | 1.2 | h30_40_0_0 | 766.48 | 781.078 | 1163.52 | 397.04 | 2.1 |
| h30_20_0_1 | 584.3 | 585.54 | 795.58 | 211.28 | 1.1 | h30_40_0_1 | 825.43 | 825.43 | 1249.28 | 423.85 | 2.1 |
| h30_20_0_2 | 567.59 | 567.59 | 818.36 | 250.77 | 1.6 | h30_40_0_2 | 795.25 | 795.25 | 1274.94 | 479.69 | 3.0 |
| h30_20_0_3 | 565.78 | 565.78 | 767.94 | 202.16 | 1.1 | h30_40_0_3 | 774.98 | 783.276 | 1179.64 | 404.66 | 2.2 |
| h30_20_0_4 | 568.84 | 569.57 | 777.32 | 208.48 | 1.2 | h30_40_0_4 | 783.67 | 785.854 | 1190.08 | 406.41 | 2.2 |
| h30_20_1_0 | 599.59 | 606.06 | 816.72 | 217.13 | 1.1 | h30_40_1_0 | 813.49 | 824.638 | 1253.14 | 439.65 | 2.4 |
| h30_20_1_1 | 626.85 | 626.85 | 866.12 | 239.27 | 1.2 | h30_40_1_1 | 854.28 | 863.888 | 1328.04 | 473.76 | 2.5 |
| h30_20_1_2 | 608.59 | 608.59 | 808.28 | 199.69 | 1.0 | h30_40_1_2 | 836.25 | 836.25 | 1273.12 | 436.87 | 2.2 |
| h30_20_1_3 | 602.05 | 605.834 | 785.9 | 183.85 | 0.9 | h30_40_1_3 | 817.26 | 824.532 | 1224.14 | 406.88 | 2.0 |
| h30_20_1_4 | 609.84 | 610.996 | 808.82 | 198.98 | 1.0 | h30_40_1_4 | 824.67 | 826.168 | 1217.12 | 392.45 | 1.8 |
| h30_20_2_0 | 681.03 | 682.112 | 874.98 | 193.95 | 0.8 | h30_40_2_0 | 886.1 | 898.558 | 1298.9 | 412.8 | 1.7 |
| h30_20_2_1 | 698.62 | 698.804 | 923.62 | 225 | 1.0 | h30_40_2_1 | 938.82 | 941.416 | 1413.4 | 474.58 | 2.0 |
| h30_20_2_2 | 680.59 | 680.59 | 894.72 | 214.13 | 0.9 | h30_40_2_2 | 908.25 | 908.25 | 1348.32 | 440.07 | 1.9 |
| h30_20_2_3 | 678.78 | 678.78 | 849.02 | 170.24 | 0.7 | h30_40_2_3 | 889.25 | 896.53 | 1306.24 | 416.99 | 1.8 |
| h30_20_2_4 | 681.84 | 681.98 | 883.88 | 202.04 | 0.8 | h30_40_2_4 | 900.97 | 905.218 | 1247.14 | 346.17 | 1.2 |

Table G.3. ALNS-STD solutions and break-even ratios for 50-patient instances.

| **Instance** | **Best-found** | **Avg.** | **ALNS-VS** | **ADD** | ***BER*** | **Instance** | **Best-found** | **Avg.** | **ALNS-VS** | **ADD** | ***BER*** |
|---|---|---|---|---|---|---|---|---|---|---|---|
| h50_10_0_0 | 746.2 | 748.994 | 909.82 | 163.62 | 0.6 | h50_30_0_0 | 1127.57 | 1134.43 | 1569.1 | 441.53 | 1.3 |
| h50_10_0_1 | 737.77 | 740.464 | 930.92 | 193.15 | 0.7 | h50_30_0_1 | 1113.81 | 1114.642 | 1598.98 | 485.17 | 1.5 |
| h50_10_0_2 | 750.24 | 751.108 | 942.04 | 191.8 | 0.7 | h50_30_0_2 | 1146.2 | 1148.628 | 1679.3 | 533.1 | 1.7 |
| h50_10_0_3 | 728.15 | 728.83 | 889.4 | 161.25 | 0.6 | h50_30_0_3 | 1083.84 | 1083.84 | 1504.66 | 420.82 | 1.3 |
| h50_10_0_4 | 743.15 | 743.412 | 913.76 | 170.61 | 0.6 | h50_30_0_4 | 1126.6 | 1126.748 | 1621.08 | 494.48 | 1.6 |
| h50_10_1_0 | 883.15 | 885.794 | 1070.26 | 187.11 | 0.5 | h50_30_1_0 | 1264.09 | 1270.016 | 1723.9 | 459.81 | 1.1 |
| h50_10_1_1 | 877.17 | 879.13 | 1072.84 | 195.67 | 0.6 | h50_30_1_1 | 1245.67 | 1249.818 | 1746.08 | 500.41 | 1.3 |
| h50_10_1_2 | 886.61 | 888.376 | 1053.3 | 166.69 | 0.5 | h50_30_1_2 | 1282.39 | 1286.464 | 1808.54 | 526.15 | 1.4 |
| h50_10_1_3 | 866.48 | 866.726 | 1021 | 154.52 | 0.4 | h50_30_1_3 | 1220.84 | 1220.84 | 1625.76 | 404.92 | 1.0 |
| h50_10_1_4 | 880.15 | 880.15 | 1030.02 | 149.87 | 0.4 | h50_30_1_4 | 1263.6 | 1263.748 | 1744.46 | 480.86 | 1.2 |
| h50_10_2_0 | 1035.15 | 1039.146 | 1211.42 | 176.27 | 0.4 | h50_30_2_0 | 1415.99 | 1417.878 | 1836.42 | 420.43 | 0.8 |
| h50_10_2_1 | 1028.57 | 1029.176 | 1219.42 | 190.85 | 0.5 | h50_30_2_1 | 1398.58 | 1400.946 | 1938.74 | 540.16 | 1.3 |
| h50_10_2_2 | 1038.61 | 1040.694 | 1216.24 | 177.63 | 0.4 | h50_30_2_2 | 1435.2 | 1437.736 | 1980.06 | 544.86 | 1.2 |
| h50_10_2_3 | 1016.5 | 1017.502 | 1183 | 166.5 | 0.4 | h50_30_2_3 | 1372.84 | 1372.84 | 1764.08 | 391.24 | 0.8 |
| h50_10_2_4 | 1032.15 | 1032.462 | 1207.66 | 175.51 | 0.4 | h50_30_2_4 | 1415.97 | 1415.972 | 1931.36 | 515.39 | 1.1 |

Table G.3. ALNS-STD solutions and break-even ratios for 50-patient instances (cont.).

| Instance | Best-found | Avg. | ALNS-VS | ADD | $BER$ | Instance | Best-found | Avg. | ALNS-VS | ADD | $BER$ |
|---|---|---|---|---|---|---|---|---|---|---|---|
| h50_20_0_0 | 937.06 | 943.964 | 1267.5 | 330.44 | 1.1 | h50_40_0_0 | 1320.89 | 1331.308 | 1871.68 | 550.79 | 1.4 |
| h50_20_0_1 | 926.25 | 927.85 | 1299.14 | 372.89 | 1.3 | h50_40_0_1 | 1312.33 | 1314.804 | 1971.42 | 659.09 | 2.0 |
| h50_20_0_2 | 945.51 | 947.582 | 1301.56 | 356.05 | 1.2 | h50_40_0_2 | 1344.03 | 1347 | 2012.02 | 667.99 | 2.0 |
| h50_20_0_3 | 902.41 | 904.42 | 1176.6 | 274.19 | 0.9 | h50_40_0_3 | 1260.57 | 1260.57 | 1774.16 | 513.59 | 1.4 |
| h50_20_0_4 | 932.87 | 932.992 | 1280.86 | 347.99 | 1.2 | h50_40_0_4 | 1316.06 | 1316.356 | 1950.78 | 634.72 | 1.9 |
| h50_20_1_0 | 1074.06 | 1078.708 | 1406.52 | 332.46 | 0.9 | h50_40_1_0 | 1455.25 | 1461.442 | 2047.46 | 592.21 | 1.4 |
| h50_20_1_1 | 1062.65 | 1065.992 | 1382.12 | 319.47 | 0.9 | h50_40_1_1 | 1442.72 | 1449.804 | 2071.16 | 628.44 | 1.5 |
| h50_20_1_2 | 1081.11 | 1084.184 | 1465.96 | 384.85 | 1.1 | h50_40_1_2 | 1480.74 | 1485.04 | 2144.36 | 663.62 | 1.6 |
| h50_20_1_3 | 1041.12 | 1041.964 | 1331.9 | 290.78 | 0.8 | h50_40_1_3 | 1397.57 | 1397.57 | 1914.2 | 516.63 | 1.2 |
| h50_20_1_4 | 1069.87 | 1070.314 | 1383.84 | 313.97 | 0.8 | h50_40_1_4 | 1456.25 | 1456.724 | 2080.46 | 624.21 | 1.5 |
| h50_20_2_0 | 1232.82 | 1234.888 | 1591.36 | 358.54 | 0.8 | h50_40_2_0 | 1607.25 | 1614.2 | 2287.3 | 680.05 | 1.5 |
| h50_20_2_1 | 1214.65 | 1217.674 | 1588.36 | 373.71 | 0.9 | h50_40_2_1 | 1596.03 | 1605.192 | 2292.64 | 696.61 | 1.5 |
| h50_20_2_2 | 1232.91 | 1235.188 | 1602.74 | 369.83 | 0.9 | h50_40_2_2 | 1629.6 | 1639.234 | 2352.22 | 722.62 | 1.6 |
| h50_20_2_3 | 1191.41 | 1192.21 | 1465.66 | 274.25 | 0.6 | h50_40_2_3 | 1549.86 | 1550.606 | 2056.96 | 507.1 | 1.0 |
| h50_20_2_4 | 1221.87 | 1222.142 | 1544.46 | 322.59 | 0.7 | h50_40_2_4 | 1608.71 | 1608.71 | 2290.88 | 682.17 | 1.5 |

Table G.4. ALNS-STD solutions and break-even ratios for 100-patient instances.

| Instance | Best-found | Avg. | ALNS-VS | ADD | $BER$ | Instance | Best-found | Avg. | ALNS-VS | ADD | $BER$ |
|---|---|---|---|---|---|---|---|---|---|---|---|
| h100_10_0_0 | 1581.67 | 1589.786 | 1975.94 | 394.27 | 0.7 | h100_30_0_0 | 2365.8 | 2373.264 | 3273.16 | 907.36 | 1.2 |
| h100_10_0_1 | 1562.52 | 1565.392 | 2051.56 | 489.04 | 0.9 | h100_30_0_1 | 2330.76 | 2339.942 | 3398.24 | 1067.48 | 1.7 |
| h100_10_0_2 | 1580.38 | 1585.93 | 2041.64 | 461.26 | 0.8 | h100_30_0_2 | 2370.46 | 2382.284 | 3241.28 | 870.82 | 1.2 |
| h100_10_0_3 | 1578.2 | 1580.2 | 2013.56 | 435.36 | 0.8 | h100_30_0_3 | 2352.21 | 2373.962 | 3191.18 | 838.97 | 1.1 |
| h100_10_0_4 | 1601.76 | 1604.676 | 1967.94 | 366.18 | 0.6 | h100_30_0_4 | 2413.79 | 2422.82 | 3398.52 | 984.73 | 1.4 |
| h100_10_1_0 | 1914.49 | 1923.402 | 2382.64 | 468.15 | 0.6 | h100_30_1_0 | 2688.06 | 2712.556 | 3702.12 | 1014.06 | 1.2 |
| h100_10_1_1 | 1902.53 | 1905.818 | 2326.82 | 424.29 | 0.6 | h100_30_1_1 | 2668.54 | 2686.552 | 3713.72 | 1045.18 | 1.3 |
| h100_10_1_2 | 1917.37 | 1921.808 | 2388.9 | 471.53 | 0.7 | h100_30_1_2 | 2698.58 | 2711.902 | 3647.28 | 948.7 | 1.1 |
| h100_10_1_3 | 1920.82 | 1924.34 | 2327.8 | 406.98 | 0.5 | h100_30_1_3 | 2700.64 | 2713.372 | 3586.44 | 885.8 | 1.0 |
| h100_10_1_4 | 1945.25 | 1947.224 | 2377.38 | 432.13 | 0.6 | h100_30_1_4 | 2756.39 | 2767.222 | 3735.9 | 979.51 | 1.1 |
| h100_10_2_0 | 1999.24 | 2007.968 | 2492.14 | 492.9 | 0.7 | h100_30_2_0 | 2774.43 | 2785.114 | 3875.58 | 1101.15 | 1.3 |
| h100_10_2_1 | 1985.6 | 1988.07 | 2413.36 | 427.76 | 0.5 | h100_30_2_1 | 2761.99 | 2768.014 | 3829.38 | 1067.39 | 1.3 |
| h100_10_2_2 | 2000.53 | 2004.33 | 2480.8 | 480.27 | 0.6 | h100_30_2_2 | 2770.29 | 2796.816 | 3692.94 | 922.65 | 1.0 |
| h100_10_2_3 | 1998.01 | 2003.274 | 2457.34 | 459.33 | 0.6 | h100_30_2_3 | 2781.12 | 2793.006 | 3736.7 | 955.58 | 1.0 |
| h100_10_2_4 | 2014.43 | 2023.398 | 2477.34 | 462.91 | 0.6 | h100_30_2_4 | 2822.4 | 2839.694 | 3783.46 | 961.06 | 1.0 |
| h100_20_0_0 | 1969.03 | 1981.914 | 2657.8 | 688.77 | 1.1 | h100_40_0_0 | 2746.65 | 2768.802 | 3872.08 | 1125.43 | 1.4 |
| h100_20_0_1 | 1934.91 | 1948.976 | 2681.96 | 747.05 | 1.3 | h100_40_0_1 | 2689.58 | 2713.964 | 3937.38 | 1247.8 | 1.7 |
| h100_20_0_2 | 1970.49 | 1979.106 | 2611.18 | 640.69 | 1.0 | h100_40_0_2 | 2772.4 | 2788.034 | 4027.06 | 1254.66 | 1.7 |
| h100_20_0_3 | 1969.34 | 1977.528 | 2629.6 | 660.26 | 1.0 | h100_40_0_3 | 2758.07 | 2768.694 | 3882.64 | 1124.57 | 1.4 |
| h100_20_0_4 | 1994.69 | 2011.222 | 2717.9 | 723.21 | 1.1 | h100_40_0_4 | 2813.14 | 2845.296 | 4122.08 | 1308.94 | 1.7 |
| h100_20_1_0 | 2304.3 | 2325.934 | 3024.92 | 720.62 | 0.9 | h100_40_1_0 | 3091.39 | 3099.684 | 4447.4 | 1356.01 | 1.6 |
| h100_20_1_1 | 2281.08 | 2293.752 | 3003.9 | 722.82 | 0.9 | h100_40_1_1 | 3061.43 | 3079.758 | 4386.1 | 1324.67 | 1.5 |
| h100_20_1_2 | 2311.35 | 2319.314 | 3036.4 | 725.05 | 0.9 | h100_40_1_2 | 3120.62 | 3139.544 | 4396.34 | 1275.72 | 1.4 |
| h100_20_1_3 | 2303.47 | 2315.992 | 3078.32 | 774.85 | 1.0 | h100_40_1_3 | 3090.52 | 3098.836 | 4295.88 | 1205.36 | 1.3 |
| h100_20_1_4 | 2340.27 | 2346.596 | 3090.96 | 750.69 | 0.9 | h100_40_1_4 | 3167.35 | 3179.958 | 4479.1 | 1311.75 | 1.4 |
| h100_20_2_0 | 2393.9 | 2406.694 | 3175.98 | 782.08 | 1.0 | h100_40_2_0 | 3171.75 | 3193.064 | 4545.96 | 1374.21 | 1.5 |
| h100_20_2_1 | 2362.11 | 2370.542 | 3134.16 | 772.05 | 1.0 | h100_40_2_1 | 3136.95 | 3164.082 | 4480.66 | 1343.71 | 1.5 |
| h100_20_2_2 | 2399.61 | 2403.952 | 3139.24 | 739.63 | 0.9 | h100_40_2_2 | 3188.39 | 3212.272 | 4612.04 | 1423.65 | 1.6 |
| h100_20_2_3 | 2385.63 | 2400.682 | 3082.96 | 697.33 | 0.8 | h100_40_2_3 | 3151.92 | 3170.768 | 4369.82 | 1217.9 | 1.3 |
| h100_20_2_4 | 2417.54 | 2427.548 | 3173.98 | 756.44 | 0.9 | h100_40_2_4 | 3220.08 | 3245.284 | 4668.38 | 1448.3 | 1.6 |

Table G.5. The ANOVA table for break-even ratios.

| *Source* | *DF* | *Adj SS* | *Adj MS* | *F-Value* | *p-Value* |
|---|---|---|---|---|---|
| $\boldsymbol{noP}$ | 3 | 73.65 | 24.55 | 111.17 | 0.000 |
| $\boldsymbol{ra}$ | 3 | 76.40 | 25.46 | 115.32 | 0.000 |
| $\boldsymbol{dd}$ | 2 | 5.53 | 2.76 | 12.51 | 0.000 |
| $\boldsymbol{noP * ra}$ | 9 | 16.84 | 1.87 | 8.47 | 0.001 |
| $\boldsymbol{noP * dd}$ | 6 | 1.26 | 0.21 | 0.95 | 0.460 |
| $\boldsymbol{ra * dd}$ | 6 | 0.61 | 0.10 | 0.46 | 0.835 |
| $\boldsymbol{no * ra * dd}$ | 18 | 1.24 | 0.07 | 0.31 | 0.997 |
| *Error* | 192 | 42.40 | 0.22 | | |
| *Total* | 239 | 217.92 | | | |

Table G.6. Multiple comparison test results for $BER$ according to the problem features.

| $noP$ | Mean $BER$ | $ra$ | Mean $BER$ | $dd$ | Mean $BER$ |
|---|---|---|---|---|---|
| 10 | 2.41 | 40 | 2.15 | 0 | 1.65 |
| 30 | 1.30 | 30 | 1.78 | 1 | 1.46 |
| 100 | 1.08 | 20 | 1.28 | 2 | 1.28 |
| 50 | 1.06 | 10 | 0.64 | | |